\documentclass[aps,superscriptaddress,prd, twocolumn,showpacs,preprintnumbers,nofootinbib, amsmath,amssymb,
aps, superscriptaddress]{revtex4-1} %prl

\usepackage{physics}
\usepackage{graphicx}
\usepackage{amsmath,amssymb}
\usepackage{amsfonts}
\usepackage{xspace} 
\usepackage[usenames]{color}
\usepackage{dcolumn}
\usepackage{bm}
\usepackage{mathrsfs}
\usepackage{multirow}
\usepackage[colorlinks=true]{hyperref}
\usepackage[all]{hypcap} 
\usepackage[utf8]{inputenc} 
\usepackage{lipsum}
\usepackage{etoolbox}
\usepackage{tikz}
\usepackage{url}
\usepackage{subfigure}
\usepackage{adjustbox}
\usepackage{caption}
\usepackage{slashed}
\usepackage{multirow}
\usepackage{array}
\usepackage{booktabs}
\usepackage{siunitx}
\usepackage{verbatim}
\usepackage{enumitem}
\usepackage{lineno}

\usepackage[normalem]{ulem} 

\usepackage{orcidlink}

\newcommand{\Kc}{K_{\textrm{\tiny CMC}}}

\def\be{\begin{equation}}
\def\ee{\end{equation}}
\def\bea{\begin{eqnarray}}
\def\eea{\end{eqnarray}}

\newcommand{\beq}{\begin{eqnarray}}
\newcommand{\eeq}{\end{eqnarray}}
\usepackage{calrsfs}
\DeclareMathAlphabet{\pazocal}{OMS}{zplm}{m}{n}

\newcommand{\adr}{\textcolor{black}}
\newcommand{\avv}{\textcolor{black}}%{\textcolor{blue}} % changed colour for the corrections
\newcommand{\pbb}{\textcolor{black}}
\newcommand{\scri}{\mathrsfs{I}}
\newcommand{\p}[1]{\partial#1}

\begin{document}

%\title{A numerical approach to Hawking radiation in analogue-gravity models}

%\title{Particle creation from numerical toy model as test for Hawking radiation}

\title{A numerical approach to particle creation in accelerating toy models}

\author{Pedro Duarte-Baptista\,\orcidlink{0009-0007-0568-5239}} \email{pedroduartebaptista@tecnico.ulisboa.pt}
\affiliation{CENTRA, Departamento de F\'{\i}sica, Instituto Superior T\'ecnico -- IST, Universidade de Lisboa -- UL,
Avenida Rovisco Pais 1, 1049 Lisboa, Portugal}

\author{Alex Va\~{n}\'o-Vi\~{n}uales\,\orcidlink{0000-0002-8589-006X}}\email{alex.vano@uib.es}
\affiliation{CENTRA, Departamento de F\'{\i}sica, Instituto Superior T\'ecnico -- IST, Universidade de Lisboa -- UL,
Avenida Rovisco Pais 1, 1049 Lisboa, Portugal}
\affiliation{Departament de Física, Universitat de les Illes Balears, IAC3, Carretera Valldemossa km 7.5, E-07122 Palma, Spain}

\author{Adri\'an del R\'{\i}o\,\orcidlink{0000-0002-9978-2211}}
\email{adrdelri@math.uc3m.es}
  \affiliation{Universidad Carlos III de Madrid, Departamento de Matem\'aticas.\\ Avenida de la Universidad 30 (edificio Sabatini), 28911 Legan\'es (Madrid), Spain.}
\begin{abstract}

The formation of black holes by the gravitational collapse of stars  is known to spontaneously excite particle pairs out of the quantum vacuum. 
For the canonical vacuum state  at past null infinity, the expected number of particles received at future null infinity 
%and the frequency spectrum 
can  be obtained in full closed form at sufficiently late times. %and only depends on the properties of the final black hole. 
However, for intermediate times, or for more complicated astrophysical processes (e.g. binary black hole mergers), the problem is technically challenging and has not yet been resolved. %numerical techniques are required. 
We develop here a numerical approach to study  scattering problems of massless quantum fields in asymptotically flat spacetimes, based on the hyperboloidal slice method used in numerical relativity and perturbation theory. This promising approach can reach both past and future null infinities, and therefore it has the potential to address the Hawking scattering problem  more rigorously than  evolution on the usual Cauchy slices. {In this first work}, we test this approach with some dynamical toy models in Minkowski using effective potentials that mimic the effects of gravity, and compute the spectrum of particles received at future null infinity. We finally discuss future prospects for applying this framework in more relevant gravitational scenarios. %realistic astrophysical scenarios. %for (i) Minkowski spacetime. 

%We develop a numerical approach to evaluate the particle creation 

{\let\clearpage\relax  \let\sectionname\relax \tableofcontents}
\end{abstract}
\maketitle

\section{Introduction}

Strong  gravitational fields can  alter vacuum fluctuations of quantum fields in such a way that particle pairs can be physically created in the spacetime. This effect  was first predicted in expanding universes in the pioneering works by Parker \cite{PhysRev.183.1057, PhysRevD.3.346}, and lays the physical foundations in modern cosmology for generating cosmic anisotropies in the early inflationary universe \cite{Mukhanov:1981xt, HAWKING1982295, PhysRevLett.49.1110, STAROBINSKY1982175, PhysRevD.28.679}. Similarly, this quantum mechanism is  pivotal in our understanding of black holes. As found by Hawking \cite{Hawking:1974rv, cmp/1103899181}, when a star undergoes gravitational collapse and forms a black hole, an observer at infinity will receive, at sufficiently late times, a thermal flux of particles \cite{cmp/1103899393, PhysRevD.12.1519}, whose temperature is fixed by the  parameters of the final black hole. This interplay between quantum theory, thermodynamics and gravity allowed a statistical-mechanics interpretation of the horizon area in terms %of entropy,
of microscopic degrees of freedom,
 which has been greatly influential in modern research on quantum gravity \cite{Wald:1999vt}.

While the implications of particle creation can be easily estimated in cosmological backgrounds, using merely analytical or simple numerical calculations \cite{Parker:2009uva}, for black holes   explicit results are only available for gravitational collapse and, furthermore, only for sufficiently late times, when the black hole has formed and settled down to equilibrium\footnote{%{\color{red}
For some quantum-information related studies, see \cite{PhysRevD.104.105020, PhysRevD.107.104004, PhysRevD.111.044040, BELFIGLIO20251, doi:10.1142/S0219887819501147}.}%}
. The reason is simple. First, in cosmology the background spacetime is determined by the Friedmann-Lemaitre-Robertson-Walker family of spacetime metrics. These are highly symmetric (homogeneous and isotropic) and conformally static backgrounds, on which the field equations, e.g. the Klein-Gordon (KG) equation, reduce to an ordinary  differential equation for the time evolution of the field modes.  All the gravitational dynamics is encoded in the cosmological scale factor, and the initial data (i.e., choice of the quantum state) can be specified on a 3-dimensional spacelike Cauchy hypersurface, where the spacetime is presumably stationary. In sharp contrast, the spacetime of a collapsing star is significantly more complex. In this case, the background {metric} is unknown a priori, and needs to be solved first using numerical techniques {given a physical model for the star}. In particular, the metric is not simply determined by one function, but it depends on all the parameters and functions modeling the physics of the star, as well as the exterior vacuum spacetime if the star is rotating. {In addition}, the spacetime is only asymptotically flat, and the initial quantum state needs to be specified on past null infinity, a 3-dimensional null hypersurface where all radially ingoing null geodesics asympotically stem from. %, together with regularity conditions at the origin of the star.  
This initial data has then to be evolved through the dynamical spacetime, and evaluated back at future null infinity after scattering with the collapsing star. {Finally}, the presence of an event horizon adds another layer of difficulty, as part of the information  falls through towards the singularity.   

For sufficiently late times, one can estimate the flux of particles received at future null infinity by approximating the {evolution of the} field modes by null geodesics, and tracking back their propagation along the collapsing scenario \cite{cmp/1103899181, doi:10.1142/p378}. In this case, the extremely high redshift experienced by these curves after passing very closely by the event horizon, screens all  physical details concerning the collapsing star. As a result, the frequency spectrum only depends on the final  state of the background, i.e., on the mass and angular momentum of the resulting black hole.  %essentially due to Hawking, for .  
This peculiarity is what allows us to obtain a %result in full closed form, and the thermal 
specific prediction \adr{regardless of the particular physical model of the star}, which is robust despite the approximations involved \cite{cmp/1104180137}.

Because theoretical derivations of particle creation are only possible under idealized settings, simulating quantum field effects numerically ---even with simplified models--- can give insights on the nature of particle creation and spark new questions in the field of semiclassical gravity. As a matter of fact, if one wishes to study the process of particle creation during the full process of gravitational collapse, and explore what is the full spectrum, including dependence with the details of the star, a numerical framework is needed. A  similar framework is  required if one poses the question of particle creation by e.g. mergers of compact astrophysical binaries. In this article, we try to take a step forward in this direction. 

Unfortunately, numerical simulations involving QFT in curved spacetimes can be notoriously challenging. In particular, to carry out this program we need to set out  a  scattering problem for a given field equation  in asymptotically flat spacetimes \cite{Geroch1977}. This is, after fixing initial data for the field modes on past null infinity $\mathcal J^-$, and evolving this data in the background spacetime, we need a way to extract this information back at future null infinity $\mathcal J^+$. Cauchy foliations, traditionally used in numerical relativity \cite{10.1093/acprof:oso/9780199205677.001.0001}, are spacelike hypersurfaces that are unable to reach these asymptotic regions of spacetime where radiation is unambiguously defined. Consequently, these hypersurfaces are inherently unsuitable for radiation extraction without some degree of systematic error, as post processing methods need to be used, such as direct extrapolation or Cauchy-characteristic extraction \cite{Bishop:1996gt,Taylor:2013zia,Moxon:2021gbv}. 

An alternative approach, \adr{available for any asymptotically flat spacetime}, is the evolution on hyperboloidal slices \cite{friedrich1983}, which are spacelike hypersurfaces that asymptotically reach null infinity. Spatially compactifying the slice and using a scri-fixing gauge \cite{Zenginoglu:2007jw} allows us to include the location of $\mathcal J$ in the computational grid.
%is to adopt a scri-fixing gauge, where one compactifies spacetime and fixes the location of $\mathcal J$ in the computational grid. This is typically achieved with the help of a conformal factor $\Omega$, and by foliating the spacetime with hyperboloidal slices - a family of spacelike hypersurfaces that asymptotically reach null infinity. 
%This setup gives access to this assymptotic region of spacetime, and makes possible 
This enables the unambiguous extraction of radiation within a finite grid in a rather simple and elegant setup%\cite{}
. The hyperboloidal approach is widely used in black hole perturbation theory \cite{Zenginoglu:2011zz,Harms:2014dqa,Zenginoglu:2010zm,Ansorg:2016ztf,PanossoMacedo:2018hab,Jaramillo:2020tuu,Destounis:2021lum} and has seen significant progress in numerical relativity \cite{Rinne:2009qx,Vano-Vinuales:2014koa,Morales:2016rgt,Vano-Vinuales:2017qij,Vano-Vinuales:2023pum,Frauendiener:2023ltp,Vano-Vinuales:2024tat,Peterson:2024bxk,Frauendiener:2025xcj}.
This is the approach that we will adopt in this work.

%To carry out this program we need to set out  a  scattering problem for a given field equation  in asymptotically flat spacetimes \cite{Geroch1977}. This is, after fixing initial data for the field modes on past null infinity, and evolving this data in the background spacetime, we need a way to extract this information back at future null infinity. In numerical relativity, a particularly convenient scheme  is provided by the hyperboloidal slice framework \cite{}. This method works by compactifying the spacetime and allows to reach both past and future null infinities. This property makes this method ideal to study the evolution of field modes from past to future null infinities.

Although, from a physical viewpoint, the most interesting problem to study is that of particle creation  of a quantum, massless scalar field in a gravitational collapse scenario, such dynamics imply solving both the massless KG equation and the Einstein Field Equations (EFE). The evolution of  fields in a dynamical spacetime  introduces significant complexity, especially when using hyperboloidal slicings of spacetime, which present additional challenges in propagating radiation from past to future null infinity. Because of this, in this work we will adopt a simplified approach. Specifically, we will simulate the propagation of  massless, scalar field modes in a fixed Minkowski background from $\mathcal J^-$ to $\mathcal J^+$  via hyperboloidal slicing, subject to a suitable effective potential. With the introduction of this potential ---which can be made time-dependent--- we can mimic the necessary gravitational dynamics for producing particle creation \cite{Mart_n_Caro_2024,moore1970quantum,Wilson_2011,physics2010007}. As a result, we can work with a toy model that captures the essential aspects of the phenomenon, while at the same time  keeping calculations relatively simple. By extracting the evolved signal at $\mathcal J^+$ , we will analyze the effects of a dynamical potential in the creation of particles by calculating the Bogoliubov coefficients \cite{birrell_davies_1982}. 

The rest of the article is organized as follows. First of all, in Sec.~\ref{Section:Theor} we review the fundamental theoretical aspects underlying Hawking radiation and particle creation by collapsing stars, including formulas for the Bogoliubov coefficients and related identities. Then, in Sec. \ref{Sec:Methodology} we will describe the methodology of the numerical framework employed, and specify the assumptions. In Sec. \ref{Section:Results}  we will report the numerical results obtained for different stationary and dynamical settings, where we test our numerical approach. Finally, in Sec. \ref{Section: Conclusion} we provide a summary of the key findings, describe limitations, and discuss future lines of work.

As for conventions, we will work with geometrized units, $G=c=1$, and keep  Planck's constant $\hbar$ explicit. The metric signature is fixed to $(-, +, +, +)$, and $\nabla_a$  represents the associated Levi-Civita connection.

%Since the original Hawking calculation involved specifying the vacuum state at past null infinity, a convenient numerical scheme is needed that can manage to fix initial data on this null hypersurface. 

\section{Theoretical framework}
\label{Section:Theor}

In this section we will {review} the basic theoretical tools regarding quantum fields in a curved spacetime. {We will first introduce Bogoliubov transformations and the topic of particle creation,} which will be relevant to understand the numerical procedure to be given later. Some standard references on this subject are \cite{DEWITT1975295, birrell_davies_1982, Fulling_1989, Wald:1995yp, Parker:2009uva, doi:10.1142/S0217751X13300238}, and we refer the reader to these for further details. {After this, we will review the standard Hawking calculation for particle creation during black hole  formation, and then discuss how to simulate this effect in accelerating toy models. }

\subsection{Quantization of a  scalar field on a {general} curved spacetime}

Let us consider a massless, minimally coupled scalar field $\phi$ propagating on a globally hyperbolic spacetime $(M,g_{ab})$, with manifold $M$ and  metric $g_{ab}$. The dynamics of this field is governed by the KG equation 
\bea
g^{ab}\nabla_a \nabla_b \phi=0\, . \label{kg}
\eea
If the spacetime is globally hyperbolic we can foliate $M$ by spacelike Cauchy hypersurfaces $\Sigma$, $M\simeq \mathbb R\times \Sigma$, and the Cauchy problem of (\ref{kg}) is well-defined given some regular initial data on some $\Sigma$ \cite{Wald:1984rg}.  Since the differential operator involved in (\ref{kg}) is linear, the corresponding space of solutions has the structure of a vector space. %We can endow this vector space with the following product:

Canonical quantization consists of finding a Hilbert space $\mathbb F$ and a (densely defined) linear operator $\hat\phi: \mathbb F\to \mathbb F$ that satisfies the equal-time canonical commutation relations on each leaf $\Sigma_t$: 
\bea
\int_{\Sigma_t} d\Sigma[\hat \phi(t,\vec x),\pi (t,\vec y)]=i\hbar \,\mathbb I\, , \label{commutationrelations}
\eea
where $t\in \mathbb R$ labels each $\Sigma_t$, and $\pi$ is the canonically conjugate operator of $\hat \phi$.
 This  construction can be carried out by using 
 %This naturally leads us to 
 a Fock representation of $\hat \phi(t,\vec x)$ in terms of creation $a_{\vec k}^{\dagger}$ and annihilation operators $a_{\vec k}$, of the form
\bea
\hat \phi(t,\vec x)= \sqrt{\hbar}\sum_{\vec k'} [ a_{\vec k}  \phi_{\vec k}(t,\vec x) + a_{\vec k}^{\dagger} \overline{ \phi_{\vec k}(t,\vec x)}]  \, ,\label{fockrep}
\eea
where $\phi_{\vec k}(t,\vec x)$ are (properly normalized) solutions of the field equation (\ref{kg}), and $[a_{\vec k},a_{\vec k'}^{\dagger}]=\delta_{\vec k,\vec k'}$, $[a_{\vec k},a_{\vec k'}]=[a_{\vec k}^{\dagger},a_{\vec k'}^{\dagger}]=0.$ %canonical commutation relations are satisfied for the creation and annihilation operators.
This operator(-valued distribution) defines a singularized state $|0\rangle$  via $a_{\vec k}|0\rangle=0$, $\forall \vec k\in \mathbb R$, which 
% The vacuum state $|0\rangle$ is a singularized state defined via $a_{\vec k}|0\rangle=0$, and 
physically represents the lowest energy state of the theory, and it is called the vacuum state. The rest of states of the Hilbert space are obtained by acting successively with the creation operator $a_{\vec k}^{\dagger}$, and are  physically interpreted as states containing quanta. The resulting Hilbert space $\mathbb F$ is a Fock space.

As one can easily notice from the above construction, Fock quantization leaves an inherent ambiguity on $\hat \phi(t,\vec x)$. This is, for each solution $\phi_{\vec k}(t,\vec x)$ of the KG equation, obtained from a specific choice of initial data on some $\Sigma$,  we have a possibly different Fock representation (\ref{fockrep}) that equally satisfies (\ref{commutationrelations}), and therefore a possibly different Fock  space of states.  Consequently, Fock quantization gives rise to infinitely-many Hilbert space representations of the canonical commutation relations, and for each of these representations there is a vacuum state that may differ from the others. To finish the quantization, we need to specify one such representation. In some cases of high degree of symmetry, like in Minkowski, there is a more ``natural'' choice of vacuum state, consisting of taking the field modes $\phi_{\vec k}(t,\vec x)$ as invariant under the full Poincare group of isommetries. In more general spacetimes, no such possibility is available, and one has to fix the vacuum state by imposing suitable initial data based on physical considerations.

A direct consequence of this ambiguity in the vacuum state of the theory is the prediction of particle creation.  If  two given Hilbert spaces of two different  representations of the commutation relations are unitarily inequivalent, then the two theories represent physically inequivalent theories. If, on the contrary, the two Hilbert spaces are unitarily equivalent, it is possible to find a one-to-one relation that maps the vacuum state of one Hilbert space with a particle state of the other space. This   is what  occurs in dynamical situations, where we have a Fock space at late times that differs from the Fock space at early times, but  the time-evolution operator acts as the unitary transformation between the two. 

Let us make this idea more concrete. Suppose that we have a spacetime that is  stationary for $t<t_1$ and $t>t_2$ with timelike killing vector field $K=\frac{\partial}{\partial t}$, while it fails to be stationary during the intermediate regime. Suppose now that we have a Fock representation of the quantum field at early times given by
\bea
\hat \phi(t,\vec x)=\sqrt{\hbar} \sum_{\vec k'} [ a^{\rm in}_{\vec k}  \phi^{\rm in}_{\vec k}(t,\vec x) + a_{\vec k}^{{\rm in}\dagger} \overline{ \phi^{\rm in}_{\vec k}(t,\vec x)}]  \label{fockrep1}\, ,
\eea
where $\phi^{\rm in}_{\vec k}(t,\vec x)$ is a basis on the subspace of solutions of (\ref{kg}) which satisfy $K^a \nabla_a \phi^{\rm in}_{\vec k} = -i \omega \phi^{\rm in}_{\vec k}$  for $t<t_1$. This is, the in modes are of ``positive-frequency'' with respect to $K$ at early times.  Similarly, we can think of another Fock representation of the quantum field at late times determined by
\bea
\hat \phi(t,\vec x)= \sqrt{\hbar}\sum_{\vec k'} [ a^{\rm out}_{\vec k}  \phi^{\rm out}_{\vec k}(t,\vec x) + a_{\vec k}^{{\rm out}\dagger} \overline{ \phi^{\rm out}_{\vec k}(t,\vec x)}]  \label{fockrep2}\, ,
\eea
where $\phi^{\rm out}_{\vec k}(t,\vec x)$ is a basis on the subspace of positive-frequency solutions of (\ref{kg}),  $K^a \nabla_a \phi^{\rm out}_{\vec k} = -i \omega \phi^{\rm out}_{\vec k}$, now  for $t>t_2$. The two equations are equally valid to study our problem. Now, since $\{\phi^{\rm out}_{\vec k},  \overline{ \phi^{\rm out}_{\vec k}} \}$ form vector basis on the full space of solutions of (\ref{kg}), we can expand the elements of the in basis in terms of the elements of the out basis:
\bea \label{bogo1}
\phi^{\rm in}_{\vec k} = \sum_{\vec k'} \alpha_{\vec k\vec k'}\phi^{\rm out}_{\vec k'} + \beta_{\vec k\vec k'}\overline{\phi^{\rm out}_{\vec k'} }\, ,
\eea
and similarly otherwise:
\bea
\phi^{\rm out}_{\vec k} = \sum_{\vec k'} \tilde\alpha_{\vec k\vec k'}\phi^{\rm in}_{\vec k'} + \tilde\beta_{\vec k\vec k'}\overline{\phi^{\rm in}_{\vec k'} }\, , \label{bogo1}
\eea
for some complex-valued coefficients. These expressions tell us that, in general, the in modes will fail to continue having a well-definite frequency at late times. Instead, they are given by a linear combination of positive- and negative-frequency modes at late times. 

These relations are called Bogoliubov transformations, and define the unitary transformation that relate the in and out Hilbert spaces of the quantum field at early and late times, respectively. Plugging in these expressions in (\ref{fockrep1}) and (\ref{fockrep2}), it is possible to find similar relations between creation and annihilation operators of the two representations:
\bea \label{bogoliubov}
a^{\rm out}_{\vec k'} = \sum_{\vec k} \alpha_{\vec k\vec k'}a^{\rm in}_{\vec k} + \overline{\beta_{\vec k\vec k'}} {a^{\rm in,\dagger}_{\vec k} }\, .
\eea
Further, it is possible to derive the following identities:
\bea \label{Eq:identity_alphabeta_1}
\sum_{\vec k} \left(\alpha_{\vec n \vec k} \overline{\alpha_{\vec k^{\prime} \vec k}}-\beta_{\vec n \vec k} \overline{\beta_{\vec k^{\prime} \vec k}}\right)=\delta_{\vec n \vec k^{\prime}}\, .
\eea

If we now evaluate the  expectation value of the number operator of the out representation in the in vacuum state, one finds a non-trivial result:
\bea
\langle {\rm in} | a_{\vec k}^{{\rm out},\dagger} a_{\vec k}^{\rm out}|{\rm in} \rangle = \sum_{\vec k'}|\beta_{\vec k \vec k'}|^2\neq 0\, .
\eea
This is, in the out Hilbert space representation, the vacuum state of the Fock space at early times fails to be the ground state of the theory. Instead of this, it contains a certain amount of particles. Physically, the dynamical transition of the spacetime background has spontaneously excited particle pairs out of the initial quantum vacuum, as seen by observers at late times. The total amount of particles for all $\vec k$ is
\bea
N= \sum_{\vec k} \langle {\rm in} | a_{\vec k}^{{\rm out},\dagger} a_{\vec k}^{\rm out}|{\rm in} \rangle = \sum_{\vec k}\sum_{\vec k'}|\beta_{\vec k \vec k'}|^2\, . \label{totalN}
\eea
It is not difficult to see that (\ref{bogoliubov}) represents a unitary transformation between the early- and late-time Fock spaces if and only if $N<\infty$, i.e., if the dynamical spacetime has only produced a ``finite amount'' of particles. \adr{If the Bogoliubov transformation (\ref{bogo1}) is originated from time evolution, which is unitary in free scalar field theories such as (\ref{kg}), then (\ref{totalN}) will be finite.} 

%\adr{The backgrounds that will be considered later in this work will depart from a stationary regime, evolve for a finite amount of time, and then return to a stationary configuration. In all these cases the time evolution of the field modes, as determined by a similar field equation to (\ref{kg}), connect the two in and out basis of field modes via the Bogoliubov transformation.}

The Bogoliubov coefficients can be  computed in terms of the field modes using explicit formulas. First, let us introduce the following product on the space of solutions of the KG equation:
\bea
(\phi_{\vec k}, \phi_{\vec k'})=-i \int_{\Sigma} d\Sigma \, n^a\left[ \phi_{\vec k} \nabla_a \overline{\phi_{\vec k'}} -\overline{\phi_{\vec k'}} \nabla_a \phi_{\vec k} \right]\, .
\eea
where $n^a$ is the unit timelike vector normal to $\Sigma$.
A straightforward calculation shows that this product is time-independent, thus independent of the choice of leaf $\Sigma$. %(see Appendix~\ref{appendix})
Furthermore, it is easy to check that positive- and negative-frequency solutions are ``orthogonal'', $(\phi^{\rm in}_{\vec k}, \overline{\phi^{\rm in}_{\vec k'}})=(\phi^{\rm out}_{\vec k}, \overline{\phi^{\rm out}_{\vec k'}})=0$ (notice that $n^a=K^a$ at both early and late times). Therefore, by ``projecting'' equation (\ref{bogo1}) with $\overline{\phi^{\rm out}_{\vec k'} }$ we can isolate the $\beta$ coefficient as
\bea
 \beta_{\vec k\vec k'} = \frac{(\phi^{\rm in}_{\vec k},\overline{\phi^{\rm out}_{\vec k'} })}{(\overline{\phi^{\rm out}_{\vec k} },\overline{\phi^{\rm out}_{\vec k'} })}\, .
\eea
evaluated at any $\Sigma$ of interest. 
It is customary to normalize the field modes by $(\overline{\phi^{\rm out}_{\vec k} },\overline{\phi^{\rm out}_{\vec k'} })=-\delta_{\vec k\vec k'}$ and $({\phi^{\rm out}_{\vec k} },{\phi^{\rm out}_{\vec k'} })=\delta_{\vec k\vec k'}$.

\subsection{Particle creation by a collapsing star}

We will illustrate now the idea discussed above with the well-known example of particle creation by a spherically-symmetric collapsing star. The original calculation for a moderately realistic scenario was  presented by Hawking in \cite{cmp/1103899393, PhysRevD.12.1519}. In this subsection, we will review instead the computation given in \cite{doi:10.1142/p378} using the Vaidya spacetime, which represents the simplest possible scenario. 

Let us consider the line element:
\bea
d s^{2}=-\left(1-\frac{2 M(v)}{r}\right) d v^{2}+2 d v d r+r^{2} d \Omega^{2}\, .
\eea
where  $M(v)$ is a mass function in terms of advanced time $v$.

\begin{figure}
    \centering
    \includegraphics[width=0.8\linewidth]{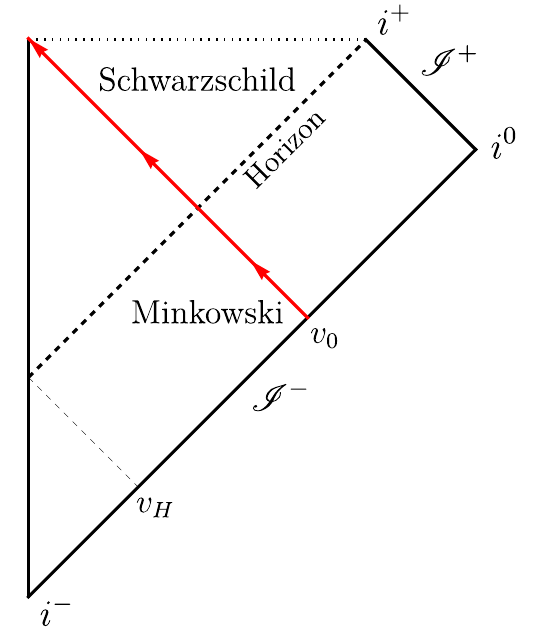}
    \caption{{Penrose diagram illustrating the propagation of the modes in a Vaidya spacetime, similar to Figure~3.3 in~\cite{doi:10.1142/p378}.}}
    \label{vaidya}
\end{figure}

This is an exact solution of the Einstein's field equations with stress-energy tensor given by $T_{ab}=\frac{\dot M}{4\pi r^2}\delta_{av}\delta_{bv}$. Physically, it represents the spacetime of a purely ingoing radial flux of energy described by the luminosity $L\equiv \dot M$. For $M=0$ we recover Minkowski spacetime, while for $M(v)=M_0$ we obtain the Schwarzschild spacetime in Eddington-Finkelstein advanced coordinates. If we take $M(v)$ as a monotonically increasing function from $0$ to $M_0$, then this spacetime  can  be used to model a simplified version of a realistic gravitational collapse, where all physical details of the collapsing star are neglected.  This is enough to derive the Hawking radiation. For simplicity, we will further assume the simplest of all cases, namely an infinitely thin incoming shock wave at the instant $v_0$, with $L(v)=M\delta(v-v_0)$ and $M(v)=M_0\Theta(v-v_0)$. In this case we simply have a patch of Minkowski spacetime (early times) attached non-smoothly to a portion of the Schwarzschild spacetime (late times), see Figure~\ref{vaidya}. 

Now, let us assume a minimally coupled, massless scalar field $\phi$ on this background. The spherical symmetry allows us to expand the field modes in a basis of spherical harmonics as
\bea
\phi(t,r,\theta,\phi)\sim \sum_{\ell m}\frac{\phi_{ \ell}(t,r)}{r}Y_{\ell m}(\theta,\phi)\, , \label{sph}
\eea
where the dependence on $m$ in $\phi_{ \ell}(t,r)$ disappears by virtue of the spherical symmetry, and the $1/r$ factor is introduced for convenience to account for the field's asymptotic decay. With this decomposition, the KG equation reduces to an effective two-dimensional partial differential equation, \adr{known as the Regge-Wheeler equation}:
\bea\label{reggewheeler}
\left(-\frac{\partial^{2}}{\partial t^{2}}+\frac{\partial^{2}}{\partial r^{* 2}}-V_{\ell}(r)\right) \phi_{\ell}(t, r)=0\, ,
\eea
where $V_{\ell}(r)$ is an effective potential and $r^*$ is the so-called tortoise coordinate. At early times, $r^*=r$ and $V_\ell=\frac{\ell(\ell+1)}{r^2}$, while at late times $r^*=r+2M\log \left(1-\frac{r}{2M}\right)$ and $V_{\ell}(r)=\left(1-\frac{2 M}{r}\right)\left[\frac{l(l+1)}{r^{2}}+\frac{2 M}{r^{3}}\right]$.

To proceed further in this derivation we will  work in the {\it geometric optics approximation}. This is, we will assume that the propagation of  massless scalar waves $\phi(t,r,\theta,\phi)$ can be well modeled by the propagation of null geodesics on the spacetime. This means that the field modes are dominated by the $\ell=0$ contribution in (\ref{sph}) (also called ``s-wave'' component). Additionally, we will solve the PDE above assuming the regularity condition at the origin, $\phi_{ \ell}(t,r=0)=0$.

In order to define the ``in'' and ``out'' Fock spaces of the quantum theory, it is convenient to switch to null coordinates. This is, at early times we write the Minkowski line element as $ds^2=-du_{\rm in}dv+r_{\rm in}^2 d\Omega^2$, for some radial function $r_{\rm in}=r_{\rm in}(u_{\rm in},v)$ while at late times the Schwarschild metric yields $d s^{2}=-\left(1-\frac{2 M}{r_{o u t}}\right) d u_{o u t} d v+r_{o u t}^{2} d \Omega^{2}$, for another  function $r_{\rm out}=r_{\rm out}(u_{\rm out},v)$. By imposing $r_{\rm in}(u_{\rm in},v)=r_{\rm out}(u_{\rm out},v)$, and in particular continuity of the metric at $v=v_0$, we obtain the relation between the in and out retarded times: 
\bea
u_{\rm out}(u_{\rm in})=-4M \log \left|\frac{v_0-u_{\rm in}}{4M}-1\right|\, . \label{inout}
\eea

We shall define the ``in'' modes as those solutions of the KG equation satisfying the following initial data at past null infinity:
\bea \label{inmodes}
\lim_{\substack{r\to \infty \\ v={\rm const}}}\phi^{\rm in}_{\omega 0}(r,v)\sim \frac{e^{-i\omega v}}{4\pi \sqrt{\omega}}\, .
\eea
These are ingoing waves with well-definite, positive frequency with respect to a null congruence measuring time $v$. They form a complete orthonormal vector basis on $L^2(\mathbb R\times \mathbb S^2)$, with the KG product.  Similarly, we define the  ``out'' modes as those solutions of the KG equation satisfying the  ``final'' data at future null infinity:
\bea \label{outmodes}
\lim_{\substack{r\to \infty \\ u_{\rm out}={\rm const}}}\phi^{\rm out}_{\omega 0}(r,u_{\rm out})\sim \frac{e^{-i\omega u_{\rm out}}}{4\pi \sqrt{\omega}}\, .
\eea
These outgoing waves have well-definite, positive frequency with respect to a null congruence measuring time $u$, and, again, they form a complete orthonormal vector basis on $L^2(\mathbb R\times \mathbb S^2)$, with the KG product.

The background  radial flux of incoming energy is expected to excite particle pairs out of the in quantum vacuum. As a result, when the in modes are evolved from past to future null infinity, they will be given by a linear combination of positive- and negative frequency solutions with respect to time $u$. From a computational viewpoint in this derivation, it is much more convenient to reverse the reasoning:  when propagating back in time, the out modes will  become a linear combination  of positive- and negative-frequencies at past null infinity. The relevant Bogoliubov coefficient can then be obtained via
\bea
\beta_{\omega \omega^{\prime}}&=&-\left(\phi_{\omega}^{\rm o u t}, \phi_{\omega^{\prime}}^{\rm i n *}\right)\nonumber\\
&=&i \int_{\mathcal J^{-}} d v  d \Omega\left(\phi_{\omega}^{\rm o u t} \partial_{v} \phi_{\omega^{\prime}}^{\rm i n}-\phi_{\omega^{\prime}}^{\rm i n} \partial_{v} \phi_{\omega}^{\rm o u t}\right)\, .\label{betabogo}
\eea 
It is possible to obtain an explicit prediction for sufficiently late times at future null infinity.  This is because,  in the Minkowski patch of our spacetime, the solution of the KG equation with final data (\ref{outmodes}) and regular condition at $r=0$ can be easily obtained from (\ref{inout}), and yields
\bea
 \phi_{\omega 0}^{\rm o u t}=\frac{1}{4 \pi \sqrt{\omega}}\left(e^{-i \omega u_{\rm o u t}\left(u_{\rm i n}\right)}-e^{-i \omega u_{\rm o u t}(v)} \theta\left(v_{H}-v\right)\right)\, ,\nonumber
\eea
where $v_H=v_0-4M$. At early times, $v\to-\infty$, we have $u_{\rm o u t}(v)\sim v\to -\infty$ and the expression above reduces to 
\bea
\phi^{\rm out}_{\omega 0}(r,v)\sim \frac{e^{-i\omega v}}{4\pi \sqrt{\omega}}\, ,
\eea
similar as (\ref{inmodes}). Thus, the out modes for $u_{\rm out}\to-\infty$ are still of pure positive-frequency with respect to $v$. This is, at early times, when the background incoming energy flux has not yet arrived, there is no particle emission reaching future null infinity. However, the situation changes dramatically for $u_{\rm out}\to +\infty (v\to v_H)$.  Avoiding technical complications (see \cite{doi:10.1142/p378} for details) it is possible to find the identity $\left|\alpha_{\omega, \omega^{\prime}}\right|^{2}=e^{8 \pi M \omega'}\left|\beta_{\omega, \omega'}\right|^{2}$, which, by virtue of the Bogoliubov identities, eventually produces:
\bea
%\langle {\rm i n}| N^{\rm o u t}|{\rm i n}\rangle=\int_0^{\infty}d\omega \int_{0}^{\infty}  d \omega^{\prime}\left|\beta_{\omega, \omega^{\prime}}\right|^{2}=\int_0^{\infty}d\omega\frac{1}{e^{8 \pi M_0 \omega}-1}\, . \nonumber
\langle {\rm i n}| N_{\omega}^{\rm o u t}|{\rm i n}\rangle= \int_{0}^{\infty}  d \omega^{\prime}\left|\beta_{\omega \omega^{\prime}}\right|^{2}=\frac{1}{e^{8 \pi M_0 \omega}-1}\, .  \label{hawkingspectrum}
\eea
which is the celebrated Hawking thermal density spectrum.

\subsection{Simulating particle creation with time dependent potential barriers}

\adr{As outlined in the Introduction, the Hawking's frequency spectrum (\ref{hawkingspectrum})  depends only on the physical parameters of the final black hole ---namely, its mass--- and is entirely independent of the initial configuration of the collapsing matter. This stems from the fact that the underlying calculation of the Bogoliubov coefficients $\beta_{\omega\omega'}$  assumes the late time limit $u_{\rm out}\to +\infty$ to relate the in and out mode basis. To obtain the full frequency spectrum, one must have complete knowledge   of the evolved modes $\phi_{\omega0}^{\rm out}$ at past null infinity, in order to  compute the integral  in (\ref{betabogo}) properly.}

\adr{Unfortunately, this calculation cannot be carried out purely by analytical means, and a numerical framework is required. The hyperboloidal slicing method introduced in the Introduction provides a promising avenue to address this problem, since it is able to reach past and future null infinities in any asymptotically flat spacetime. However, the  involved numerical infrastructure and the general approach  to  particle creation must  first  be tested  in simpler settings. For this reason, as a first step in this program, we will simply focus on analogue models in Minkowski spacetime, designed to simulate the effects of particle creation by gravitational fields.}

\adr{As discussed in the previous subsection, the dynamics of the scalar-field modes in any spherically-symmetric spacetime geometry can be reduced to an effective (1+1)-dimensional wave equation in flat space, subject to a  potential barrier $V_{\ell}(r)$, given by equation (\ref{reggewheeler}). %This is, at the end of the day, the effects of the spacetime geometry on a scalar field are encoded in an effective potential. 
Remarkably, the same equation arises if, instead of a curved geometry, one simply works in Minkowski spacetime and introduces a suitable effective potential in the Klein-Gordon equation (\ref{kg}),  %restricted to spherically symmetric configurations:
\bea
\left[\eta^{\mu\nu}\nabla_{\mu} \nabla_{\nu}+V_\ell(t,x)\right]\phi_\ell(t,x)=0\, , \label{kg2}
\eea
where $\eta_{\mu\nu}$  now denotes the flat  metric. At early times during the collapse, the background is effectively Minkowski, with $V_\ell(x)=\frac{\ell(\ell+1)}{x^2}$, while at late times  it tends towards the Regge–Wheeler potential, $V_{\ell}(x)=\left(1-\frac{2 M}{r(x)}\right)\left[\frac{l(l+1)}{r^{2}(x)}+\frac{2 M}{r^{3}(x)}\right]$, associated with the Schwarzschild geometry. To simulate the phenomenon of particle creation during such a collapse, one could introduce a suitable, effective time-dependent potential $V_\ell(t,x)$ that interpolates between the two limits.}

\adr{Motivated by this observation, in this work we will numerically study particle creation for scalar fields in Minkowski spacetime, induced by time-dependent effective potentials. This approach enables us to test the applicability of the hyperboloidal-slicing numerical framework for scattering problems in asymptotically flat spacetimes. In this context, the role of the “background” is played by the effective potential $V$, which is made time-dependent so as to be able to excite particles out of the vacuum. To ensure that the total particle number given in (\ref{totalN}) is finite, this effective potential will only be dynamical during a finite amount of time.}

\adr{To properly address this study, we need to restrict to specific families of effective potentials that can be adequately treated in our numerical framework. A distinctive feature of the Regge–Wheeler potential $V_\ell(r)$ is that, while it is mostly concentrated around $r = 3M$, it has  support over the entire real line, decaying only polynomially as $r \to \infty$. This poses a challenge for the hyperboloidal slicing framework, which can efficiently handle only functions with compact support. To overcome this limitation, we introduce simplified toy models for the potential barrier—namely, effective potentials of compact support—and consider time-dependent oscillations of this barrier, both in its amplitude and in its radial position, so as to keep the potential barrier sufficiently confined in space during the entire numerical evolution of the field modes. Although these are idealized models, they retain a clear physical analogy with e.g. oscillating compact stars, and we therefore expect them to reproduce the same qualitative features of particle creation. Other effects such as friction, heat dispersion etc are not relevant for studying the phenomenon of particle creation and shall be neglected in this work.}

\section{Methodology}
\label{Sec:Methodology}

To achieve the objectives outlined {above}, this section introduces the steps taken to propagate a classical complex  scalar field $\phi$ ---representing each of the {massless} field modes of a quantum field--- through a dynamical {background}, as well as the necessary analysis to ascertain particle creation. We use a setup that treats this phenomenon numerically. Because we need to access past null infinity $\scri^-$ and future null infinity $\scri^+$, as radiation  is only unambiguously defined at null infinity, we must resort to a spacetime foliation that allows us to reach these two asymptotic regions. The use of hyperboloidal slices with compactification is ideal in this scenario. The field $\phi$ is propagated from $\scri^-$ to $\scri^+$ and the main procedure is divided into 5 steps:
\begin{itemize} [leftmargin=*]
    \item Provide \textbf{initial data at $\scri^-$} on a signal (i.e., the field modes) to be propagated, $\phi$. \adr{This  serves to specify the choice of the in vacuum state in the quantum theory.}
    \item Evolve the initial data on \textbf{hyperboloidal slices that extend towards $\scri^-$ (``ingoing")} -- first half of the evolution.
    \item \textbf{Translate/interpolate} the data from the ``ingoing" slices to a hyperboloidal slice that extends towards $\scri^+$ (``outgoing"). \adr{This is the main tool to study the scattering problem between $\scri^-$ and $\scri^+$.}
    \item Propagate the signal through these \textbf{outgoing hyperboloidal slices} -- second half of the evolution -- and extract $\phi$ at $\scri^+$.

    \item Compute the \textbf{Bogoliubov coefficients $\alpha_{\omega\omega'}$ and $\beta_{\omega\omega'}$} {by comparing with the out basis of field modes}. If the background is dynamical, the $\beta_{\omega\omega'}$ coefficients will be non-zero, i.e.,~initial positive-frequency modes {at $\scri^-$} become a linear combination of positive- and negative-frequency modes {at $\scri^+$}. This certifies the effect of particle creation. 
\end{itemize}

\begin{figure}
    \centering
    \includegraphics[width=0.8\linewidth]{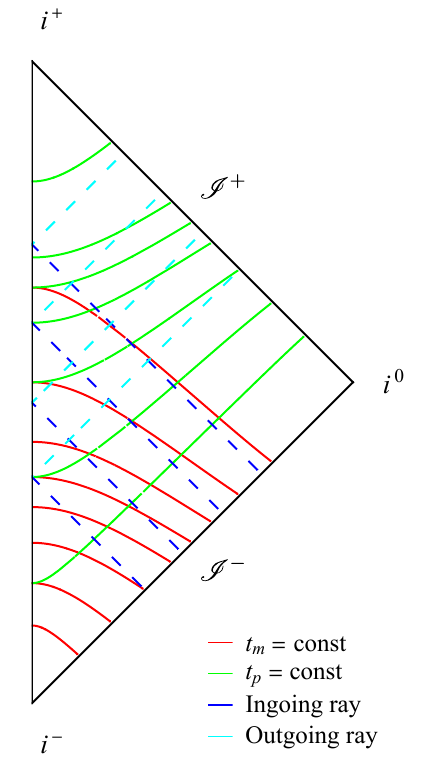}
    \caption{Penrose diagram of Minkowski spacetime depicting radiation propagating from $\scri^-$ to $\scri^+$, and the two foliations of spacetime using ingoing (constant $t_m$) and outgoing (constant $t_p$) hyperboloidal slices.}
    \label{fig:SetupScheme}
\end{figure}

\noindent For a schematic of the numerical setup,  see Figure \ref{fig:SetupScheme}.

\subsection{{Hyperboloidal 3+1 foliation of the} Klein-Gordon equation in Minkowski spacetime}

%We shall begin by treating the equation to be solved numerically, i.e, the massless Klein-Gordon equation with a \adr{generic, spherically-symmetric} potential $V(t,r)$:

Let us consider the KG equation of a massless, scalar field $\tilde{\phi}$, subject to a generic potential $V(x)$ \adr{with spatial compact support}:
\begin{equation}\label{Eg:waveq}
   % \left[\square + V(t,r) \right]\tilde{\phi} = 0 \Leftrightarrow  
   [\eta^{\mu \nu}\nabla_\mu \nabla_\nu + V(x)] \tilde{\phi}=0,
\end{equation}
where $\eta_{\mu\nu}$ is taken to be the flat Minkowski metric, and $x^\gamma$ denotes a generic spacetime point. This is the equation to be solved numerically in this work.
For simplicity, we restrict ourselves to spherical symmetry. This is, if  $(\tilde{t},\tilde{r},\theta,\varphi)$ represent the usual polar coordinates, we will assume  $\tilde{\phi}=\tilde{\phi}(\tilde{t},\tilde{r})$ and $V=V(\tilde t,\tilde r)$.

The line element in these coordinates reads 
\begin{equation}
    ds^2 = -d\tilde{t}^2 + d\tilde{r}^2 + \tilde{r}^2(d\theta^2+\sin^2\theta d\varphi^2)\, .
    \label{Eq: Line Element PolarCoord}
\end{equation}
%with $d \sigma^2 = (d\theta^2+\sin^2\theta d\varphi^2)$ the solid angle element in spherical coordinates. 
Because  $\tilde{t}$ and $\tilde{r}$ are the usual Minkowski coordinates, constant $\tilde{t}$ hypersurfaces represent infinite spacelike Cauchy slices. In order to reach past and future null infinity, we foliate Minkowski spacetime along hyperboloidal slices, each one labelled by a given hyperboloidal time $t$, and compactified with a radial coordinate $r$ using a suitable compactification factor $\Omega$. More precisely, we perform the following coordinate transformation from $(\tilde{t},\tilde{r})$ to $(t,r)$:
\begin{align}
    \tilde{t}&=t + h(r),\\
    \tilde{r}&=\frac{r}{\Omega(r)}.
\end{align}
\adr{The precise form of the height and conformal factors, $h(r)$ and $\Omega(r)$ respectively, depend on the particular properties of the chosen asymptotically flat spacetime.}
In our \adr{Minkowski} setup, we choose $\Omega=\frac{|\Kc|}{6}(1-r^2)$ and $h(r)=\pm \sqrt{(3/\Kc)^2+r^2/\Omega^2}$, for some given constant $|\Kc|$ \cite{Zenginoglu:2007jw}, \avv{which represents  a constant-mean-curvature (CMC) slice of Minkowski spacetime.} The specific sign of the height function $h(r)$ depends on wether we are creating hyperboloidal slices that extend towards $\scri^+$ ($+$, $\Kc<0$) or towards $\scri^-$ ($-$, $\Kc>0$). In these coordinates, the origin is represented by $r=0$ whereas null infinity is located at $r=1$ on the computational grid, precisely where the compactification factor vanishes, $\Omega=0$. To better distinguish between the first and second evolutions, we denote the time coordinate corresponding to $\scri^-$ slices as $t_m$ and that to $\scri^+$ slices as $t_p$. \par
The next step implies writing the equations in the new coordinate system $(t,r,\theta,\varphi)$. For this, we rewrite the Minkowski line element by computing the differentials
\begin{align}
    d\tilde{t}&=dt+h'(r)dr,
    \label{Eq:dt}\\
    d\tilde{r}&=\left(\frac{\Omega-r\Omega'}{\Omega^2}\right)dr, \label{Eq:dr}
\end{align}
and substitute them on \eqref{Eq: Line Element PolarCoord} to obtain
\begin{align}
    ds^2=&-dt^2-2h'(r)dtdr+\left[\left(\frac{\Omega-r\Omega'}{\Omega^2}\right)^2-h'(r)^2\right]dr^2 \nonumber \\
    &+\frac{r^2}{\Omega^2}(d\theta^2+\sin^2\theta d\varphi^2).
\end{align}
By inspection we can identify the metric components:
\begin{equation}
    \eta_{\mu\nu}=\begin{pmatrix}
        -1 & -h'(r) & 0 & 0 \\
        -h'(r) & \left(\frac{\Omega-r\Omega'}{\Omega^2}\right)^2-h'(r)^2 & 0 & 0 \\
        0 & 0 & \frac{r^2}{\Omega^2} & 0\\
        0 & 0 & 0 & \frac{r^2}{\Omega^2}\sin^2\theta
    \end{pmatrix} .%,
\end{equation}
\begin{comment}
with its inverse as
\begin{equation}
    g^{\mu\nu}=\begin{pmatrix}
        -1+ \frac{\Omega^4h'^2}{(\Omega-rh'(r))^2} & -\frac{\Omega^4h'(r)}{(\Omega-r\Omega'^2} & 0 & 0 \\
         -\frac{\Omega^4h'(r)}{(\Omega-r\Omega')^2} & \frac{\Omega^4}{(\Omega-r\Omega')^2} & 0 & 0 \\
        0 & 0 & \frac{\Omega^2}{r^2} & 0\\
        0 & 0 & 0 & \frac{\Omega^2}{r^2\sin^2\theta}
    \end{pmatrix}.
\end{equation}
\end{comment}
%
%By looking at Eq \eqref{Eq:KGfirststep}, we see that one must also compute the Christoffel symbols $\Gamma_{\mu\nu}^\lambda$ on this new coordinate system. Because we are working on spherical symmetry, all partial derivatives with respect to the angular coordinates $\theta$ and $\varphi$ are null and we only need to compute $\Gamma_{\mu\nu}^t$ and $\Gamma_{\mu\nu}^r$. The only non-zero Christoffel symbols are
\begin{comment}
\begin{align}
    \Gamma_{rr}^t&=h''(r)+h'(r) \left(\frac{2 \Omega '}{\Omega (r)}+\frac{r \Omega ''}{\Omega -r \Omega '}\right),\\
    \Gamma_{\theta \theta}^t &= \frac{r \Omega h'(r)}{\Omega-r \Omega '},\\
    \Gamma_{\varphi \varphi}^t &=\frac{r \sin ^2(\theta ) \Omega h'(r)}{\Omega -r \Omega '},\\
    \Gamma_{rr}^r &=\frac{r \Omega ''}{r \Omega '-\Omega }-\frac{2 \Omega '}{\Omega },\\
    \Gamma_{\theta \theta}^r &=\frac{r \Omega}{r \Omega '-\Omega },\\
    \Gamma_{\varphi \varphi}^r &=\frac{r \sin ^2(\theta ) \Omega }{r \Omega '-\Omega },\\
\end{align}
\end{comment}
Introducing this into \eqref{Eg:waveq}, the KG equation yields
\begin{equation}
\begin{split}
&\p_t^2\tilde{\phi}= \\
&\quad \left(-\frac{\Omega ^4 \p_t\tilde{\phi}}{r \left(r \Omega '-\Omega \right)^3} \left[r \left(h'(r) \left(r \Omega ''-2 \Omega '\right)-r h''(r)\Omega ' \right)\right.\right.\\
&\left.\left. \quad + \Omega \left(r h''(r)+2 h'(r) \right)\right] + \frac{2 \Omega^4 h'(r) \p_t\p_r\tilde{\phi}}{\left(\Omega -r \Omega '\right)^2} \right.\\
& \left. \quad +\frac{\Omega ^4 \p_r\tilde{\phi}\left(r \left(r \Omega ''-2 \Omega '\right)+2 \Omega \right)}{r \left(r \Omega '-\Omega \right)^3} \right.\\
&\left. \quad - V(t,r) \tilde{\phi} -\frac{\Omega^4 \p_r^2\tilde{\phi}}{\left(\Omega -r \Omega '\right)^2} \right) /\left(\frac{\Omega ^4 h'(r)^2}{\left(\Omega -r \Omega '\right)^2}-1\right).
\end{split}
\label{Eq:KGequation_2T2S}
\end{equation}

Equation \eqref{Eq:KGequation_2T2S} is a 2nd order in time, 2nd order in space differential equation with mixed derivatives. However, by introducing a new variable $\tilde{\Pi}=\p_t \tilde{\phi}$, we can rewrite it as a 1st order in time, 2nd order in space (FT2S) set of partial differential equations (PDEs):
\begin{subequations}
\begin{align}
    &\p_t\tilde{\phi}=\tilde{\Pi}\\
    &\p_t\tilde{\Pi}= \nonumber\\
    &\quad \left(-\frac{\Omega ^4 \p_r\tilde{\phi}}{r \left(r \Omega '-\Omega \right)^3} \left[r \left(h'(r) \left(r \Omega ''-2 \Omega '\right)-r h''(r)\Omega ' \right)\right.\right. \nonumber \\
    &\left.\left. \quad + \Omega \left(r h''(r)+2 h'(r) \right)\right] + \frac{2 \Omega^4 h'(r) \p_t\p_r\tilde{\phi}}{\left(\Omega -r \Omega '\right)^2} \right.\nonumber \\
    & \left. \quad +\frac{\Omega ^4 \p_r\tilde{\phi}\left(r \left(r \Omega ''-2 \Omega '\right)+2 \Omega \right)}{r \left(r \Omega '-\Omega \right)^3} \right. \nonumber \\
    &\left. \quad - V(t,r) \tilde{\phi} -\frac{\Omega^4 \p_r^2\tilde{\phi}}{\left(\Omega -r \Omega '\right)^2} \right) /\left(\frac{\Omega ^4 h'(r)^2}{\left(\Omega -r \Omega '\right)^2}-1\right).
\end{align}
\end{subequations}
Due to spherical symmetry $\tilde{\phi}=\tilde{\phi}(t,r)$, our PDE system contains no derivatives on the angular coordinates and we reduce our spatial dimensions to one -- our radial coordinate -- allowing a 1+1 decomposition framework.
\par

% This is true, but both things are unrelated %Even though the structure of the PDE system is complete, it is now convenient for us 
It is further convenient to rescale the scalar field $\tilde{\phi}$, because it decays as $\sim 1/\tilde r\to 0$ asymptotically. If unrescaled, we would have trouble providing given data through $\scri^-$, and extracting the field at $\scri^+$ would yield no results that could be analyzed straightforwardly. We thus perform a rescaling of the scalar field to counteract this decay, by using the conformal factor $\Omega$, which satisfies $\Omega\sim 1/\tilde r$ asymptotically:
\begin{equation}
    \tilde{\phi}=\Omega \phi, \quad \tilde{\Pi}=\Omega\Pi,
    \label{eq:RescaleField}
\end{equation}
with $\phi$ and $\Pi=\p_t\phi$ our rescaled complex scalar field and its respective time derivative. As mentioned above, the main difference between the two evolutions (reaching $\scri^-$ and $\scri^+$, respectively) is the sign of the height function $h(r)$ and of $K_{CMC}$, and this difference is totally encoded in the evolution equations at the level of the radial derivatives of the height function.  Taking into account that $\Omega'=-r(\abs{K_{CMC}}/3)$, $\Omega''=-\abs{K_{CMC}}/3$ and the first radial derivative of $h(r)$,
\begin{equation}
    h'(r)=-K_{CMC}\frac{(r/\Omega)}{\sqrt{9+(r^2K^2_{CMC}/\Omega^2)}},
    \label{Eq: DerivativeHeightFunction}
\end{equation}
we can rewrite the evolution equations for our rescaled scalar field as
\begin{subequations}
\begin{align}
    &\p_t \phi= \Pi, \label{Eq:KleinGordon1T2Sphi}\\
    &\p_t \Pi = \left(\frac{K_{CMC} \Omega }{\sqrt{K_{CMC}^2 r^2+9 \Omega^2}}+\frac{2 K_{CMC}}{3}\right) \Pi \nonumber\\
    &\quad +\frac{2}{3} K_{CMC} r \p_r\Pi +\Omega ^2 \p_r^2 \phi \nonumber\\
    & \quad+\frac{\Omega\left(9 \Omega  \sqrt{K_{CMC}^2 r^2+9 \Omega^2}-2 K_{CMC}^2 r^2-9 \Omega^2\right) \p_r\phi}{3 r \sqrt{K_{CMC}^2 r^2+9 \Omega^2}} \nonumber\\
    & \quad +\frac{\Omega \left(3 \Omega  \sqrt{K_{CMC}^2 r^2+9 \Omega ^2}-2 K_{CMC}^2 r^2-9 \Omega^2\right) \phi}{3 r^2 \sqrt{K_{CMC}^2 r^2+9 \Omega ^2}} \nonumber\\
    &  \quad -V \phi. \label{Eq:KleinGordon1T2SPi}
\end{align}
\end{subequations}
Equations \eqref{Eq:KleinGordon1T2Sphi}and \eqref{Eq:KleinGordon1T2SPi} work for both $\{t_m,r\}$ (first) and $\{t_p,r\}$ (second) evolutions as the type of hyperboloidal slice is totally encoded in the sign of $K_{CMC}$.
\par
In the following, we numerically solve these equations by applying the Method of Lines, discretizing space using 4th order accurate centered Finite Differences and integrating the corresponding ordinary differential equation system in time using the 4th order Runge-Kutta method. To compute spatial derivatives at $r=0$ and $r=1$ ($\scri$), i.e., at the edges of the grid, we introduce four ghost points, two on each side, so that the grid is fully described by $x_j=j\Delta x$ with $j=-2,-1,0,...,N,N+1,N+2$, with $N\in \mathbb N$. 

\subsection{Construction of the initial vacuum state}

To study the problem of particle creation we first need to define an ``in'' vacuum state. This amounts to specifying suitable initial data for the field modes at $\scri^-$, \adr{which  in our numerical grid  is located at $r=1$}. Specifically, we require the field modes to have positive-frequency $\omega$ with respect to the \adr{advanced time coordinate $v$ at $\scri^-$, which coincides with the coordinate $t$ introduced above when $r=1$}. 

Although this condition is typically implemented  using plane waves  (see e.g.~\eqref{inmodes}), in  numerical settings it is impossible to work with signals that \avv{extend} infinitely in the time coordinate $v$. 
\avv{Likewise, the  wave packets employed in Hawking's original calculation (see e.g.  section 3.3.2 of~\cite{doi:10.1142/p378}), which are localized in time but also form a complete orthonormal basis, still have non-negligible amplitude over an infinite time range, posing similar numerical difficulties. While their amplitude decays with time, the decay is very slow.}

Consequently, in this work we will define the initial data for the field modes as plane waves modulated by a rapidly decaying envelope, effectively yielding compact support. More precisely, we impose the following initial conditions for the field modes: 
\begin{subequations}\label{Eq:initialPhiPi}
\begin{align}
   %\phi_{initial} &= 
   \phi_{\omega_0}(t,r=1)&=f(t)e^{-i\omega_0t}\, , \quad {t\geq 0}
   \label{Eq:InitialPhi}\\
   %\Pi_{initial} &= 
   \Pi_{\omega_0}(t,r=1) &= \p_t \left[ f(t)e^{-i\omega_0t}\right]\, , \quad {t\geq 0}
   \label{Eq:initialPi}
\end{align}
\end{subequations}
where $\omega_0$ is the signal's initial frequency %This form of the initial signal ensures that $\phi_{\omega_0}$ is of positive-frequency during almost all $t$ in our grid. 
(here onwards, we denote~$\phi$ and~$\Pi$ with the subindex~${\omega_0}$ to denote the initial frequency they carry).
The function $f(t)$ is taken to be a smooth envelope defined by the following Gaussian profile
\begin{equation}
    f(t)=t^2 e^{-\frac{(t^2-t_0^2)^2}{4 \sigma^4}},
    \label{Eq:GaussianFunction}
\end{equation}
where the $t^2$ factor is introduced to ensure the signal is null and smooth enough at the start of the evolution {at $t=0$}. Lack of smoothness is problematic at the numerical level, as it hinders convergence and thus the reliability of our results. {For $t\geq 0$ this $f(t)$ simply represents a bump-like function confined around $t_0$ and width $\sigma$.}
%The constant parameter $t_0$  controls the ``center'' of the initial signal while $\sigma$  controls its width. 
We expect that this choice of envelope will have an effect on the recovery of the frequencies in the analysis, while other constructions \avv{(like modified wave packets with compact support)} may provide cleaner Bogoliubov coefficients. In this first work we adopt the described numerically suitable choice of envelope, and leave testing other options for the future.

Besides providing given data at $\scri^-$, we set initial data on the hyperboloidal slice {$t=0$} as null everywhere:
\begin{subequations}
\begin{align}
    \phi_{\omega_0}({t=0}, r )&=0, \\ %
    \Pi_{\omega_0}({t=0},r)&=0. %t_0
\end{align}
\end{subequations}
Inspection of \eqref{Eq:initialPhiPi} %\eqref{Eq:InitialPhi} and \eqref{Eq:initialPi} 
can make some readers consider these equations more akin to boundary conditions than to initial conditions. In fact, these are implemented in the code at $r=1$ using their respective derivatives on the RHS of the evolution equations, yet the solution depends on those conditions as it would depend on the initial conditions. 

In this particular setup, the boundary conditions are set at the origin $r=0$ and at the outer boundary $r=1$, corresponding to the location of $\scri^-$. At the origin, $r=0$ we implement parity boundary conditions. Because $\phi_{\omega_0}$ is a scalar quantity, it has even parity at the origin and so does its time derivative $\Pi_{\omega_0}$. The boundary conditions are thus
\begin{subequations}
\begin{align}
    \phi_{\omega_0}(t,-r)&=\phi_{\omega_0}(t,r), \quad {r\sim 0}\, ,\\
    \Pi_{\omega_0}(t,-r)&=\Pi_{\omega_0}(t,r), \quad {r\sim 0}\, .
\end{align}
\end{subequations}
%and are imposed on the ghost points, which in terms of grid functions becomes
%\begin{subequations}
%\begin{align}
%    \phi_{-1}(t)&=\phi_1(t), \quad\quad \phi_{-2}(t)=\phi_2(t),  \\
%    \Pi_{-1}(t)&=\Pi_1(t), \quad\quad \Pi_{-2}(t)=\Pi_2(t).
%\end{align}
%\end{subequations} 

Due to the choice of foliations - ingoing or outgoing hyperboloidal slices - no \textit{physical} boundary conditions are necessary at null infinity. In the first half of the evolution, nothing can leave the domain through $\scri^-$ and we are prescribing our desired data there, whereas in the second half nothing can enter the domain through $\scri^+$. However, regarding the latter, even though there are no physical boundary conditions, to fill the values of the ghost points we still numerically impose \textit{outflow} boundary conditions, which means extrapolating the signal on the ghost points to represent the signal outside this boundary. These boundary conditions are computed as described by equations (109) and (110) in \cite{Calabrese_2006}.

%are implemented in our numerical scheme as follows 
%\begin{subequations}\label{Eq:outflow}
%\begin{align}
%    (\Delta x)^5 D^5_- \phi_{N+1}&=0,  \quad \quad (\Delta x)^5 D^5_- \phi_{N+2}=0 \\
%    (\Delta x)^4 D^4_- \Pi_{N+1}&=0,    \quad \quad (\Delta x)^4 D^4_- \Pi_{N+2}=0,
%\end{align}
%\end{subequations}
%where subscripts $(N+1)$ and $(N+2)$ denote the ghost points on the right side of the upper boundary of the grid $r=1$. $D^4_-$ and $D^5_-$ denote the backward finite differences of the 4th and 5th radial derivative of $\phi$ and $\Pi$, respectively.\par
As a final remark, we note that the grid used in the setup is non-staggered grid (a staggered grid would be the current grid shifted half a step to the left or right) since we require access to $r=1$ to provide the signal through $\scri^-$. However, the RHS of the KG equation at the origin $r=0$ and $\scri^+$ ($r=1$) are formally singular. Therefore, to be able to compute the RHS at those points, we evaluate a ``l'Hôpitalized'' version of those equations.

\subsection{{Translation of data  between ingoing and outgoing hyperboloidal foliations}}
\begin{figure}[h!]
    \centering
    \includegraphics[width=0.7\linewidth]{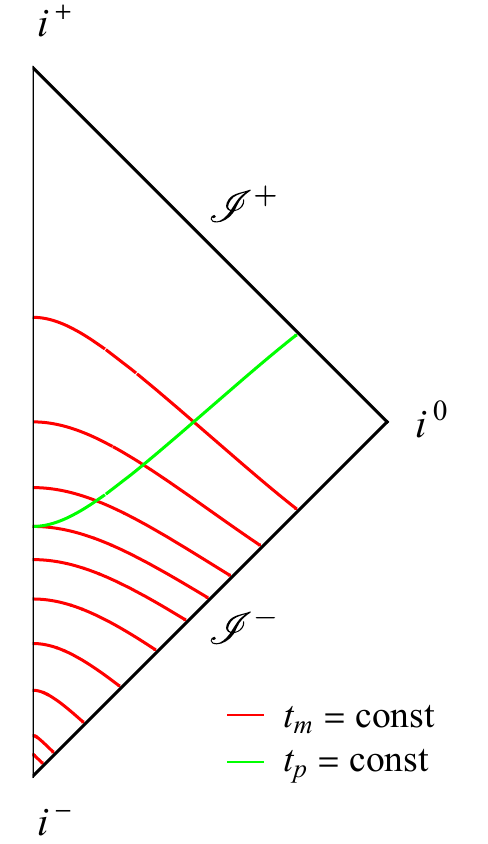}
    \caption{Penrose diagram of Minkowksi spacetime showcasing a schematic of the relation between the information on outgoing and ingoing hyperboloidal slices. Information on the $t_p$ slice is present at multiple $t_m$ slices.}
    \label{fig:Translation}
\end{figure}

We will refer to ``translation'' the procedure of populating the initial slice of the foliation reaching $\scri^+$ \adr{(``out'' foliation)}, with data from the former evolution departing from $\scri^-$, which uses another foliation \adr{(``in'' foliation)}.
For the translation we need to relate the signal described at the ingoing slices to the signal at the outgoing slices. {As shown in} Figure \ref{fig:Translation}, the data of $\phi_{\omega_0}$ (and $\Pi_{\omega_0}$) on a single outgoing $t_p=t_{p_{match}}$ slice  (green line) {intersects with} multiple ingoing slices (red lines), i.e., at different $t_m$ instants. 
\par

To perform this translation, we recall the equations 
\begin{subequations}\label{Eq:ttranslation}
\begin{align}
    t_m&=\tilde{t}+\sqrt{\left(\frac{3}{K_{CMCm}}\right)^2 + \frac{r^2}{\Omega^2}},
    \label{Eq:t_m}\\
    t_p&=\tilde{t}-\sqrt{\left(\frac{3}{K_{CMCp}}\right)^2 + \frac{r^2}{\Omega^2}} \label{Eq:t_p}.
\end{align}
\end{subequations}
Equating the values of $\tilde t$ above and choosing a specific $t_p=t_{p_{match}}$, i.e., a specific slice extending towards $\scri^+$, we can obtain the value of the radius $r$ as a function of $t_m$, $r=r(t_m)$, i.e., as a function of the slices extending towards $\scri^-$. By further requiring that $r \geq 0$ the function we obtain is
\begin{align}
    r(t_m)=&\frac{1}{2}\left(\sqrt{4+\frac{4}{-1+\frac{(t_m - t_{p_{match}})^2}{9\left(\frac{1}{\abs{K_{CMCm}}} + \frac{1}{\abs{K_{CMCp}}}\right)^2}}}\right.
    \nonumber \\ & \quad\left. - \frac{2}{\sqrt{-1+\frac{(t_m - t_{p_{match}})^2}{9\left(\frac{1}{\abs{K_{CMCm}}} + \frac{1}{\abs{K_{CMCp}}}\right)^2}}}\right).
\end{align}
This function gives, for a specific ingoing slice $t_m$, the radius at which the $t_m$ and $t_{p_{match}}$ coincide and where we have to extract the signal's value ${\phi_{\omega_0}(t_m,r)}$. Since this $r$ value is not guaranteed to be part of the original equally-spaced grid, one has to resort to interpolation of the signal to extract its value. The interpolation method is done using B-splines, which consist of piecewise polynomials functions of degree $k$, and employed in a separate \textit{Python} code connected to the main code. This degree is chosen to be \adr{of} 4th order to match the convergence order of the finite differences and the Runge-Kutta method. After obtaining all the $\{r,{\phi_{\omega_0}(t_m,r)},{\Pi_{\omega_0}(t_m,r)}\}$ values throughout all the $t_m$ slices, we can then reconstruct the signal at $t_{p_{match}}$ on the desired equispaced grid points by interpolating again.\par
The choice of $t_{p_{match}}$ and the $K_{CMC}$'s all have an effect on the length of signal we can recover. Specifically, the choice of $t_{p_{match}}$ and $S=1/\abs{K_{CMCm}} + 1/\abs{K_{CMCp}}$. Different values for these parameters lead to recovery of the signal at $t_{p_{match}}$  for different finite intervals of the grid ($r(t_m)$ values). It is not possible to recover the signal for the entirety of the space region $r \in [0,1]$, no matter the choice of parameters. Although including or excluding the origin (i.e.,~$r=0$) is optional, it is impossible to reach the region at $\scri^+$ (i.e.,~$r=1$), as that would require evolving on $\scri^-$ slices for infinite time. The region around spacelike infinity cannot be covered by hyperboloidal slices. Thus, one must carefully choose a suitable outgoing slice that maximizes coverage of the grid while also ensuring that the main body of the signal — its oscillatory part — lies within said grid. Recovering the main body of the signal close to $r=1$ is not desirable, due to the compression that an ingoing scalar field is subject to when approaching that boundary on outgoing slices, which negatively affects the resolution and convergence of our solutions. It is thus important that the initial signal is well-suited for translation, mainly being smooth enough and not too wide, so we recover the main body without loss of important information. In our particular setup, we choose $t_{p_{match}}=-6.2$, and our slices have $|K_{CMCm}|=|K_{CMCp}|=1$ which means we recover a grid $r \in [0.128037,0.677834]$.\par
This takes us to another issue: if we cannot recover the whole signal, we must reconstruct the missing parts throughout the remaining grid points. At the continuum level and for the chosen value of $t_{p_{match}}$, the signal around the origin and close to~$\scri^+$ is zero, but numerical results are imperfect and the gridfunctions take on small but non-zero values there that we need to model appropriately. We thus add an artificial decay for the initial and end tails of the signal. This means doing two more separate interpolations for these regions, where we add a finite number of decaying points on each side of the recovered grid (first and last $r_{match}$ points), besides the boundary points where the signal is forced to be null ($r=0$ and $r=1$). These points are equally spaced on the grid and, as we move away from the main body (to the left or to the right), each scalar field value at those points is 10 times smaller than at the previous point. We perform the same type of interpolation as the one used in the main body (B-splines), which does induce a wave-like behaviour near the tail, but forces a "decaying effect". Because we perform 3 interpolations separately for the 3 regions (both tails and main body), the spatial derivatives of the signal will be discontinuous at the innermost (first) and outermost (last) $r_{match}$ points. However, because at these points the signal is already very close to zero, this behaviour is not too problematic, though some loss of convergence around those regions is expected. This tradeoff is acceptable, as skipping the additional interpolations results in even worse convergence in the tails.

\subsection{Choice of the effective potential barrier}

The scattering problem between $\scri^-$ and $\scri^+$ becomes interesting provided we work with a non-trivial effective potential in (\ref{Eg:waveq}). Specifically, to obtain the phenomenon of particle creation, the potential must be time-dependent in an acceletared way so as to be able to mimic a dynamical gravitational field. If the potential vanishes or is only stationary, $V=V(r)$, the $\beta_{\omega\omega'}$ Bogoliubov coefficient is expected to vanish, meaning there is no particle creation, as with stationary spacetimes. In this work we will numerically explore four different scenarios: %(i) no potential, $V=0$, (ii) a static potential $V=V(r)$, (iii) and two time-dependent potentials that perform a fixed number of oscillations throughout a certain time period.} 

\begin{itemize}[leftmargin=*]
    \item \textbf{\adr{Vanishing} potential.} This is the simplest case, with  $V(t,r)=0$, corresponding to \adr{the scattering of spherical waves ($\ell=0$) in}  Minkowski spacetime. This case is useful as a first non-trivial test of our numerical code.

    \item \textbf{Static potential {barrier}.} We consider a time-independent potential, using a bump function for the radial profile \cite{alma9981639990001453}:
    \begin{equation}
    V(r)=
    \begin{cases}
         V_0 \, e^{-\frac{1}{1-\left(\frac{r-r_0}{\delta}\right)^2}}, &\text{if}  \,\,\,\,r_0-\delta \leq r \leq r_0 + \delta, \\
         0,  &\text{elsewhere},
    \end{cases}
    \label{Eq:RadialPot}
\end{equation}
where $V_0$, $r_0$ and $2\delta$ are, respectively, the amplitude, the center and  the width of the bump. This function is smooth at any point and has compact support. \adr{This case physically represents the presence of a spherical potential barrier located at radius $r=r_0$ in Minkowski space, which can be  understood as effectively originating from the presence of a compact, static star.}\par
\end{itemize}

In  these two cases no particle creation should take place, and we will use this property to check our numerical code. The second case is slightly more involved than the first one, because the potential is \adr{able} to scatter  incoming scalar waves. This is, if the amplitude of an incoming signal is equal or below $V_0$, a fraction should be reflected back by the potential, while the rest should get transmitted through it. Still, because of the static nature of the interaction, no ``frequency-mixing'' should arise. 

{When performing the translation of data from the in hyperboloidal foliation to the out one, it is convenient to choose a $t_p=t_{p_{match}}$ slice in the in foliation on  which the signal has not yet been affected by the potential. Otherwise, %If we choose a $t_p$ slice for which the field has already interacted with the potential, 
we might not be able to recover the whole signal in the out slice, as it has already been scattered,} and thus it can be non-zero for a wider range of compactified radii on the hyperboloidal slice.
\par
Regarding  dynamical scenarios, we propose to study two oscillating problems:

\begin{itemize}[leftmargin=*]

 \item \textbf{\adr{Pulsating potential barrier}.} In this setting, the potential is initially vanishing, and at some instant of time $t_{ON}$ its amplitude starts oscillating around a given value $V_{\max}$ for a finite amount of time, until it becomes null again for the remainder of the evolution. The time-dependent profile is chosen as
        \begin{equation}
            V(t,r)=
    \begin{cases}
         V_0(t) e^{-\frac{1}{1-\left(\frac{r-r_0}{\delta}\right)^2}}, &  r_0-\delta \leq r \leq r_0 + \delta, \\
         0,  &\text{elsewhere},
    \end{cases}
    \end{equation}
    with
    \begin{equation}
        V_0(t)= \frac{V_{max}}{2}\left(1+\sin \left[\omega_{pot}(t-t_{ON})-\frac{\pi}{2}\right]\right).
    \end{equation}

    \item \textbf{\adr{Shaking potential barrier}.} At early times the potential  $V=V(t,r)$  is static and given by \eqref{Eq:RadialPot}. Then, at some instant of time $t_{ON}$, the  potential barrier becomes time-dependent and its peak location starts oscillating radially between $r=r_{0}$ and $r=r_{0}+\Delta r$ a total number of $n$ cycles, after which it settles down to the static potential \eqref{Eq:RadialPot} once again. The dynamical regime is described by 
    \begin{equation}
            V(t,r)=
    \begin{cases}
         V_0 e^{-\frac{1}{1-\left(\frac{r-r_0(t)}{\delta}\right)^2}}, & r_0(t)-\delta \leq r \leq r_0(t) + \delta, \\
         0,  &\text{elsewhere},
    \end{cases}
    \end{equation}
    with
    \begin{equation}
        r_0(t)= r_{0} + \frac{\Delta r}{2}\left(1+\sin \left[\omega_{pot}(t-t_{ON})-\frac{\pi}{2}\right]\right).
    \end{equation}

\end{itemize}

In both cases $\omega_{pot}$ denotes the harmonic frequency of the oscillations, which start at $t=t_{ON}$ and finish at $t=t_{ON} + nT_{pot}$, where $n$ is the integer number of oscillations and $T_{pot}=(2\pi)/\omega_{pot}$ is the potential's period. The oscillations are restricted to a finite interval of simulation time, so that the spacetime remains stationary at both   ``past'' and ``future'' regions. \adr{Both effective potentials can be physically interpreted as representing the qualitative effects on the field modes by different oscillation forms of compact stars: the pulsating potential barrier may effectively describe the dynamics of stars with constant radius and variable mass; while the shaking potential barrier may describe stars with constant mass but oscillating radius.}

Since we aim to recover the signal at the hyperboloidal slice $t_p=t_{p_{match}}$ in the out foliation still unperturbed by the potential, we only  start the oscillations at the second half of the evolution (as there is no need to do it on the first half --- translation will recover the unperturbed signal for the second half). This only occurs at $t=t_{ON}$, which has to be chosen such that the signal is able to feel the effect of the dynamical potential for the finite amount of time it is oscillating.

\subsection{Computation of Bogoliubov coefficients}

To determine particle creation, we need to compute the Bogoliubov coefficients, as explained in Sec.~\ref{Section:Theor}. %\sout{However, unlike most theoretical works, our simulation follows a forward propagation approach, evolving} 
To do so we have to evolve {each} scalar field {mode} $\phi_{\omega_0}$ from $\scri^-$ to $\scri^+$. Then, the $\alpha_{\omega \omega'}$ and $\beta_{\omega \omega'}$ coefficients can be obtained from the KG product
\begin{equation}
    \alpha_{\omega \omega'}=\left ( u_\omega^{in},u_{\omega'}^{out}\right)  \quad \text{and} \quad  \beta_{\omega \omega'}=-\left( u_\omega^{in},u_{\omega'}^{out*}\right), \label{alphabeta}
\end{equation}
where $u^{in}_{\omega}$ represent the positive-frequency \textit{in} modes and $u^{out}_\omega$ the positive-frequency \textit{out} modes, to make a clear distinction between the modes and the initial signal we are propagating, $\phi_{\omega_0}$. The hypersurface where we compute the KG product is taken to be $\scri^+$. Since our initial signal constitutes an \textit{in} mode, as it is defined at $\scri^-$ in accordance to the inertial coordinates of the \textit{in} region, the evolved scalar field $\phi_{\omega_0}$ at $\scri^+$ is the resulting numerical evolution of an \textit{in} mode that arrives at the \textit{out} region. This means that, in the KG products, the \textit{in} modes $u_\omega^{in}$ and $\p_u u_\omega^{in}$ are given, respectively, by the normalized values of $\phi_{\omega_0}$ and $\Pi_{\omega_0}$ at $\scri^+$. Finally, because our initial signal is an \textit{in} mode with frequency $\omega=\omega_0$, then the Bogoliubov coefficients we are computing are actually $\alpha_{\omega_0\omega'}$ and $\beta_{\omega_0\omega'}$, given by
\begin{subequations}
\label{Eq:BogoliubovCoefficientAnalysis}
\begin{align}
     \alpha_{\omega_0\omega'}&=-i \int_{\scri^+}du r^2 d\Omega \left(\phi_{\omega_0}\p_u  u_{\omega'}^{out*} - u_{w'}^{out*} \, \Pi_{\omega_0}  \right),
     \label{Eq:AlphaNewExpansion}\\
     \beta_{\omega_0\omega'}&=i\int_{\scri^+}du r^2 d\Omega \left( \phi_{\omega_0}\p_u  u_{\omega'}^{out} - u_{\omega'}^{out} \, \Pi_{\omega_0}\right).
     \label{Eq:BetaNewExpansion}
\end{align}
\end{subequations}

\begin{figure}[h!]
    \centering
    \includegraphics[width=0.9\linewidth]{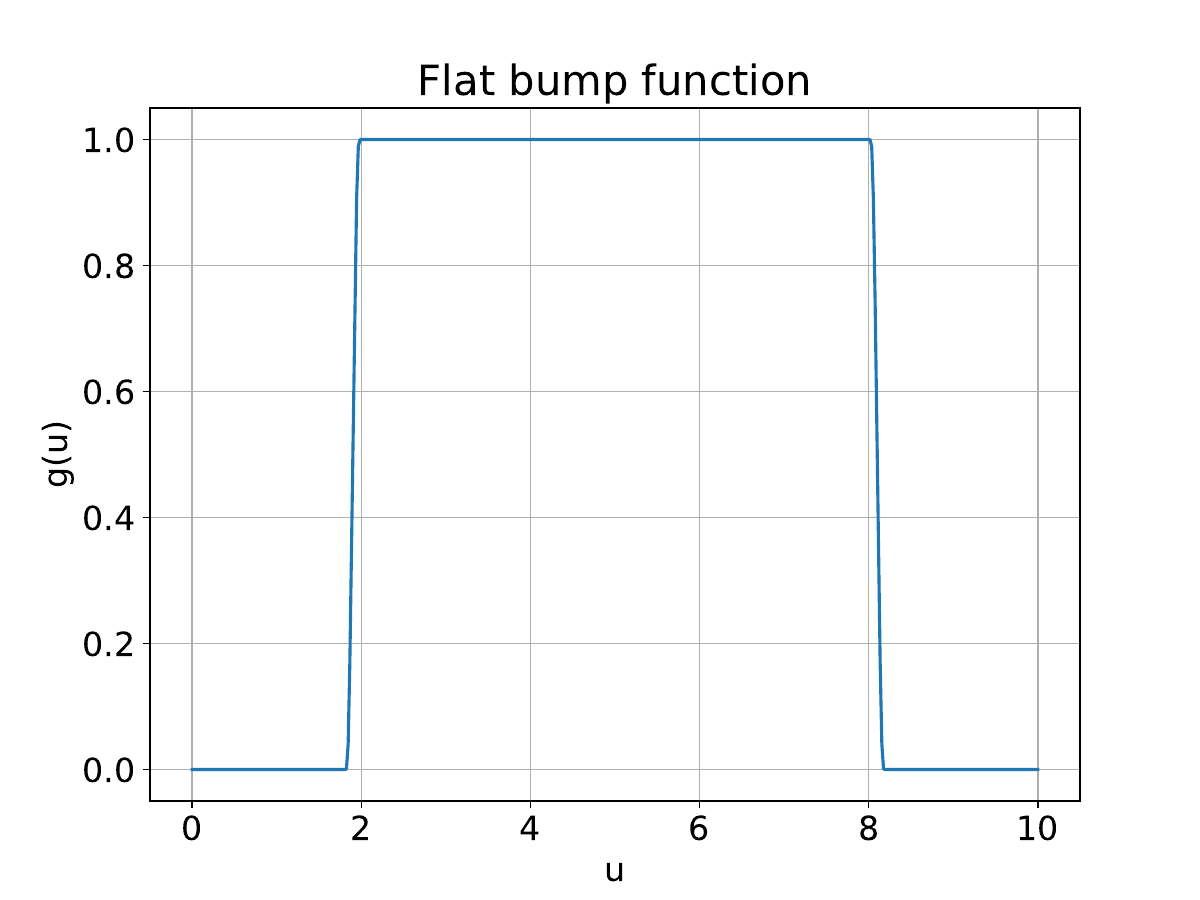}
    \caption{Illustration of the flat bump function defined in (\ref{Eq:BumpFunctionOutMode}) with parameters $q=5$, $a=3$ and $b=3.2$.}
    \label{fig:FlatBumpFunction}
\end{figure}

The $u_{\omega'}^{out}$ modes, with which we project our signal, must form an orthonormal basis. %\color{red}%
The standard practice is to use a plane-wave basis. Since the finiteness of the numerical grid leads to some technical difficulties, we  modulate the plane-wave basis %} 
by a bump function so that they are localized in time. At future null infinity, the out modes are then given by
\begin{equation}
u_{\omega'}^{out}=g(u)e^{-i\omega'u}.
\label{Eq:OutMode}
\end{equation}
Here, $g(u)$ is chosen to be a smooth $C^\infty$, flat bump function \cite{tu_introduction_2011} -- a compactly supported function that is identically 1 on some interval and transitions to 0 at the boundaries. This ensures the \textit{out} modes are compact supported and as close as possible to plane waves within the plateau region. This choice also grants a clear way to normalize the modes, as we will see later. The function is defined as:
\begin{equation}
g(u)=
    \begin{cases}
         0, & u-q \leq -b,\\
         1-\frac{e^{-\frac{b^2-a^2}{(u-q)^2-a^2}}}{e^{-\frac{b^2-a^2}{(u-q)^2-a^2}}+e^{-\frac{b^2-a^2}{b^2-(u-q)^2}}} , &-b<u-q<-a\\
         1, & -a \leq u-q \leq a,\\
         1-\frac{e^{-\frac{b^2-a^2}{(u-q)^2-a^2}}}{e^{-\frac{b^2-a^2}{(u-q)^2-a^2}}+e^{-\frac{b^2-a^2}{b^2-(u-q)^2}}} , &a<u-q<b\\         
         0,  &u-q \geq b,
    \end{cases}
    \label{Eq:BumpFunctionOutMode}
\end{equation}
where $q$, $a$ and $b$ control, respectively, the center, width and steepness of the envelope. More specifically, $b-a$ is the length of the regions where the bump function goes from 0 to 1. Therefore this difference controls how steep this passage is and when it tends to 0, the bump function gradually becomes a square window. The profile of this function is presented in Figure \ref{fig:FlatBumpFunction}. The time spacing in $u$ used to define the \textit{out} modes is the same as the one extracted from the signal at future null infinity.\par

{Formulas (\ref{Eq:BogoliubovCoefficientAnalysis}) require that} 
%Before proceeding to the calculation of the Bogoliubov coefficients it is important to normalize 
our signal $\phi_{\omega_0}$ and the \textit{out} modes {are properly normalized}. This is done by dividing our signal  and the \textit{out} mode with  {their respective KG norm}:
\begin{subequations} \label{Eq:Normalization}
\begin{align}
    \hat\phi_{\omega_0} \Big|_{\scri^+}= &\frac{\phi_{\omega_0} \Big|_{\scri^+}}{\sqrt{\left( \phi_{\omega_0},\phi_{\omega_0} \right)}}\,, \quad  \hat\Pi_{\omega_0} \Big|_{\scri^+}= \frac{\Pi_{\omega_0} \Big|_{\scri^+}}{\sqrt{\left ( \phi_{\omega_0},\phi_{\omega_0} \right)}}\,, \\\quad & \hat u_{\omega}^{out}\Big|_{\scri^+}=\frac{u_{\omega}^{out}\Big|_{\scri^+}}{\sqrt{\left( u_{\omega}^{out},u_{\omega}^{out} \right)}}\,.
\end{align}
\end{subequations}
%These normalized quantities are the ones that should be used in the computation of the Bogoliubov coefficients in equation \eqref{Eq:BogoliubovCoefficientAnalysis}. 
Because the KG product is invariant in time, the integrals in \eqref{Eq:BogoliubovCoefficientAnalysis} can be computed at any hypersurface {of the foliation}, like  $\scri^-$ or $\scri^+$. %or at any other hyperboloidal slice. %their value will remain constant, so long as the whole signal is inside of the domain. 
The value of the KG product will be the same (up to numerical errors) provided the whole signal is contained in the domain of integration. %\adr{It is also for this reason that the same KG product is used as a normalization constant for both $\phi$ and its time derivative $\Pi_{\omega_0}$}. 
Since the \textit{out} modes are defined and localized at future null infinity, their KG product is {naturally} computed there. As for the $\phi_{\omega_0}$ signal, we can compute the KG product  at $\scri^+$ so long as the simulation is run for sufficient time to ensure the whole signal is fully covered. However, to avoid errors arising from numerics, it is more efficient to compute the KG product at $\scri^-$ directly, using the initial signal. 
%This can be done because the KG product of a signal on a hyperboloidal slice will be the same (up to numerical errors) as the KG product computed at $\scri^-$ and at $\scri^+$, if the signal is fully covered in any of these hypersurfaces. 
%Therefore, the normalization is done computing~\eqref{Eq:Normalization} with the following KG products
In summary:
\begin{subequations}
\begin{align}
    \left( \phi_{\omega_0},\phi_{\omega_0} \right) &=-i\int_{\scri^-}r^2dv d\Omega\left(\phi_{\omega_0} \Pi_{\omega_0}^*-\phi_{\omega_0}^*\Pi_{\omega_0}\right),\\
    \left( u_{\omega'}^{out},u_{\omega'}^{out} \right) &=-i\int_{\scri^+}r^2du d\Omega\left(u_{\omega'}^{out} \p_u u_{\omega'}^{out*}-u_{\omega'}^{out*} \p_u u_{\omega'}^{out}\right).
\end{align}    
\end{subequations}

When projecting the \textit{out} modes over the signal one must take into account the maximum frequency that can be resolved with the time spacing we are using. A reference for this threshold is given by the Nyquist frequency $f_N$, which depends on the sampling rate $f_S$ 
\begin{equation}
    f_N=\frac{1}{2}f_S=\frac{1}{2 \Delta t} (Hz),
\end{equation}
and where the sampling rate depends on the time spacing used in the output of the simulation. Frequencies above the Nyquist frequency suffer a distortion effect known as aliasing, where the sampled signal will appear to have a lower frequency than its original frequency. This Nyquist frequency is converted to angular frequency, which is then taken as a reference for the maximum resolved frequency. The frequency spacing used is taken to be $\Delta \omega'=1/T$ where $T$ is the total time length of the processed signal.\par

After extracting and normalizing the signal from $\scri^+$, we loop over the frequency range—from $\Delta \omega'$ up to near the Nyquist frequency, in steps of $\Delta \omega'$—and, for each $\omega'$, we define the corresponding \textit{out} modes, normalize them, and compute the Bogoliubov coefficients $\alpha_{\omega_0 \omega'}$ and $\beta_{\omega_0 \omega'}$. Finally, the coefficients can be plotted as functions of $\omega'$, yielding two spectra. Regardless of the specific scenario under study, and independently of whether particle creation occurs, the coefficients must satisfy Eq.~\eqref{Eq:identity_alphabeta_1}, which we rewrite here for convenience:
%After extracting the signal from $\scri^+$ and normalizing it, we make a cycle over the frequencies (from $\Delta w'$ to near Nyquist frequency in multiples of $\Delta w'$) where, for each $w'$, we define the \textit{out} modes, normalize them and with them compute the $\alpha_{w_0w'}$ and $\beta_{w_0w'}$ coefficients. In the end, we can plot the coefficients as functions of $w'$, obtaining two spectra. Irrespective of the scenario being studied and creation of particles taking place, the coefficients should obey equation~\eqref{Eq:identity_alphabeta_1}, which we now rewrite here as
\begin{equation}
    \int_{0}^{+\infty} d\omega' \left( |\alpha_{\omega_0\omega'}|^2  - |\beta_{\omega_0\omega'}|^2  \right)=1,
    \label{equation:ImportantFinalSteprewritten}
\end{equation}
%where we changed the variables $(n=k' \rightarrow w_0)$ and $(k \rightarrow w')$ to agree with our current nomenclature. 
In the numerical implementation, an indefinite integral cannot be performed, so the integration is carried out between the minimum and maximum sampled frequencies. Most spectra decay to zero as the frequency increases; however, if the signal contains high-frequency components beyond the numerical resolution, these cannot be recovered, and the resulting spectra may therefore be incomplete.
%In the numerical context, we cannot perform an indefinite integral, so we perform the numerical integration between the minimum and maximum frequencies. Most spectra fall to zero as we increase the frequency, however if there are higher frequencies in the signal that cannot be resolved, then those will not be recovered as well and the spectra might be incomplete.\par

There are some further technicalities regarding the analysis to be explained. As mentioned before, most theoretical works use unlocalized plane waves, which are not applicable in this numerical approach. Here we need to apply some form of envelope to the plane waves we wish to use on both the initial signal and the \textit{out} modes. The width of these signals becomes entirely our choice. For the initial signal the width is somewhat restricted to our translation process - we cannot have a very wide signal that cannot be fully translated to the outgoing slices. It also seems there is no clear choice for the width of the \textit{out} modes. This matter becomes more complicated when noticing that the KG product of these modes is sensitive to their width. This, in turn, affects their normalization which ultimately changes the normalized values of the $\alpha_{\omega\omega'}$ and $\beta_{\omega\omega'}$ coefficients, hindering our ability to reproduce the result of the  condition \eqref{equation:ImportantFinalSteprewritten}. Now we present our procedure to find an unambiguous normalization.

\begin{figure}[h]
    \centering
    \includegraphics[width=0.90\linewidth]{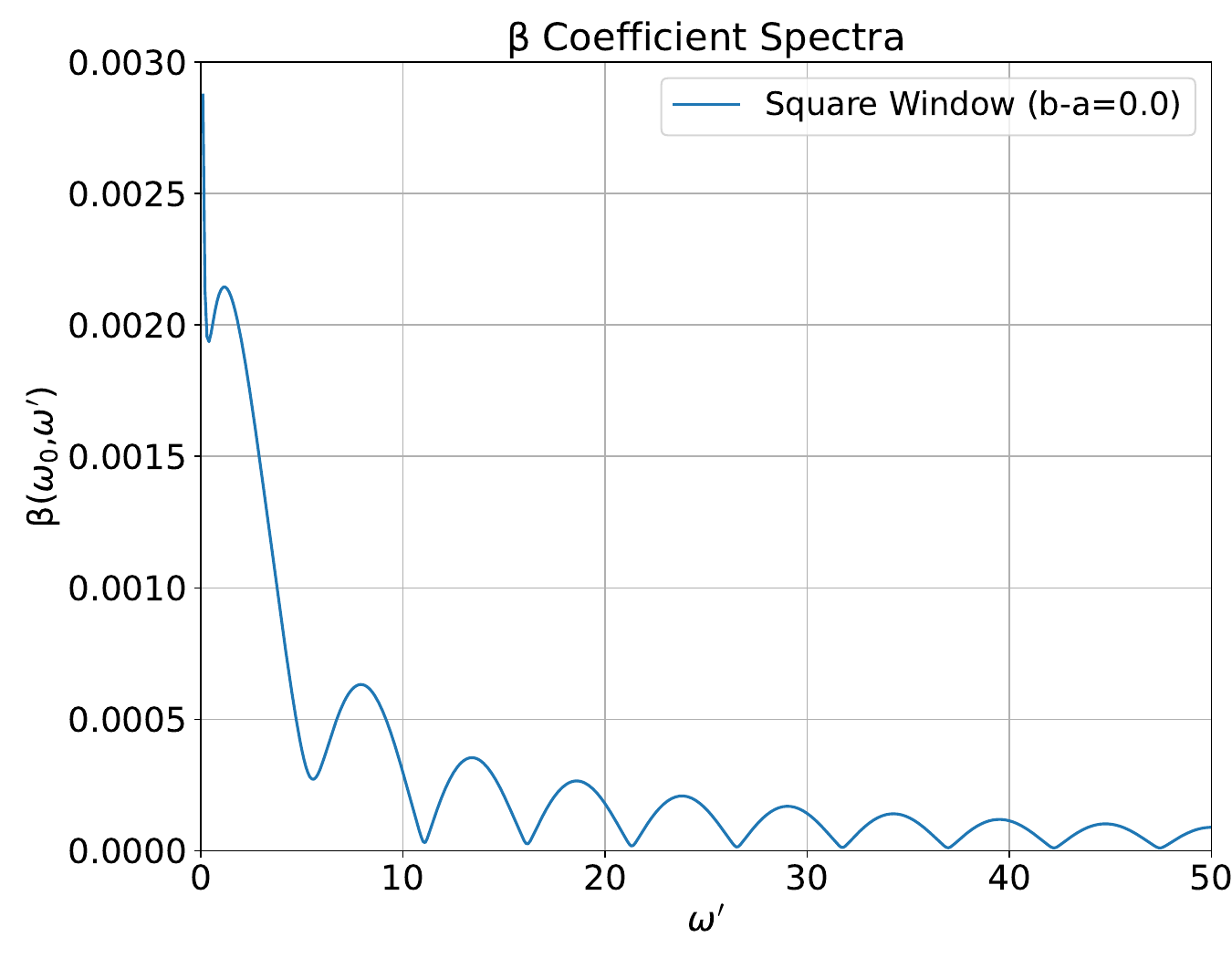}
    \includegraphics[width=0.90\linewidth]{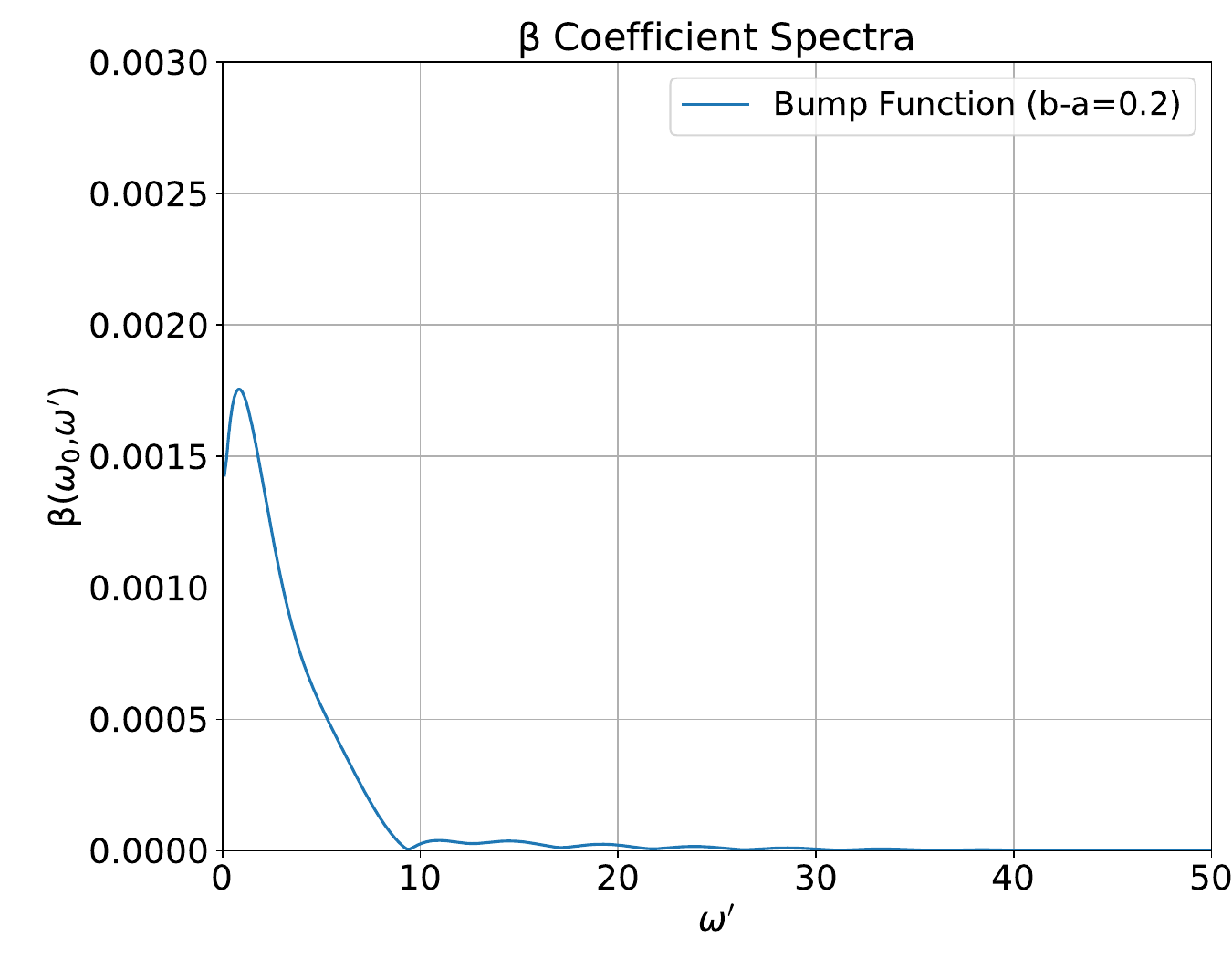}
    \caption{ {Frequency spectrum for the Bogoliubov coefficient $\beta_{\omega_0\omega}$ for fixed $\omega_0$  in the case of a vanishing potential barrier. The spectrum is computed from (\ref{Eq:BetaNewExpansion}) using a square window for the envelope of the \textit{out} modes (up) and a bump function with $b-a=0.2$ for the envelope (down). The results found are compatible with zero, which indicates no particle creation, as expected}.}
    \label{fig:EffectofSteepnessonSpectra}
\end{figure}

Firstly, the width of the \textit{out} modes ($2a$) is taken to be the width of the main body of the signal at $\scri^+$. This is found by defining a threshold (in our case $10^{-2}$) and extracting the time interval for which the normalized signal $\phi_{\omega_0}$ and $\Pi_{\omega_0}$ are above said threshold. Slightly changing this threshold does not qualitatively change our results. Once this time interval is obtained, one easily finds the center of the signal, which is then used as the center $q$ of the \textit{out} modes on the analysis. The choice of the slope's interval $b-a$ also has some effects on the analysis' results. Taking the interval to be null turns the bump function into a rectangular window and undermines the smoothness of our obtained spectra, as can be seen on Figure [\ref{fig:EffectofSteepnessonSpectra}]: for the square window, the spectrum exhibits noticeable bumps, whereas in the other case, the spectra appears smoother and more well-behaved. 

\begin{figure}[h]
    \centering
    \includegraphics[width=0.90\linewidth]{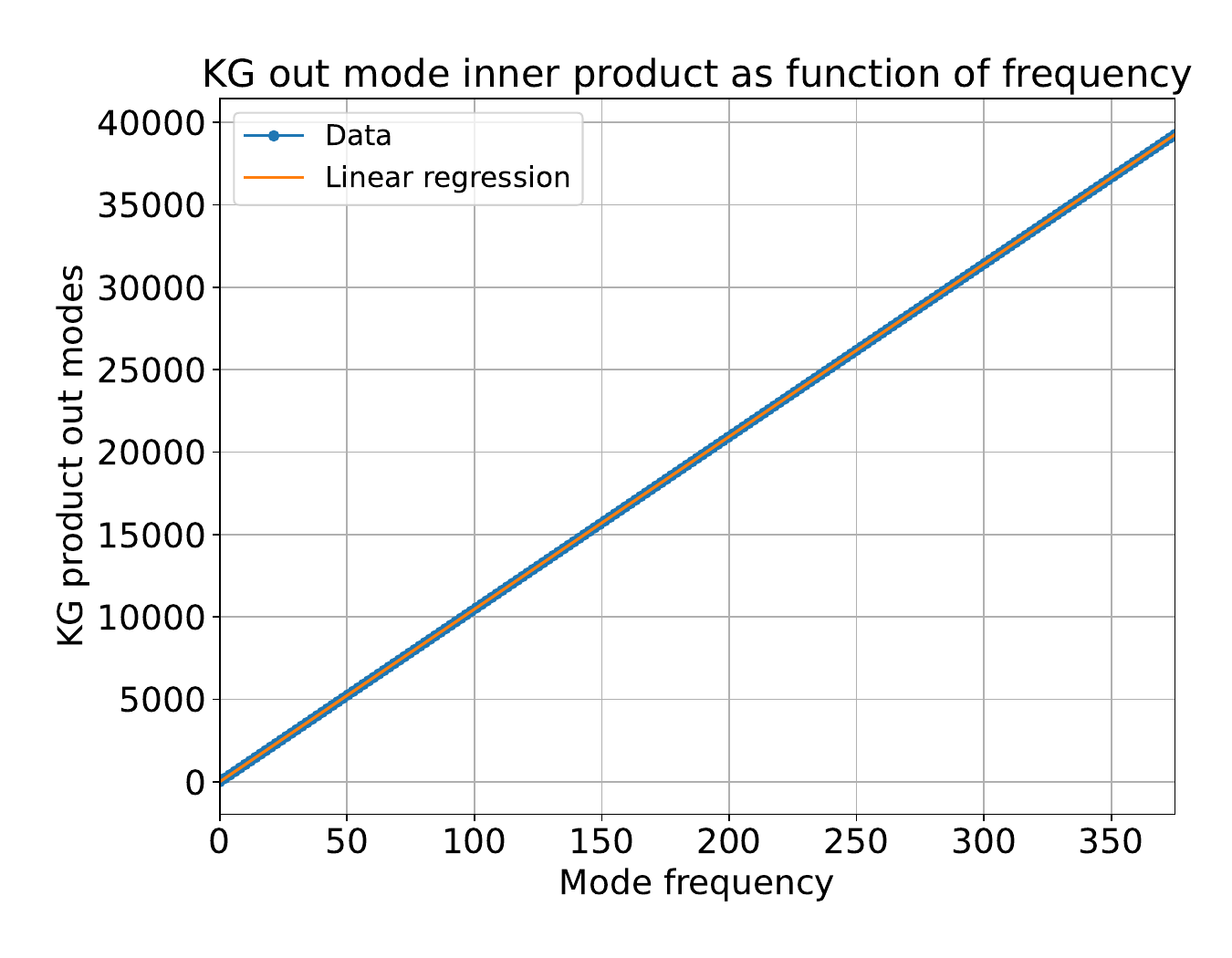}
    \includegraphics[width=0.90\linewidth]{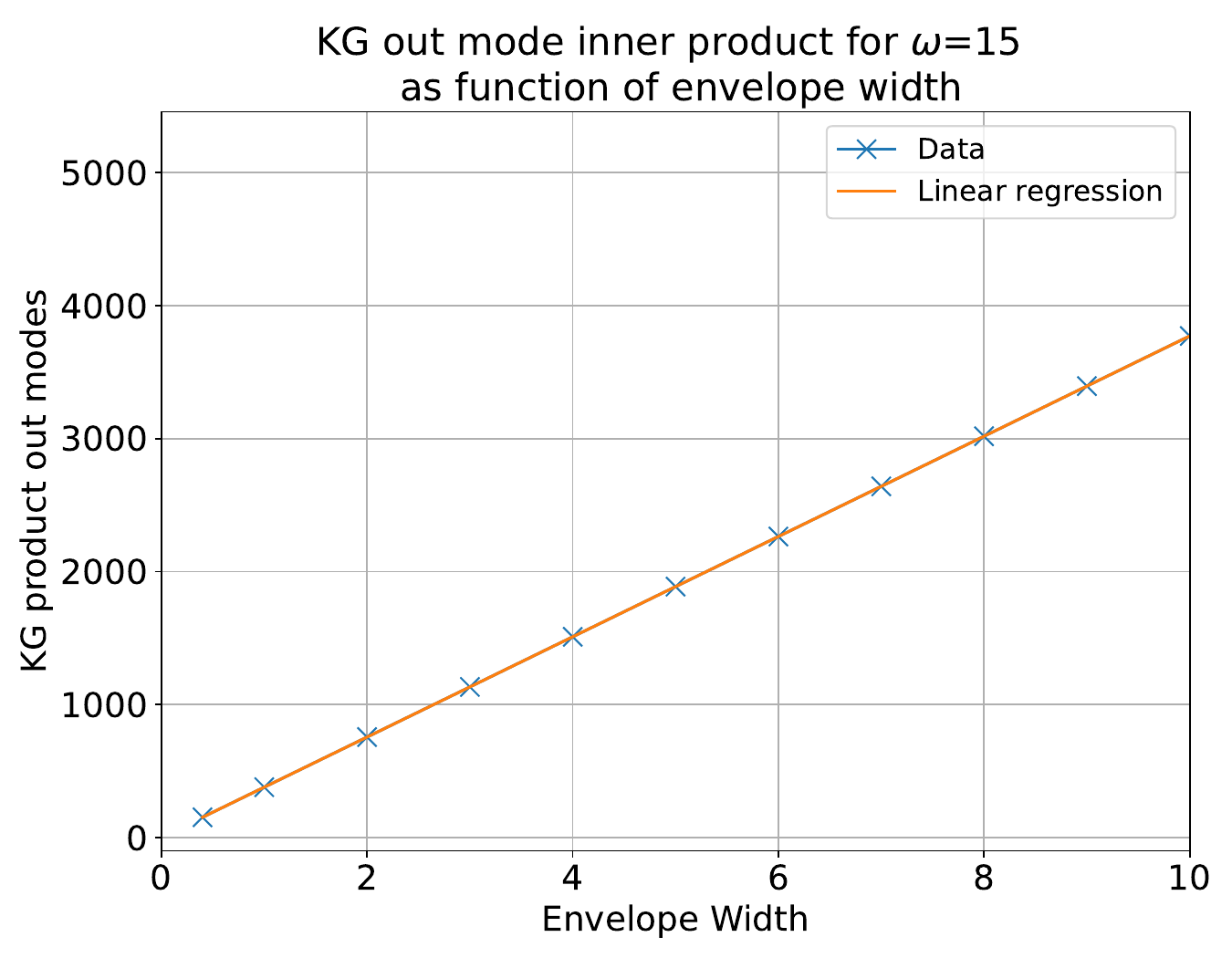}
    \caption{Plot showing the dependence of the  {norm $\left( u_{\omega}^{out},u_{\omega}^{out} \right)$} for the \textit{out} mode (\ref{Eq:OutMode}) with respect to the mode frequency $\omega$ (upper panel) and with respect to the envelope width $2a$ (lower panel). In both cases, we assume  $q=17$ and $r=b-a=0.2$, and a linear regression is performed.}
    \label{fig:KGdependenceWidthFreq}
\end{figure}

%\par
Secondly, we find that the KG product has a linear dependence on the frequency of the modes as well as on the width of the envelope. These results can be observed in Figure \ref{fig:KGdependenceWidthFreq}.
By performing linear regressions for the dependence of the KG product on the width for modes with different frequencies, we find that the slope $m$ and the y-intercept $c$ are also dependent on the frequency, that is $m=c_1\omega$ and t $c=c_2\omega$. Therefore, the KG product of the \textit{out} modes takes the form
\begin{equation}
    KG=c_1\omega(2a)+c_2\omega,%{\color{red}
    \quad c_1 \approx 25, \quad c_2 \approx 4.7.%}
\end{equation}

Further tests showed that $c_1$ and $c_2$ change with the envelope slope interval $b-a$. More precisely, as the envelope gradually becomes a square window $(b-a \rightarrow 0)$, $c_2$ decreases and the vertical shift becomes negligible.\par

To obtain a value that is independent of the width of the envelope used for the \textit{out} modes, we have to multiply the \textit{out} KG product by a scaling factor. Taking into account the dependence of the product we find, the rescaling involves first a vertical shift and then a multiplication by a scaling factor. Furthermore, this scaling factor includes a division by the width of the modes to make the product independent of their length. The KG product used for the normalization of the \textit{out} modes thus is rescaled as
\begin{equation}
      \left( u_{\omega}^{out},u_{\omega}^{out} \right) \longrightarrow \frac{ \left( u_{\omega}^{out},u_{\omega}^{out}\right)-c_2\omega}{2a} \times 6.29,
      \label{Eq:NewNormalization}
\end{equation}
where $c_2=c_2(b-a=0.2)\approx4.4659$. The $6.29$ factor appears as an almost experimental adjustment to correct the discrepancy observed during testing between the obtained results and the expected spectral behaviour, ensuring that equation (\ref{equation:ImportantFinalSteprewritten}) holds. Although this factor was found empirically, the same rescaled KG product (meaning the same values of $c_2$ and $6.29$) is used in all of the analysis for all the scenarios studied, providing consistent results aligned with theoretical predictions.

%{\color{red}
We further note that, although the windowed out modes (\ref{Eq:OutMode}) are not strictly orthogonal for different $\omega$, the support of the evolved signal at future null infinity, $(\phi_{\omega_0},\Pi_{\omega_0})$, lies entirely within the window interval $-a< u-q< a$ in (\ref{Eq:BumpFunctionOutMode}), so that the  integrals in (\ref{Eq:AlphaNewExpansion})-(\ref{Eq:BetaNewExpansion}) effectively project onto the standard plane-wave basis. For this reason, the lack of exact orthogonality of the windowed modes does not affect the  physical interpretation of the Bogoliubov coefficients in our setup. In particular, the coefficients $\beta_{\omega_0\omega'}$ can still be consistently interpreted as encoding particle creation.%}

\section{Numerical results}
\label{Section:Results}

In this section, we present the results of our numerical simulations and their corresponding analysis, organized on a case-by-case basis. Each case includes the obtained spectra of the Bogoliubov coefficients, and, in the case of the pulsating potential, the time evolution of the initial signal as well.

The parameters used for the initial data in Eqs.~(\ref{Eq:InitialPhi})–(\ref{Eq:initialPi}) are $\omega_0 = 15$, $t_0 = 2.5$, and $\sigma = 0.8$. The time step and grid spacing are $\Delta t = 0.000625$ and $\Delta x = 0.0005$, respectively, corresponding to $N = 2000$ grid points used for the numerical integration of the equations in Sec.~III.A. {The strength parameter in the Kreiss-Oliger dissipation~\cite{kreiss1973methods} included is set to $\xi = 0.02$.}

For the effective potentials, we take $V_0 = V_{\mathrm{max}} = 1000$, $r_0 = 0.1$, and $\delta = 0.1$. The dynamical potentials are switched on at $t_{\mathrm{ON}} = 14$ and perform $n$ complete oscillations with frequency $\omega_p = 8$, where $n = 6$ for the pulsating potential and $n = 3$ for the shaking potential barrier.

\subsection{{Numerical verification of Klein–Gordon inner product invariance}}
To assess the accuracy and reliability of our numerical simulations, we first verify the time-independence of the Klein–Gordon (KG) inner product for each of the background effective potentials considered. This quantity should yield the same value when evaluated on any complete spacelike hypersurface—whether defined by standard Minkowski, hyperboloidal, or advanced/retarded time coordinates—provided that the signal remains fully contained within the computational domain. Accordingly, we compute here the KG norm of the evolved scalar mode $\phi_{\omega_0}$ on different hyperboloidal slices, as well as at past and future null infinity, using, respectively,
\begin{subequations}
\begin{align}\label{normspatialslice}
    \left( \phi_{\omega_0},\phi_{\omega_0} \right) &=-i\int_{\Sigma_t} r^2 drd\Omega \frac{1}{N} \left[\phi_{\omega_0} \Pi_{\omega_0}^* -\phi_{\omega_0}^* \Pi_{\omega_0} \right.\\
    & \quad \quad \quad \quad \quad \quad \quad\left.-N^r(\phi_{\omega_0} \p_r \phi_{\omega_0}^* -\phi_{\omega_0}^* \p_r  \phi_{\omega_0})\right], \nonumber\\
    \left ( \phi_{\omega_0},\phi_{\omega_0} \right) &=-i\int_{\scri^-}r^2dv d\Omega(\phi_{\omega_0} \Pi_{\omega_0}^*-\phi_{\omega_0}^* \Pi_{\omega_0}),\\
        \left( \phi_{\omega_0},\phi_{\omega_0} \right) &=-i\int_{\scri^+}r^2du d\Omega(\phi_{\omega_0} \Pi_{\omega_0}^*-\phi_{\omega_0}^* \Pi_{\omega_0}),
\end{align}
\end{subequations}
%where $\phi_1=\phi_2=\phi_{\omega_0}$, our propagated signal.
where the lapse $N$ and the shift vector $N^\mu$ % which should not be confused with the $\alpha_{\omega \omega'}$ and $\beta_{\omega \omega'}$ coefficients, 
are given by
\begin{equation}
    N=\sqrt{\Omega^2+\left(\frac{K_{CMC}r}{3}\right)^2}, \quad \quad  N^\mu=\left(0,\frac{K_{CMC}r}{3},0,0\right)
\end{equation}
with $\Omega=\abs{K_{CMC}}(1-r^2)/6$ the {same choice of conformal factor made above}.\par

\begin{figure}[h]
    \centering
    \includegraphics[width=0.95\linewidth]{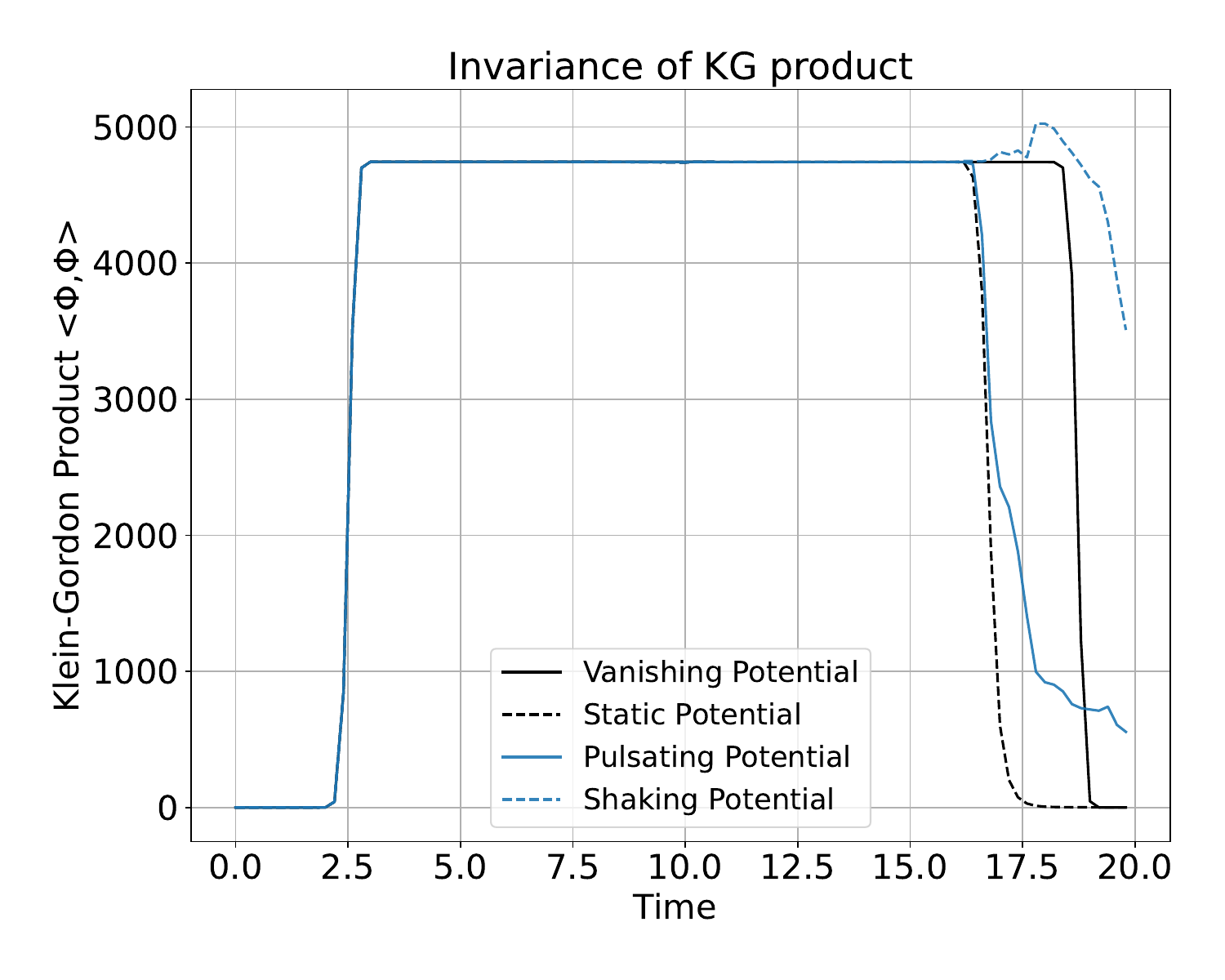}
    \caption{{Computation of the norm $ \left( \phi_{\omega_0},\phi_{\omega_0} \right)$ on each hyperboloidal slice $\Sigma_t$, using Eq.~(\ref{normspatialslice}), as a function of time $t$ throughout the entire numerical evolution of the scalar mode $\phi_{\omega_0}$ as scattered by the four different effective potentials considered in this work. }}%The plot shows a zoomed-in view around the time-interval during which the KG \adr{norm} remains largely unchanged.
    \label{fig:Inv_KG_zoomed}
\end{figure}

{Our results are summarized in Fig.~\ref{fig:Inv_KG_zoomed} and Table~\ref{tab:KGProductatnullinfty}.}
They confirm that the Klein–Gordon (KG) product remains conserved only when the entire signal is contained within the computational domain.
Any departure from strict time-independence originates from numerical inaccuracies or physical fluxes crossing the numerical domain boundaries (i.e., portions of the signal entering or leaving the grid), as illustrated in Fig.~\ref{fig:Inv_KG_zoomed}. The nearly constant value of the KG norm computed on most hyperboloidal slices agrees with that obtained along $\scri^-$ and $\scri^+$, as shown in Table~\ref{tab:KGProductatnullinfty}, further confirming that the KG product remains invariant regardless of the hypersurface on which it is evaluated. 

The case with a vanishing potential barrier shows the best KG conservation, since the signal does not scatter and remains entirely within the domain for longer. For non-trivial potential barriers, the scattering causes part of the signal to leave the domain earlier, leading to small deviations from perfect conservation.

\begin{table}[h]
    \centering
    \begin{tabular}{|c|c|c|c|c|}
    \hline
         $(\phi_{\omega_0},\phi_{\omega_0})$ at &  Vanishing $V$ & Static $V$ & Pulsating $V$  & Shaking $V$ \\
       \hline
         $ \scri^-$  & $4743.5$ & $4743.5$ &$4743.5$ &$4743.5$\\
       \hline
         $\scri^+$ & $4741.9$ & $4741.3$ & $4741.9$ & $4674.3$\\
       \hline
       Relative error & $0.03 \%$ & $0.05 \%$ &$0.04\%$ & $1.5\%$ \\
       \hline
    \end{tabular}
    \caption{KG norms {of $\phi_{\omega_0}$} computed at both $\scri^-$ and $\scri^+$ for the four effective potentials  studied in this work. The initial signal $\phi_{\omega_0}$ is the solution of the KG equation with initial data \eqref{Eq:initialPhiPi} and $\omega_0=15$, $t_0=2.5$, $\sigma=0.8$.}
    \label{tab:KGProductatnullinfty}
\end{table}

%For the scenarios where the signal gets trapped inside the potential, it is difficult to obtain a perfect match between the KG product computed at $\scri^-$ and at $\scri^+$, as one would need to evolve the system for longer for the full signal to leave the spacetime through future null infinity. This is most notable for the shaking potential barrier, whose Klein-Gordon product at $\scri^+$ does not match the KG product values at both $\scri^-$ and the hyperboloidal slices, with a relative error of  $\approx 1.460 \%$.

In scenarios where the signal becomes trapped by the potential, the KG product at $\scri^+$ cannot fully match that at $\scri^-$ without a longer evolution time. This mismatch is most notable in the shaking potential case, showing a relative error of about 1.460\%.

\adr{This test on the time-independence of the KG product verifies that our numerical framework respects the unitarity of the scalar field theory, and ensures that the total particle number predicted in equation (\ref{totalN}) will produce finite numbers.}

%\subsection{Stationary scenarios}

%We now present the results obtained when propagating an initial signal with frequency $w_0=15$, $t_0 = 2.5$ and $\sigma=0.8$. The time step used was $\Delta t=0.000625$ and the grid spacing was $\Delta x=0.0005$ which is equivalent to $N=2000$ grid points. Dissipation is set to $\xi=0.02$.
%\adr{As Alex says, I would unify sections B and C into a single one, and display the 4 scenarios one after the other.}

\subsection{Case 1: Vanishing potential}

After evolving the initial signal with initial frequency $\omega_0$ in the absence of a potential barrier, we extract the resulting scalar field mode at $\scri^+$ by recording the signal data at $r = 1$ throughout the entire second evolution. Following the normalization procedures given in Eqs.~(\ref{Eq:Normalization}) and (\ref{Eq:NewNormalization}), and the computation of the Bogoliubov coefficients in Eq.~(\ref{Eq:BogoliubovCoefficientAnalysis}), we obtain the $\alpha_{\omega\omega'}$ and $\beta_{\omega\omega'}$ spectra.

\begin{figure}[h!]
    \centering
    \includegraphics[width=0.9\linewidth]{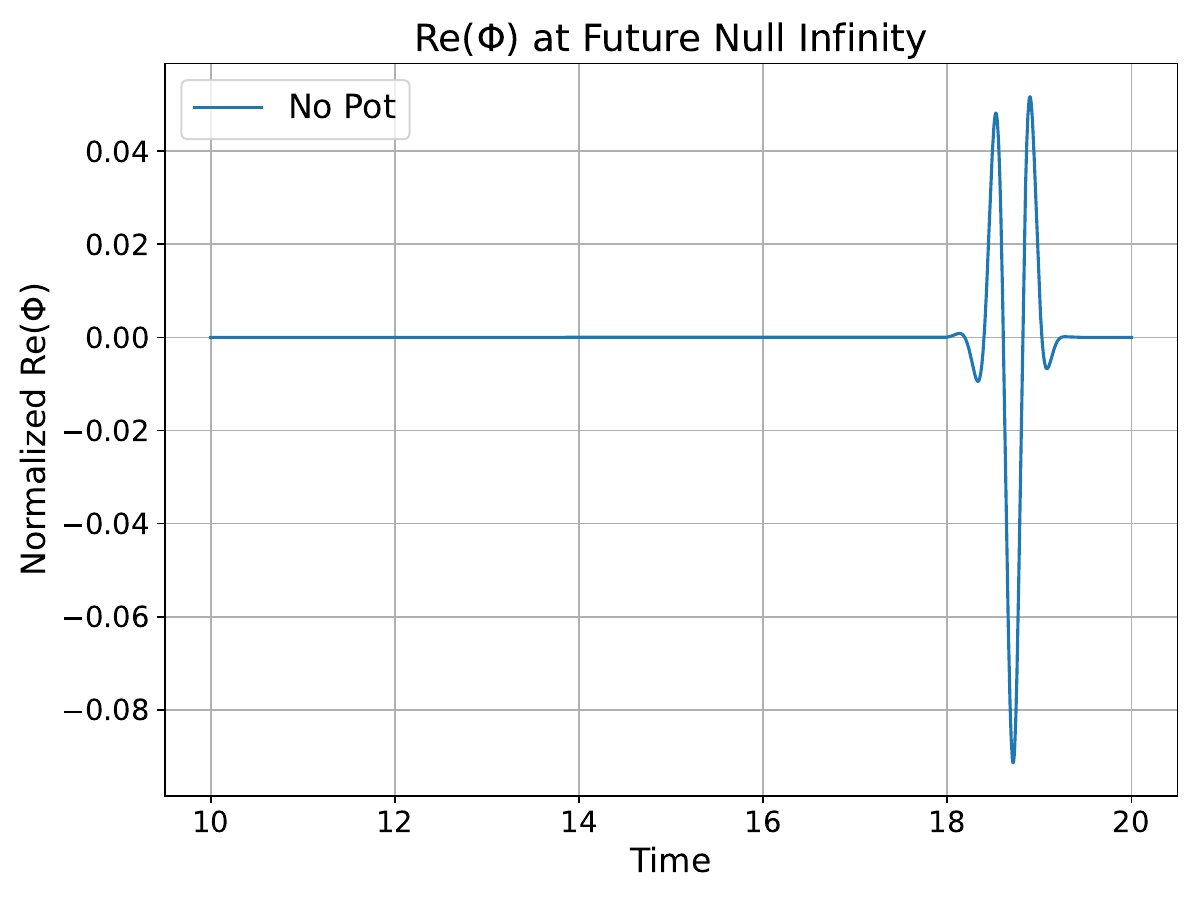}
    \includegraphics[width=0.9\linewidth]{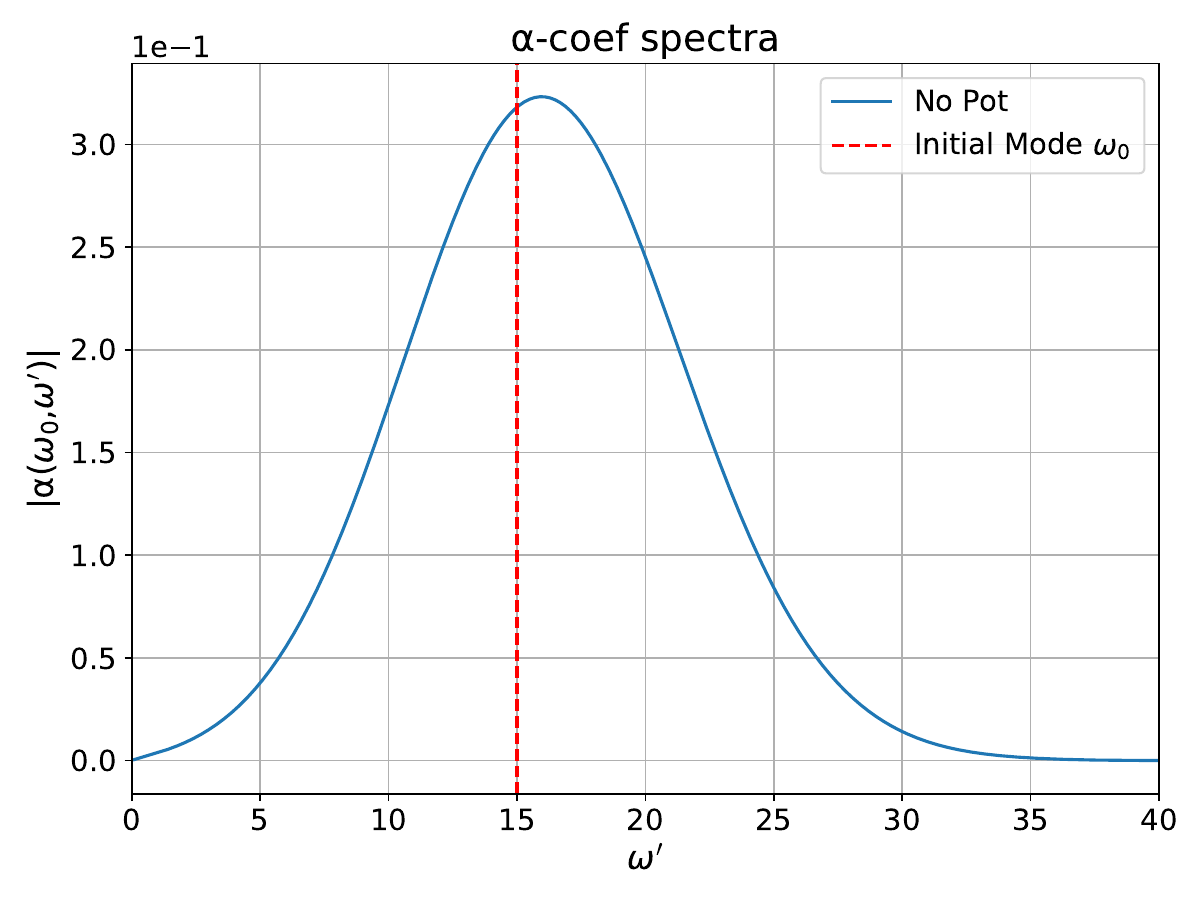}
    \includegraphics[width=0.9\linewidth]{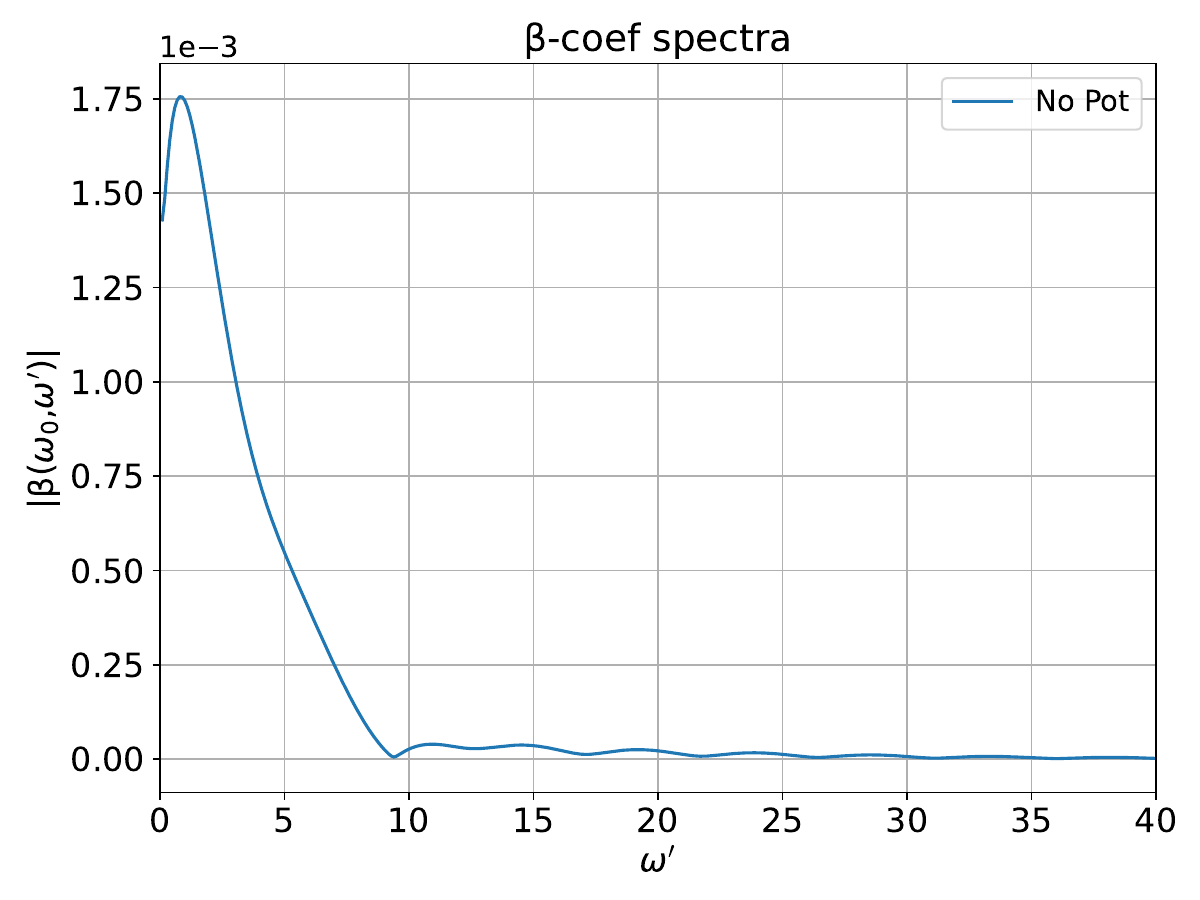}    
    \caption{ Real part of the scattered signal $\phi_{\omega_0}$ received at $\scri^+$ for a vanishing potential barrier, with initial frequency $\omega_0 = 15$ (top panel). The corresponding Bogoliubov coefficients $\alpha_{\omega_0 \omega'}$ (middle panel) and $\beta_{\omega_0 \omega'}$ (bottom panel) are shown as functions of $\omega'$. The $\alpha_{\omega_0 \omega'}$ spectrum is peaked around $\omega'=\omega_0$. The $\beta_{\omega_0 \omega'}$ spectrum is compatible with zero (in particular, two orders below the $\alpha_{\omega_0 \omega'}$ spectrum), indicating that particle creation does not occur.}
    \label{fig:NoPotBogSpectra}
\end{figure}

The evolved signal at $\scri^+$ and the corresponding spectra are displayed in Fig.~\ref{fig:NoPotBogSpectra}. As expected, the mode $\phi_{\omega_0}$ reaches $\scri^+$ unperturbed, preserving its waveform but with an inverted sign relative to the initial signal. The $\alpha_{\omega\omega'}$ coefficients peak around the initial frequency of the evolved signal, while the $\beta_{\omega\omega'}$ coefficients are negligible—approximately two orders of magnitude smaller than the $\alpha_{\omega\omega'}$ ones.

To verify the consistency of our results, we also compute condition~(\ref{equation:ImportantFinalSteprewritten}), which yields a numerical value of $1.00045$, consistent with the unitarity condition for the Bogoliubov transformation.

\subsection{Case 2: Static potential barrier}
For the static potential case, we propagate the same initial signal as before, but now in the presence of a potential of the form~(\ref{Eq:RadialPot}) with parameters $V_0 = 1000$, $r_0 = 0.1$, and $\delta = 0.1$. In contrast to the previous case, the introduction of a static potential affects the propagation of $\phi_{\omega_0}$, primarily by partially reflecting the signal as it reaches the barrier and partially confining the transmitted component. The trapped portion subsequently bounces between the origin and the potential boundary, gradually leaking out and propagating towards $\scri^+$. Consequently, to recover the full signal at future null infinity, the evolution time must be extended long enough for the trapped component to fully exit the computational domain.

\begin{figure}[h!]
    \centering
    \includegraphics[width=0.9\linewidth]{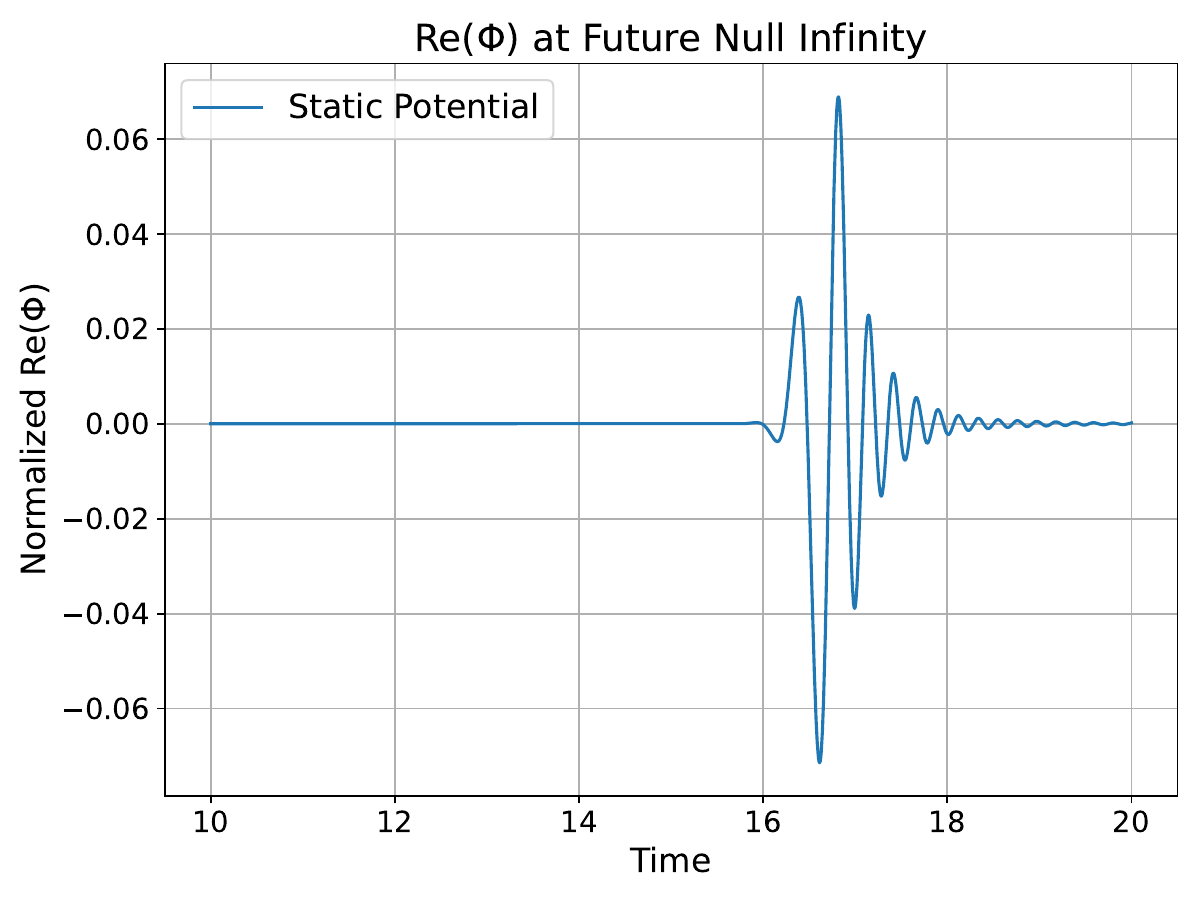}
    \includegraphics[width=0.9\linewidth]{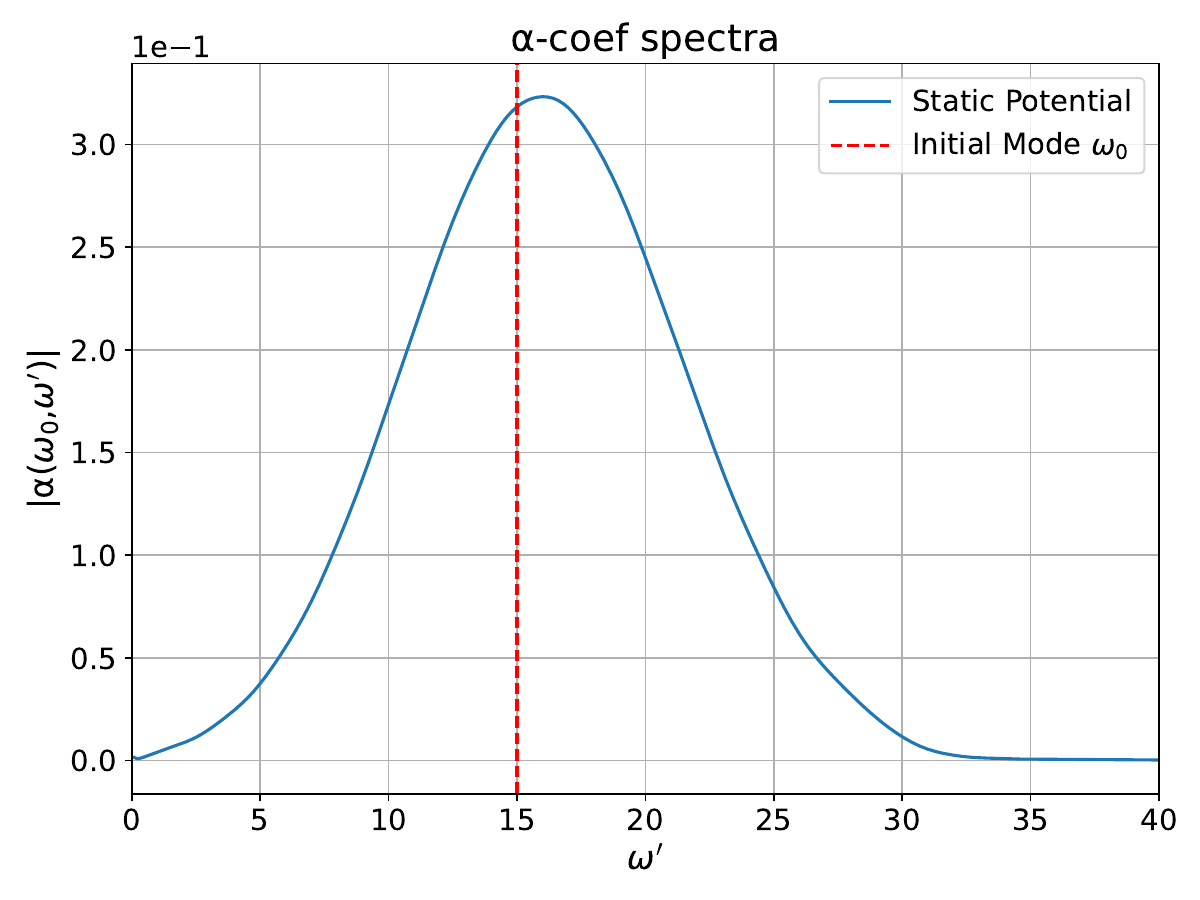}
    \includegraphics[width=0.9\linewidth]{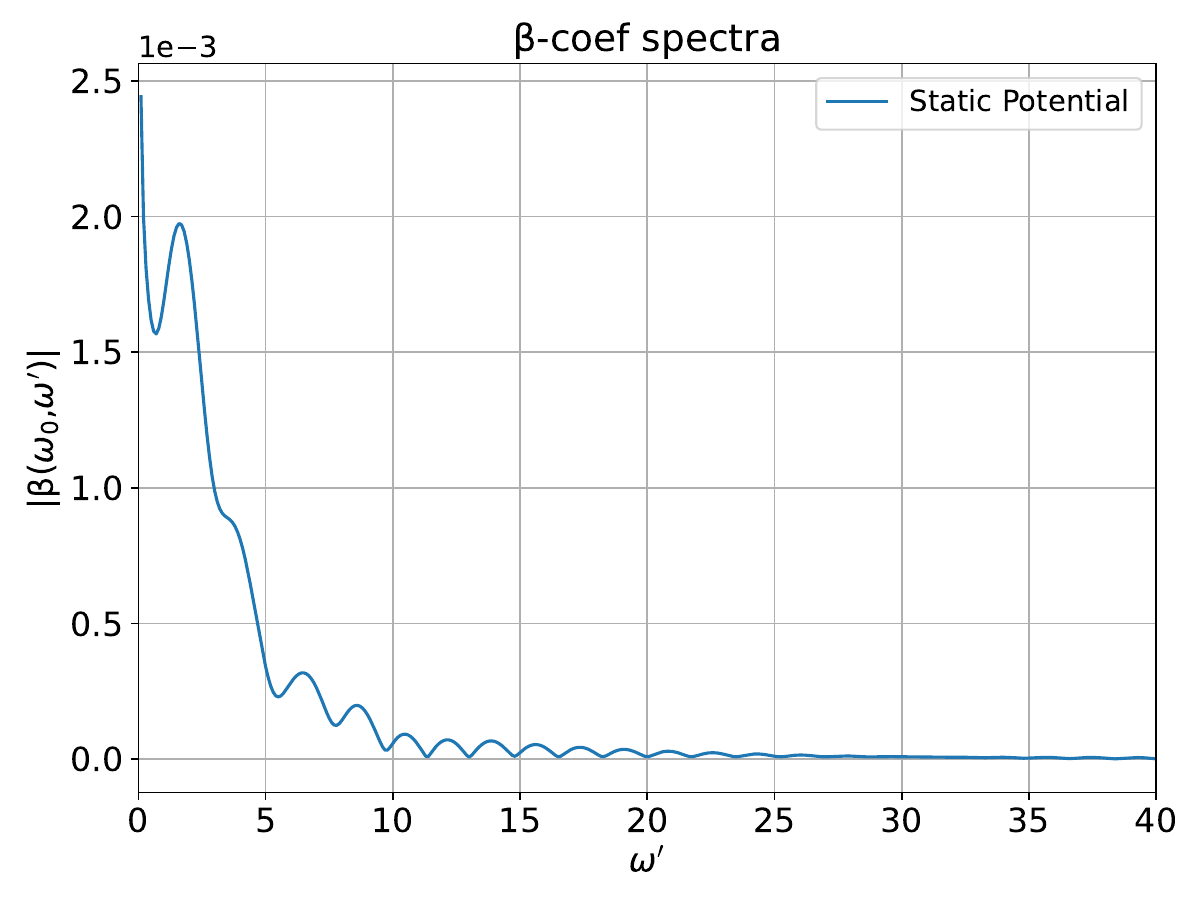}
    \caption{{Real part of the scattered signal $\phi_{\omega_0}$ received at $\scri^+$ for a static potential barrier, with initial frequency $\omega_0 = 15$ (top panel). The corresponding Bogoliubov coefficients $\alpha_{\omega_0 \omega'}$ (middle panel) and $\beta_{\omega_0 \omega'}$ (bottom panel) are shown as functions of $\omega'$. The $\beta_{\omega_0 \omega'}$ spectrum is compatible with zero, indicating that particle creation does not occur.}}
    \label{fig:RadialPotBogSpectra}
\end{figure}

Figure~\ref{fig:RadialPotBogSpectra} shows the real part of the mode $\phi_{\omega_0}$  at $\scri^+$, together with the spectra of the Bogoliubov coefficients. As expected, the signal arrives at $\scri^+$ distorted, exhibiting damped oscillations as the trapped portion slowly leaks out of the potential. However, the $\alpha_{\omega\omega'}$ spectrum displays no new frequency components, indicating that the outgoing signal roughly preserves its initial frequency. The $\beta_{\omega\omega'}$ coefficients remain negligible—roughly two orders of magnitude smaller than the $\alpha_{\omega\omega'}$ ones—which is consistent with the theoretical expectation that no particle creation occurs in a stationary background.

The numerical check of the unitarity condition~(\ref{equation:ImportantFinalSteprewritten}) yields a value of $1.001492$ for this case.

\subsection{Case 3: Pulsating potential barrier}

\begin{figure}[h!]
    \centering
    \includegraphics[width=0.9\linewidth]{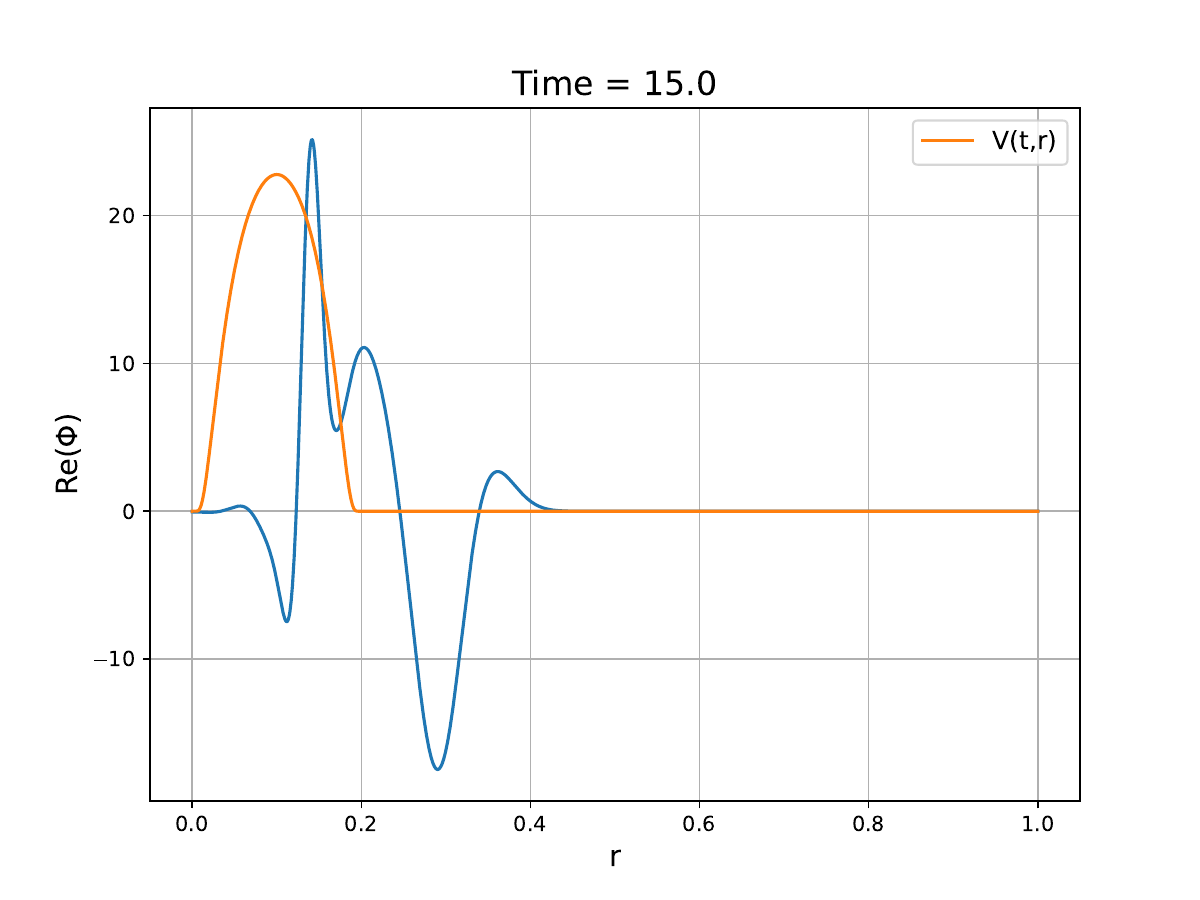}
    \includegraphics[width=0.9\linewidth]{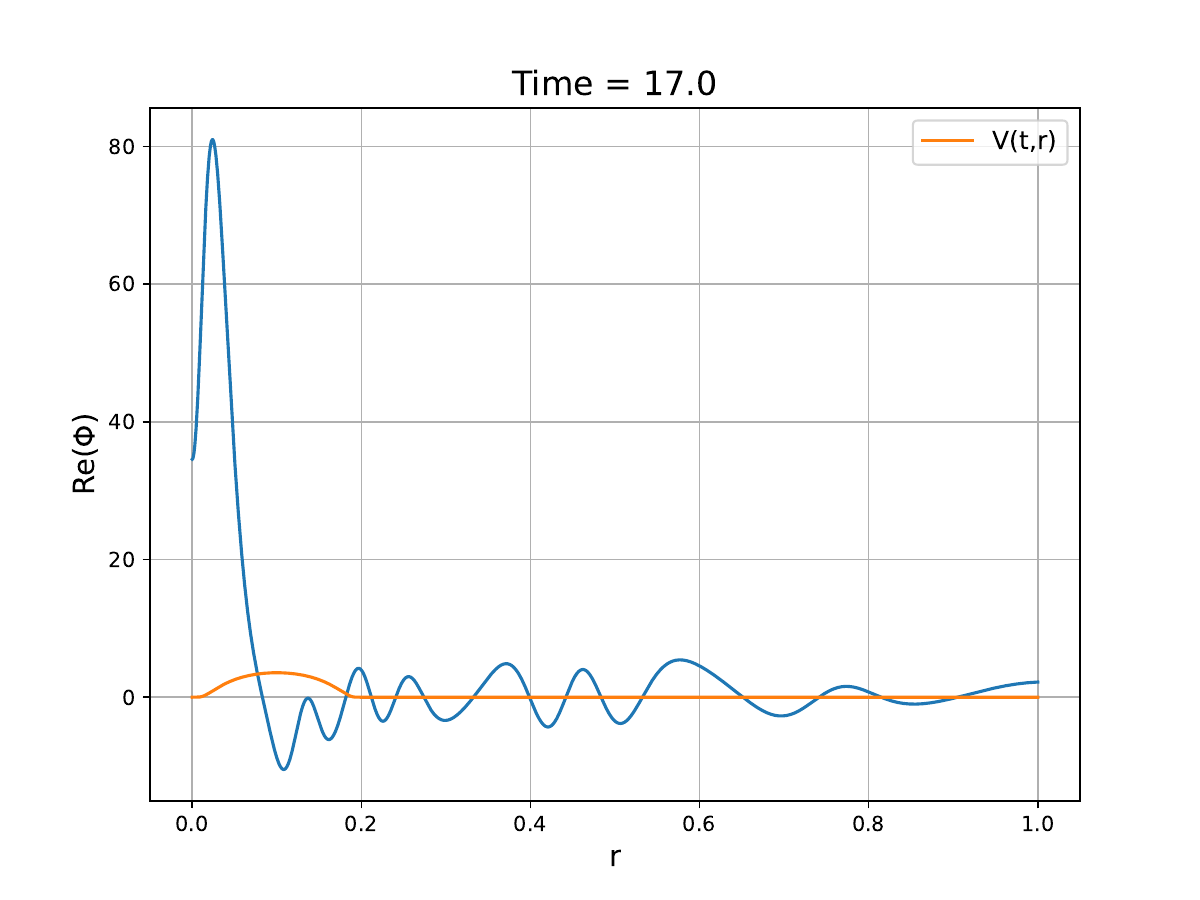}
    \includegraphics[width=0.9\linewidth]{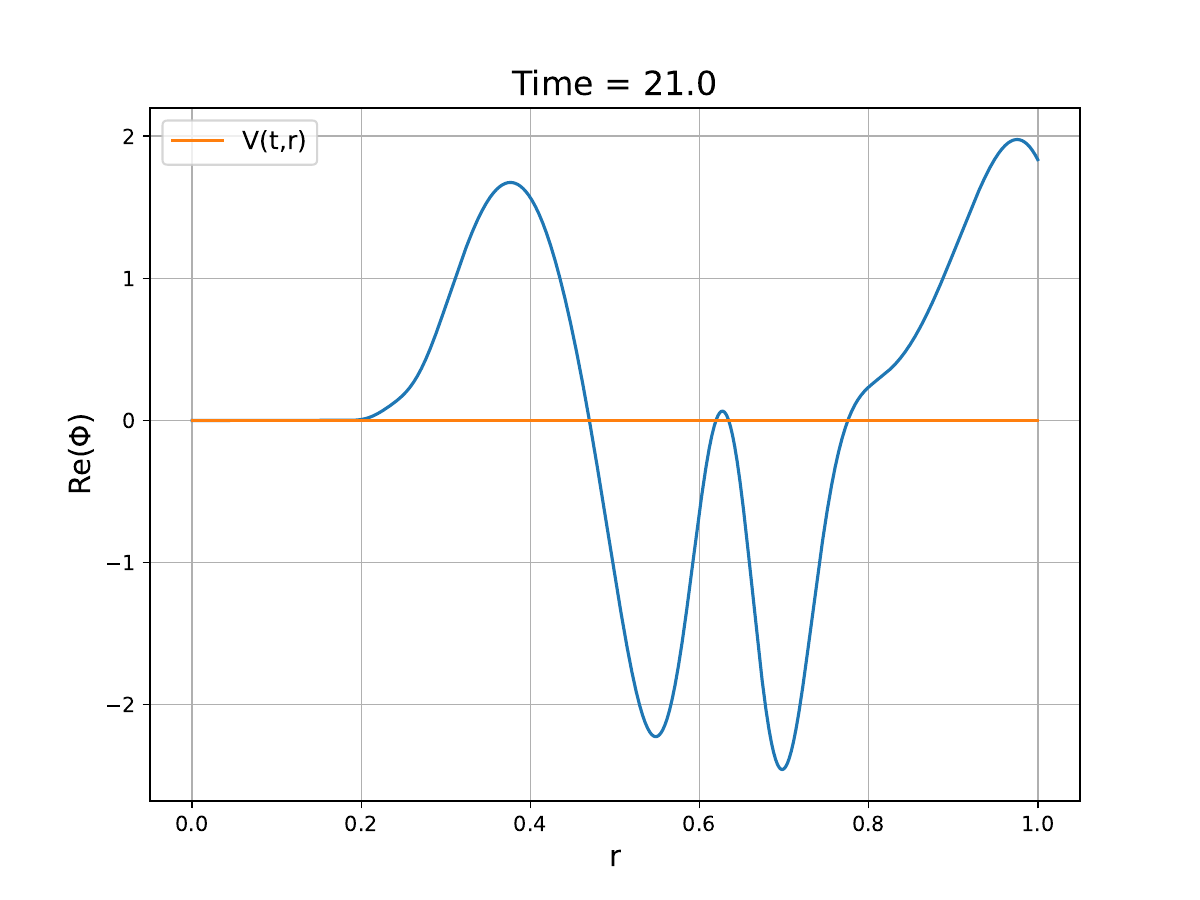}
    \caption{{Time evolution of Re($\phi_{\omega_0}(t,r))$ when the scalar field is scattered by the pulsating potential barrier (depicted by the orange line). The time evolution is illustrated with three different snapshots (from up to down), where each one shows the radial profile of Re($\phi_{\omega_0}(t,r)$). At each snapshot the potential is rescaled using the maximum value of Re($\phi_{\omega_0}$) for ease of visualization.}}
    \label{fig:AmpPotEvolution}
\end{figure}

Figure~\ref{fig:AmpPotEvolution} shows the evolution of the initial signal under the influence of a pulsating potential performing $n = 6$ oscillations. Since the translated signal at $t_p = -6.2$ is unaffected by the potential, the first half of the evolution {(finishing at $t_{\text{final}}^{(1)} = 10$)} is identical to the vanishing potential case, and we therefore only display the real part of $\phi_{\omega_0}$ after this translation. The second half of the evolution is run for a longer time {(ending at $t_{\text{final}}^{(2)} = 25$ as measured from the start of the simulation)} so that the entire signal has sufficient time to reach $\scri^+.$%, with the first and second stages ending at $t_{\text{final}}^{(1)} = 10$ and $t_{\text{final}}^{(2)} = 25$, respectively (both measured from the start of the simulation).

When the potential is switched on at $t_{ON} = 14.00$, the signal immediately begins to scatter—part of it is reflected by the barrier, while another part penetrates into the potential region. Because the potential itself oscillates, the trapped component of the signal is periodically amplified, leading to a buildup of energy within the potential region. A fraction of this trapped energy leaks out during the oscillations, while the rest remains confined until the potential returns to its stationary regime $V(t, r) = 0$, at which point it escapes and propagates towards $\scri^+$.

\begin{figure}[h!]
    \centering
    \includegraphics[width=0.9\linewidth]{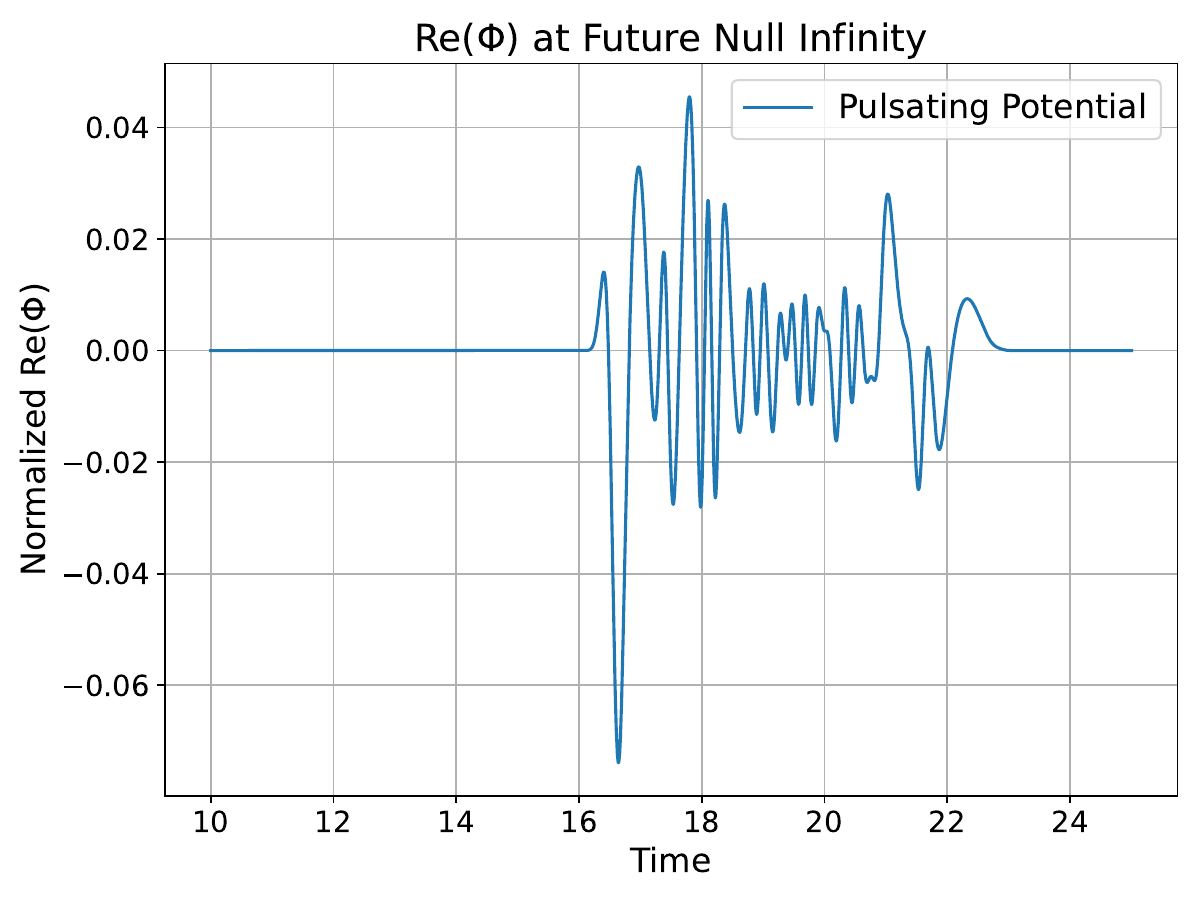}
    \includegraphics[width=0.9\linewidth]{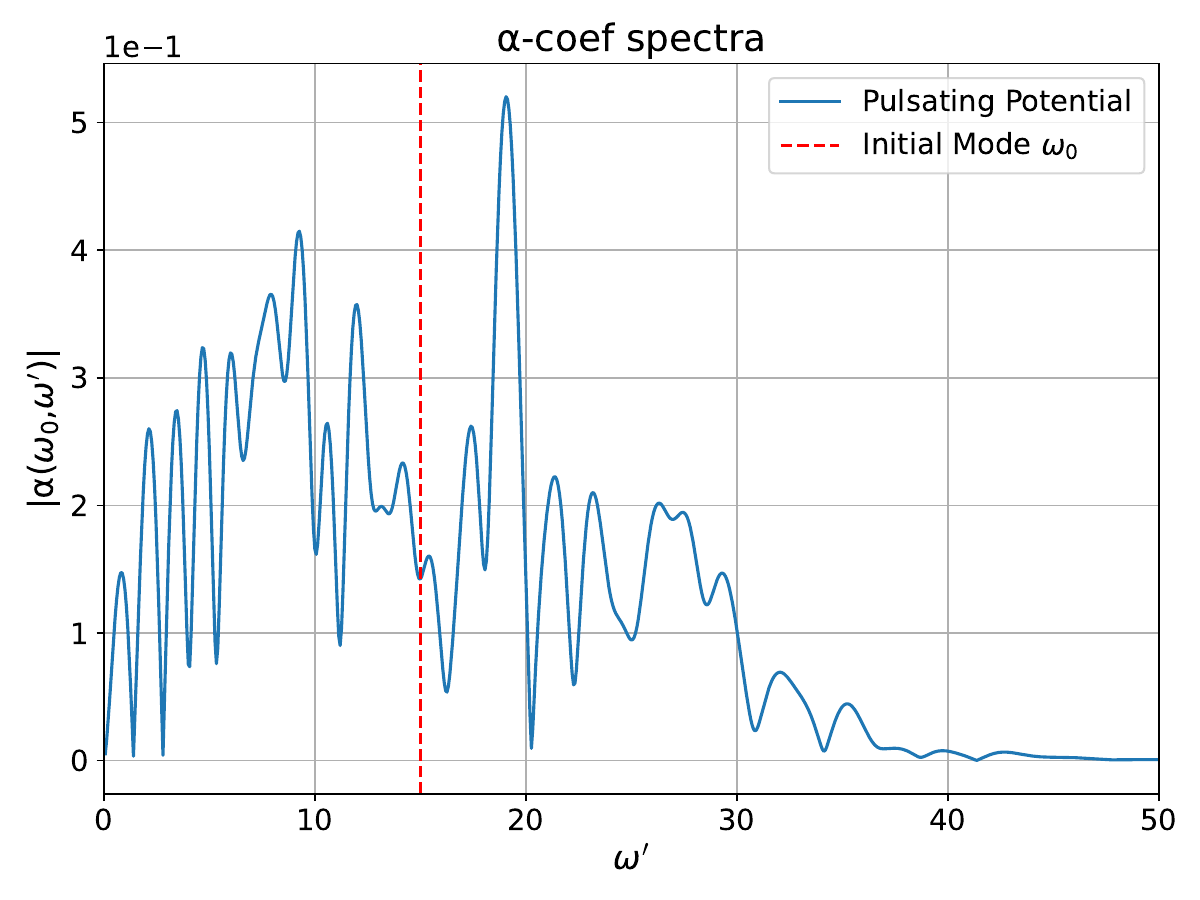}
    \includegraphics[width=0.9\linewidth]{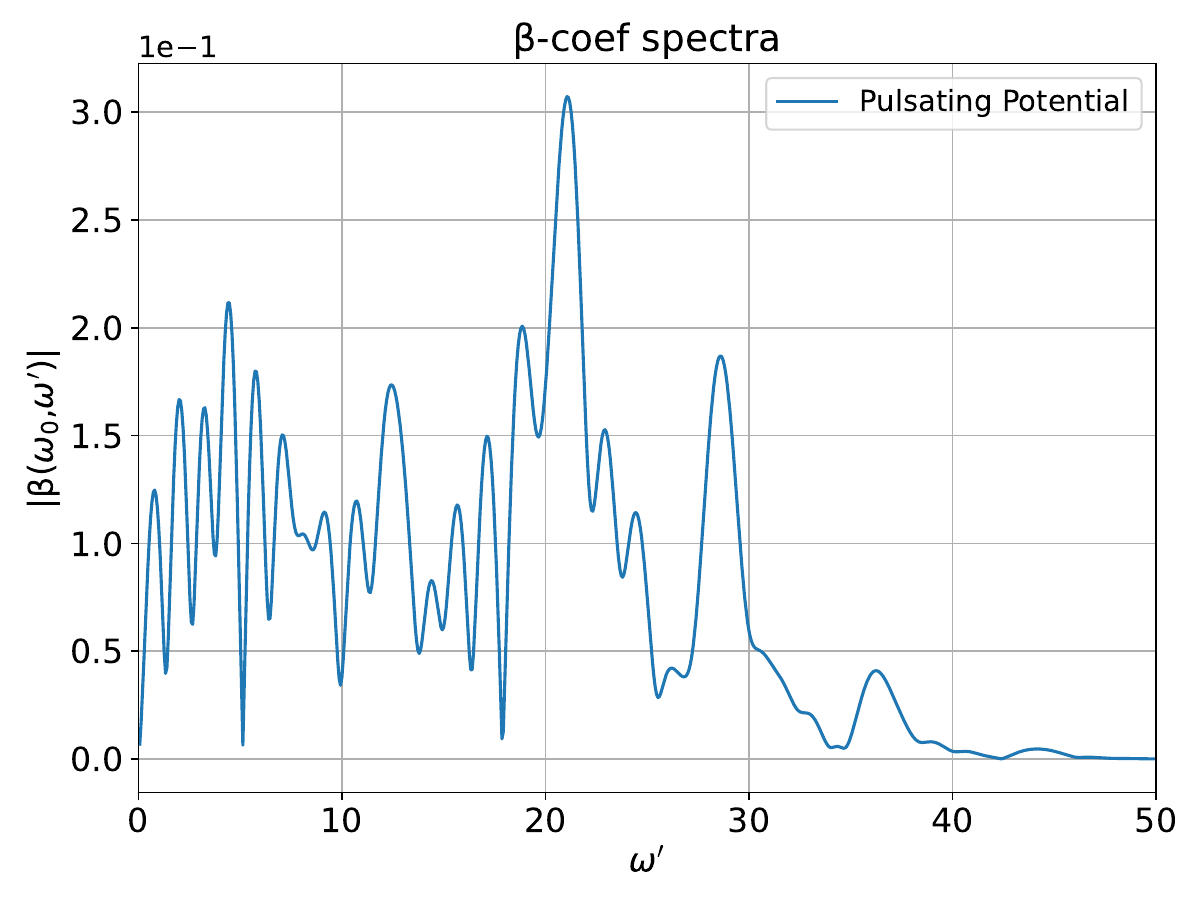}
    \caption{Real part of the scattered signal $\phi_{\omega_0}$ received at $\scri^+$ for a pulsating potential barrier, with initial frequency $\omega_0 = 15$ (top panel). The corresponding Bogoliubov coefficients $\alpha_{\omega_0 \omega'}$ (middle panel) and $\beta_{\omega_0 \omega'}$ (bottom panel) are shown as functions of $\omega'$. The $\beta_{\omega_0 \omega'}$ spectrum is non-negligible, indicating that particle creation occurs. \adr{Both spectra tend to zero as $\omega\to \infty$.}}
    \label{fig:AmpPotBogSpectra}
\end{figure}

The analysis of the corresponding modes is shown in Figure~\ref{fig:AmpPotBogSpectra}, which displays both the real part of the evolved signal at $\scri^+$ and the resulting spectra of the Bogoliubov coefficients. \adr{The unitarity check~(\ref{equation:ImportantFinalSteprewritten}) yields a value of $1.000491$, confirming that  unitarity is preserved in the numerical simulation.}

As can be seen, the $\alpha_{\omega\omega'}$ spectrum is markedly different from those obtained in the stationary scenarios, displaying the excitation of new positive-frequency modes. The presence of peaks of comparable amplitude in the $\beta_{\omega\omega'}$ spectrum clearly indicates particle creation: these correspond to negative-frequency modes spontaneously excited from the vacuum. \adr{We have verified that the Bogoliubov coefficient $\beta_{\omega_0\omega}$ decays for sufficiently large $\omega$, ensuring the convergence of the sum in Eq.~(\ref{totalN}) and thus a finite total number of created particles. This decay behaviour can be seen in Fig.~11 for $\omega_0 = 15$.}

\subsection{Case 4: Shaking potential barrier}

For the shaking potential scenario, most parameters remain unchanged. In this case, however, the potential performs $n = 3$ oscillations with frequency $\omega_p = 8$. The second stage of the simulation is extended to $t_{\text{final}}^{(2)} = 40$, following an initial stage ending at $t_{\text{final}}^{(1)} = 10$, both measured in code time units.

As in the static potential case, part of the signal is reflected while another part enters the potential region. During the dynamical phase, the trapped component is amplified due to the energy injected by the oscillating potential. Once the potential returns to its stationary configuration, the accumulated signal gradually leaks out of the potential region and propagates towards $\scri^+$. In this setup, and in contrast to the previous case, a smaller number of oscillations was deliberately chosen to limit the buildup of energy within the potential during the dynamic phase. Nevertheless, when the oscillations cease, the trapped component requires a longer time to escape, increasing the overall duration of the signal within the computational domain.

\begin{figure}[h!]
    \centering
    \includegraphics[width=0.9\linewidth]{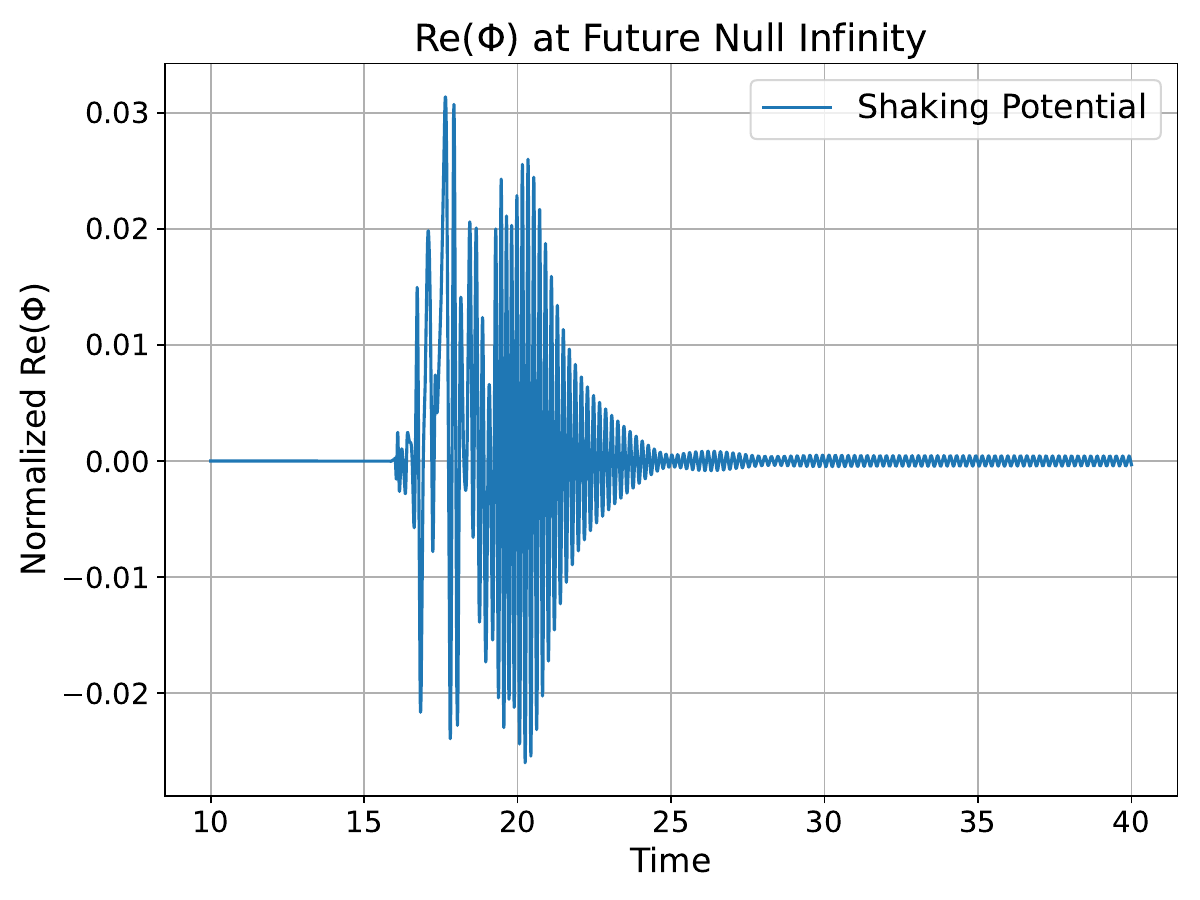}
    \includegraphics[width=0.9\linewidth]{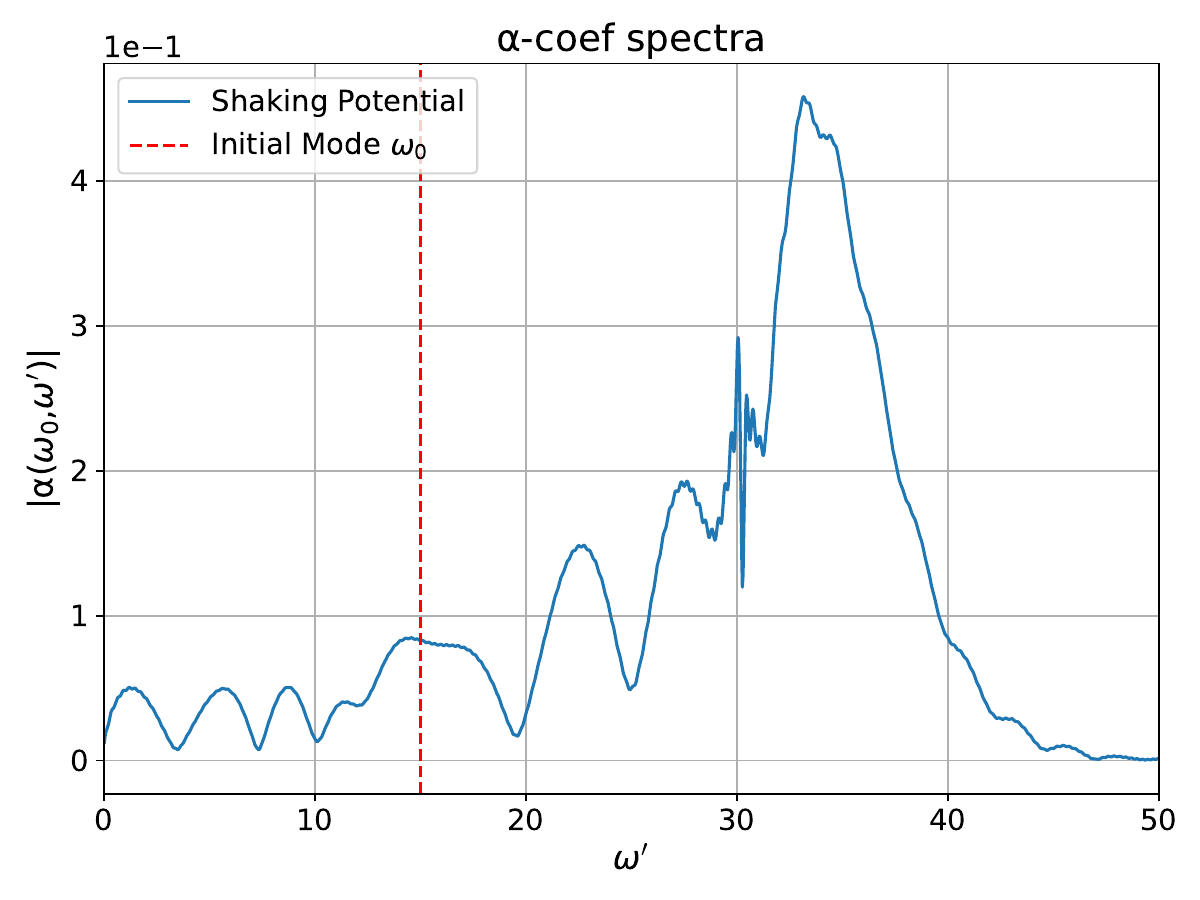}
    \includegraphics[width=0.9\linewidth]{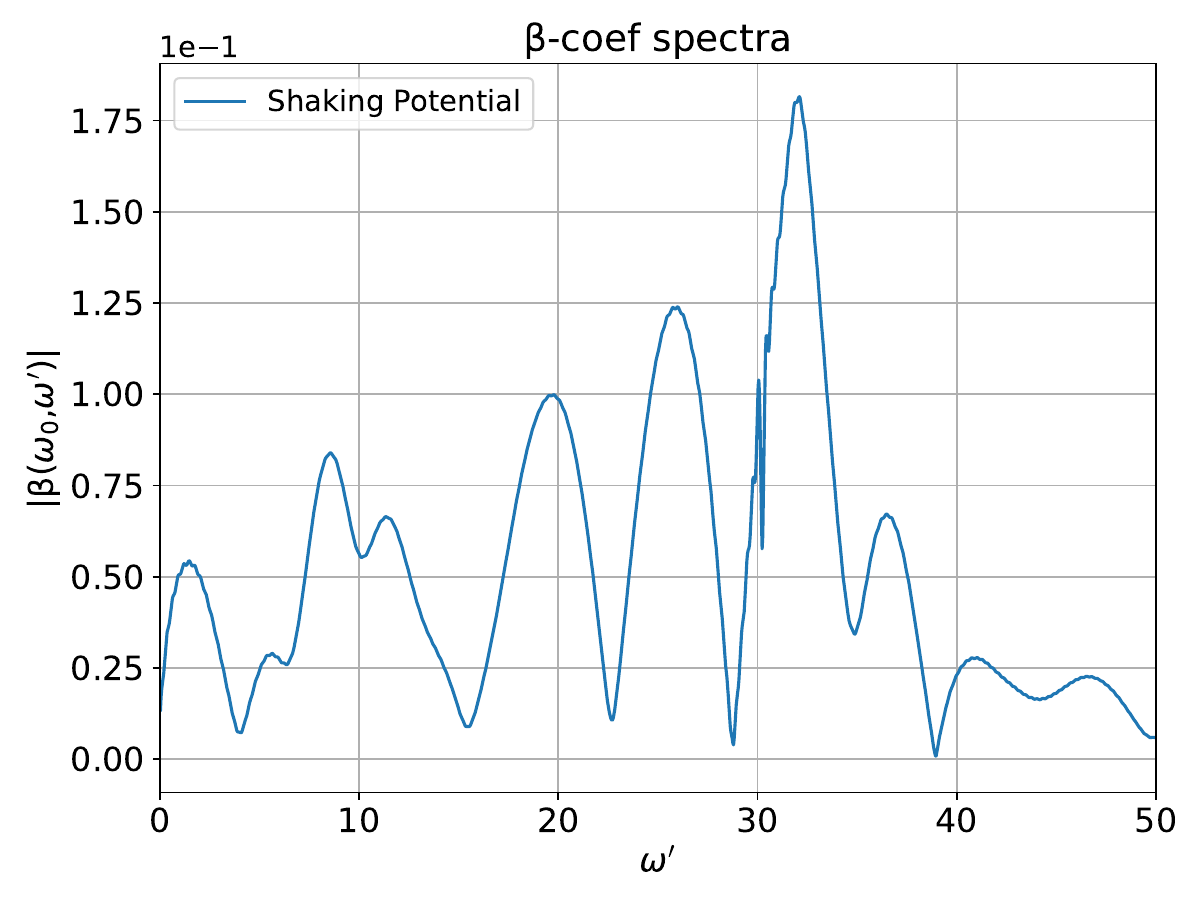}
    \caption{Real part of the scattered signal $\phi_{\omega_0}$ received at $\scri^+$ for a shaking potential barrier, with initial frequency $\omega_0 = 15$ (top panel). The corresponding Bogoliubov coefficients $\alpha_{\omega_0 \omega'}$ (middle panel) and $\beta_{\omega_0 \omega'}$ (bottom panel) are shown as functions of $\omega'$. The $\beta_{\omega_0 \omega'}$ spectrum is non-negligible, indicating that particle creation occurs. \adr{Both spectra tend to zero as $\omega\to \infty$, ensuring that the total particle number will be finite.}}
    \label{fig:CenterPotBogSpectra}
\end{figure}

The resulting Bogoliubov coefficient spectra are displayed in Fig.~\ref{fig:CenterPotBogSpectra}, together with the real part of the evolved signal at $\scri^+$. \adr{The unitarity condition~(\ref{equation:ImportantFinalSteprewritten}) yields a value of $0.983225$, slightly smaller than in previous cases, likely due to part of the signal not yet having fully reached $\scri^+$, but still good enough to trust the numerical simulation}. Both the $\alpha_{\omega\omega'}$ and $\beta_{\omega\omega'}$ spectra exhibit the excitation of new positive- and negative-frequency modes, respectively. The comparable amplitudes of the dominant peaks in both spectra indicate particle creation taking place. \adr{We have verified that the Bogoliubov coefficient $\beta_{\omega_0\omega}$ decays for sufficiently large $\omega$, ensuring the convergence of the sum in Eq.~(\ref{totalN}) and hence a finite total number of created particles. This decay behavior is shown in Fig.~12 for $\omega_0 = 15$.}

For both dynamical scenarios—the pulsating and shaking potentials—we obtain $\alpha_{\omega\omega'}$ and $\beta_{\omega\omega'}$ spectra that differ markedly from the static potential case. The $\alpha_{\omega\omega'}$ spectra reveal multiple positive-frequency modes instead of a single peak around the initial frequency signal, while the $\beta_{\omega\omega'}$ spectra show non-negligible amplitudes of comparable magnitude. These results consistently confirm that particle creation occurs in both dynamical configurations, even though the detailed spectral features differ between them.

\subsection{Convergence tests of the numerical method}
In this section we include pointwise and norm convergence tests for different scenarios to check the convergence of our solutions. The low, medium, high and highest resolution simulations are run with, respectively, $N_{\rm low}=500$, $N_{\rm medium}=1000$, $N_{\rm high}=2000$ and $N_{\rm highest}=4000$ grid points, with the time-steps resized accordingly (divided by a factor of $f=2,2^2$ and $2^3$), i.e., $\Delta t_{low}=0.0025$, $\Delta t_{\rm medium}=0.00125$,$\Delta t_{\rm high}=0.000625$ and $\Delta t_{\rm highest}=0.0003125$. The scenarios analysed are the pulsating and shaking potentials, as they are the most challenging ones. 
The convergence runs were evolved only up to $t=25.00$. Therefore, we did not inspect the full evolution of the shaking potential (which goes on until $t=40$ in our analysis).

The pointwise convergence plots contain 3 curves, each taking into account two resolutions (low-medium, medium-high, high-highest). Each curve represents the difference between the solutions at specific grid points, divided by the corresponding convergence factors. Specifically, the plotted functions are $\left(\frac{u_{\rm low}-u_{\rm medium}}{f^{2n}},\frac{u_{\rm medium}-u_{\rm high}}{f^n},u_{\rm high}-u_{\rm highest}\right)$, which, in a perfect convergence regime, should overlap perfectly.\par
\begin{figure}[h]
    \centering
    \includegraphics[width=0.90\linewidth]{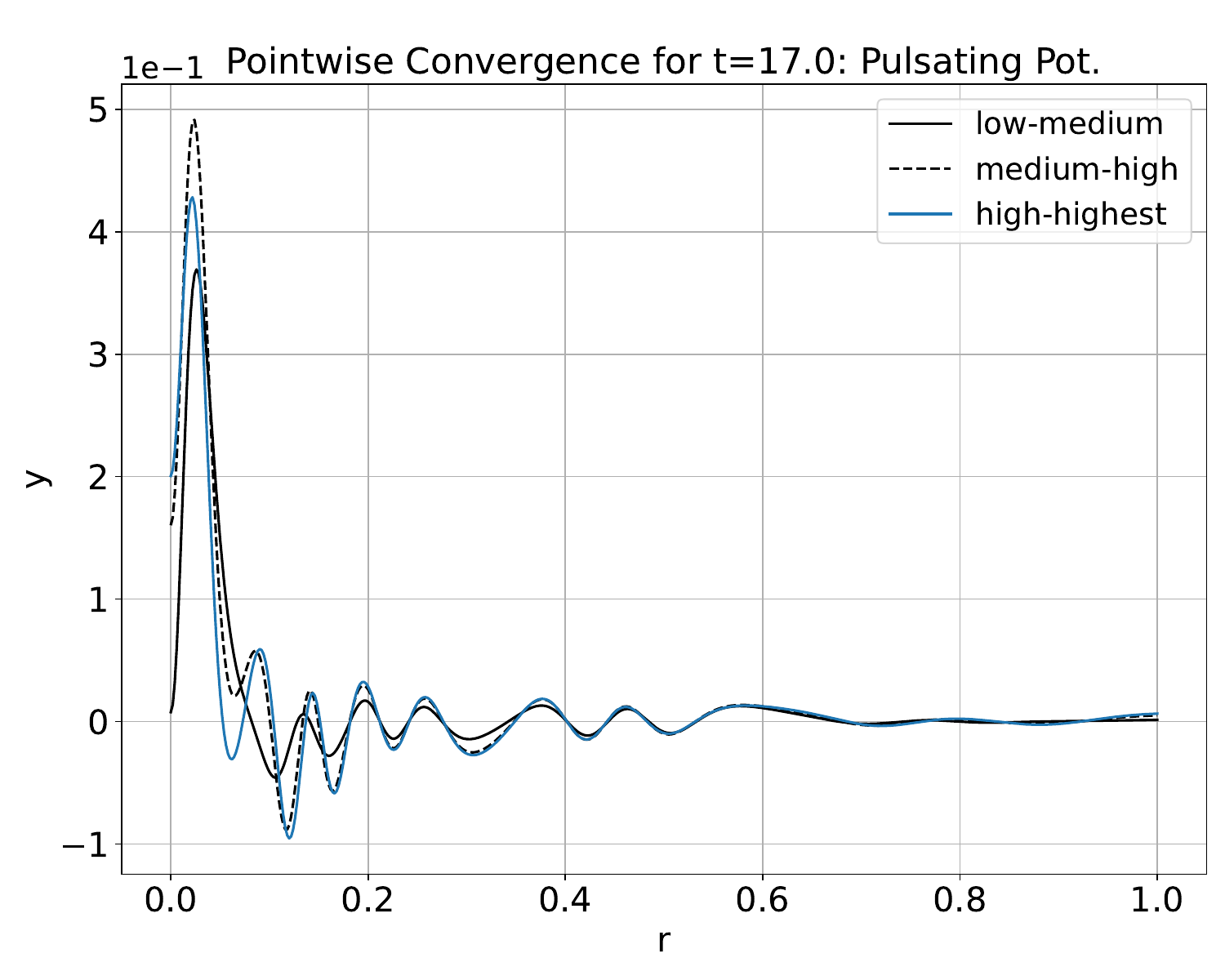}
    \includegraphics[width=0.90\linewidth]{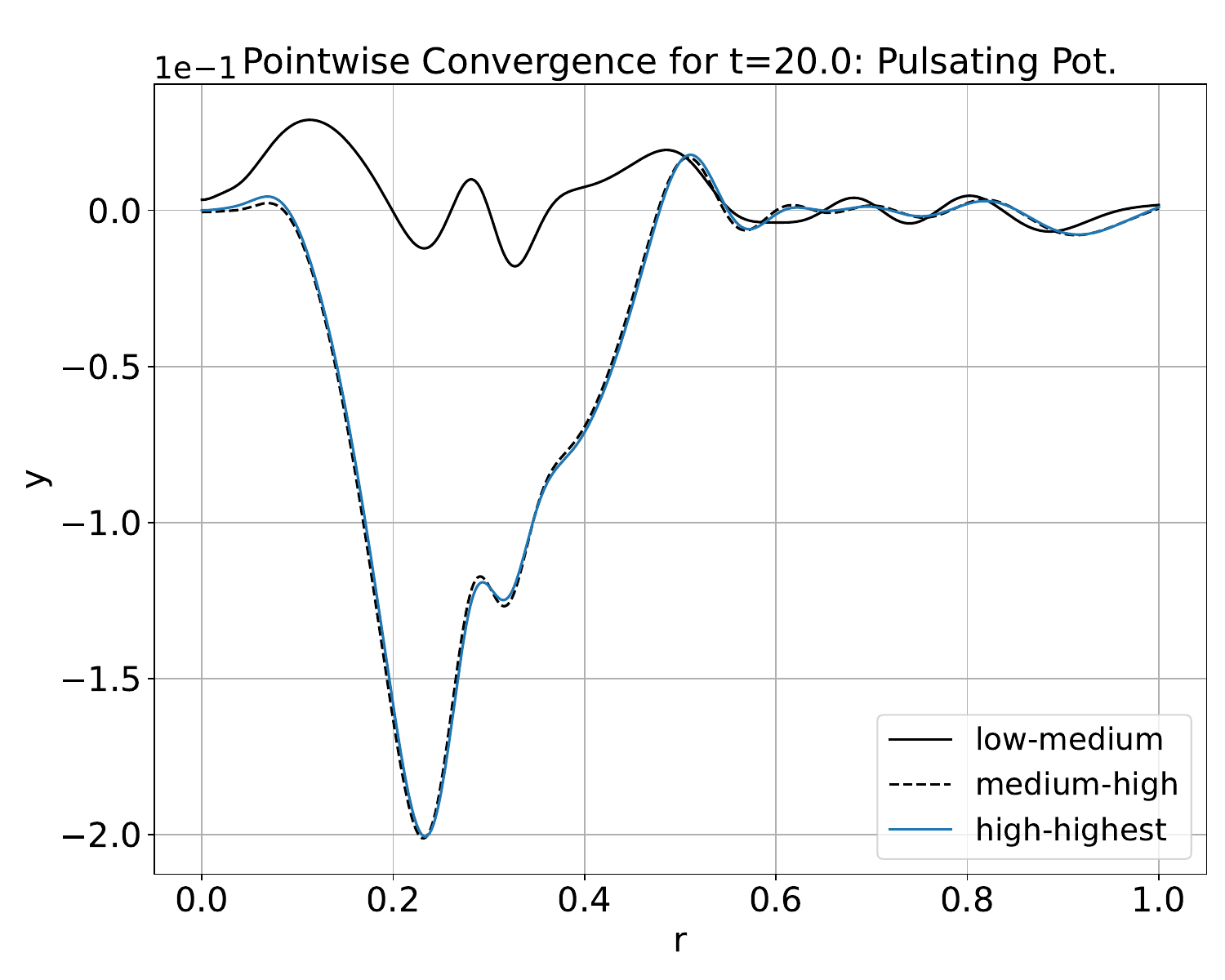}
    \includegraphics[width=0.85\linewidth]{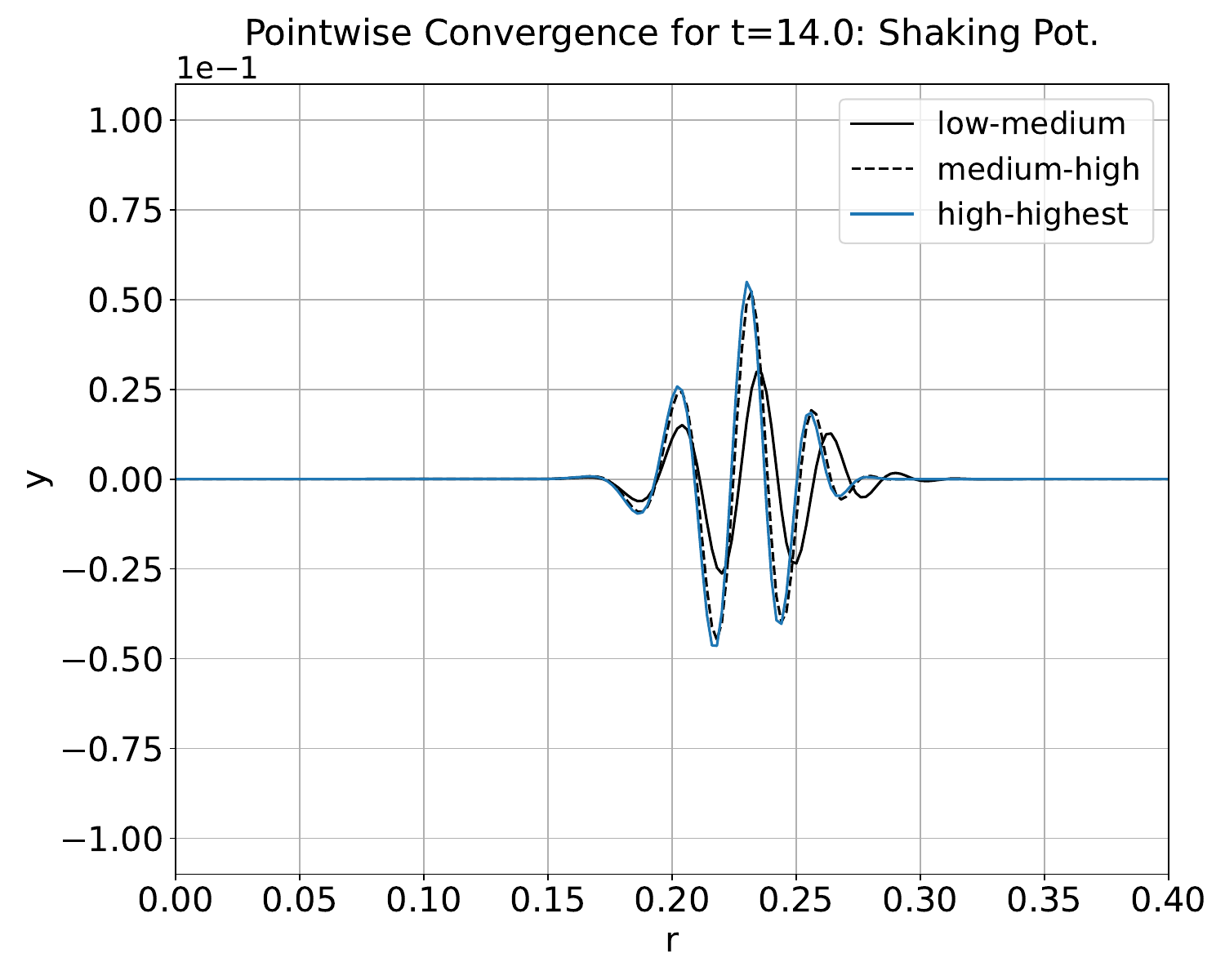}
    \caption{Pointwise convergence tests run for 2 different scenarios: pulsating potential (top and center) and shaking potential (bottom), at different time instants and using 4 different resolutions.}
    \label{fig:PointwiseConvTests}
\end{figure}

Figure~\ref{fig:PointwiseConvTests} illustrates the pointwise convergence tests for the pulsating potential at times $t=17$ and $t=20$ and for the shaking potential at $t=14$. Although only some snapshots of the convergence tests are shown here, we can confidently say that convergence is good throughout most of the first half of the evolution. Since the dynamics occur during the second half, it is particularly important to verify convergence at those later times. From the remaining plots, we observe that the low-medium resolution curve deviates noticeably from the higher resolution ones, suggesting that the low resolution runs are not in the convergence regime. In contrast, the two higher resolution curves generally agree well. Even in regions where they deviate, the curves remain close, which strongly indicates that our solutions converge reasonably. For the shaking potential at $t=14.0$, we observe a slight loss of convergence, likely caused by the translation process. Most problematic regions appear at later times, once the dynamics begin at $t=14.00$. In the pulsating scenario at $t=17.0$, convergence is overall reasonable except in the potential region ($0<r<0.2$) and near the origin. %\avv{This is why I'd like to see the result of convergence with the new l'H\^{o}pitalized RHS at the origin.} 
This phenomenon is present throughout most of the second half of evolution. The plot at $t=20.0$ shows that eventually convergence is recovered at late times.

\par

\begin{figure}[h!]
    \centering
    \includegraphics[width=0.9\linewidth]{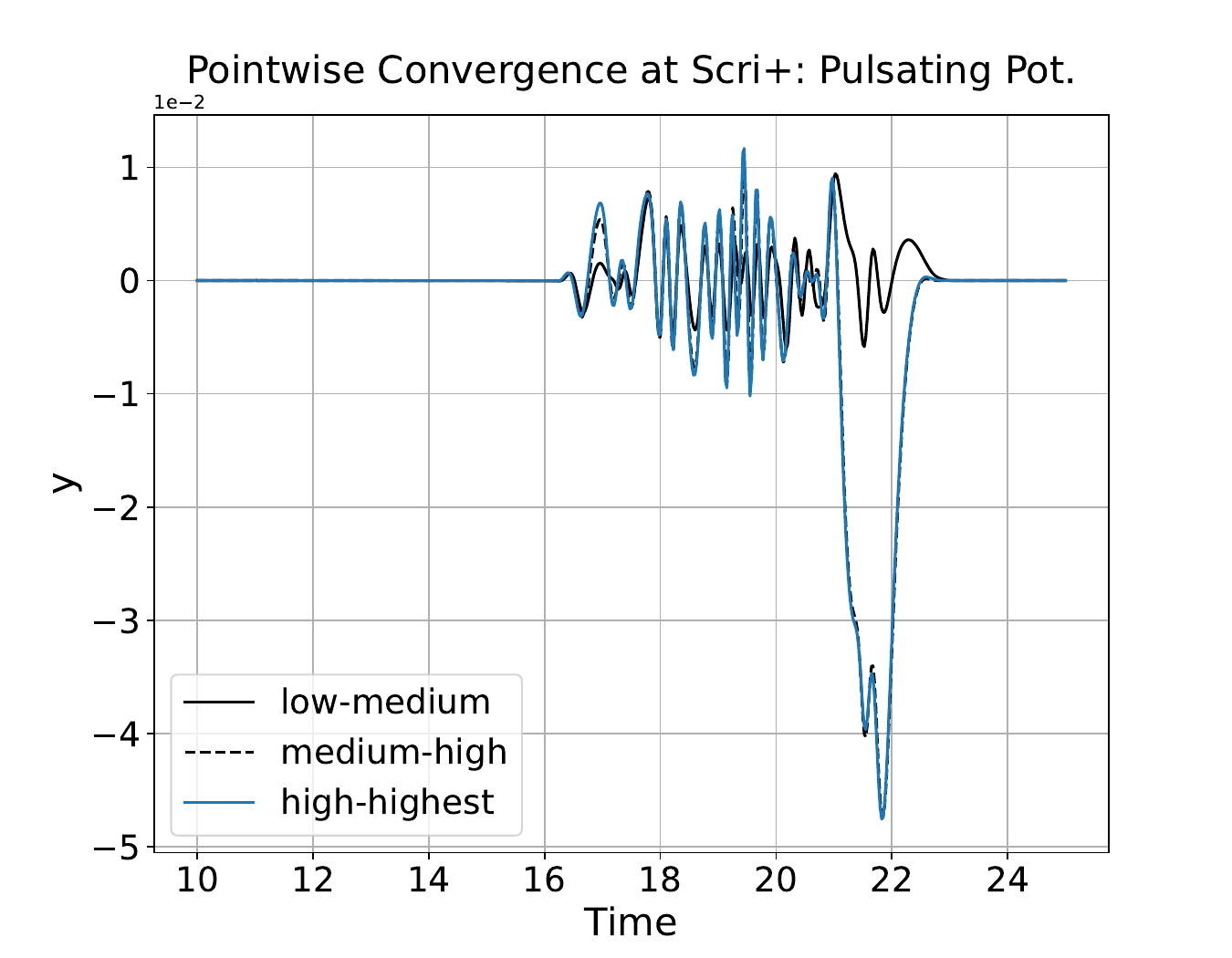}
    \caption{Pointwise convergence test of $Re(\phi_{\omega_0})$ for the pulsating potential scenario at $\scri^+$, perfomed with 4 resolution runs. The black line (low-medium resolutions) is not converging properly which means that the low resolution run is not on the convergence regime.}
    \label{fig:PointwiseConvScri}
\end{figure}

\begin{figure}[h!]
    \centering
    \includegraphics[width=0.92\linewidth]{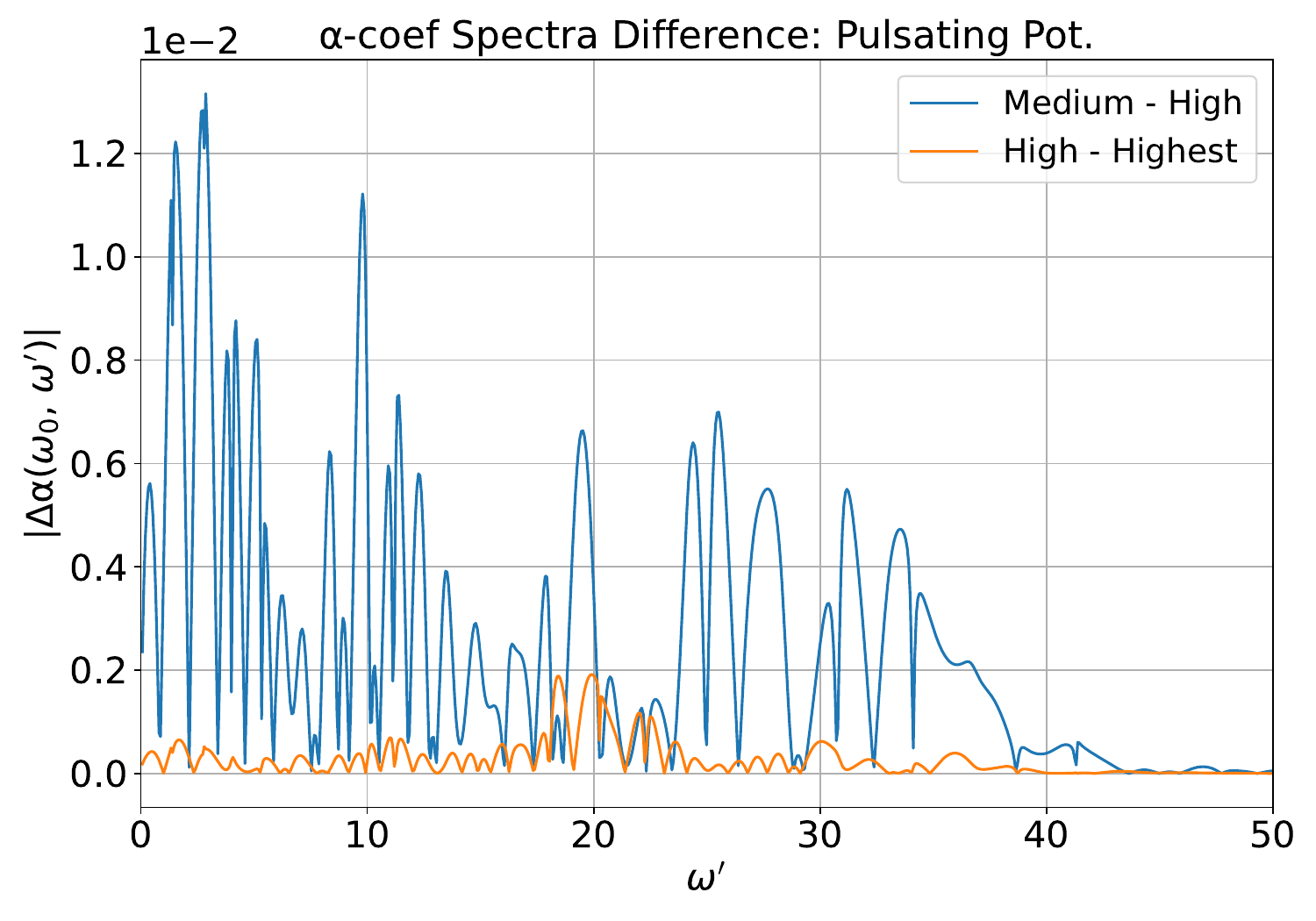}
    \includegraphics[width=0.92\linewidth]{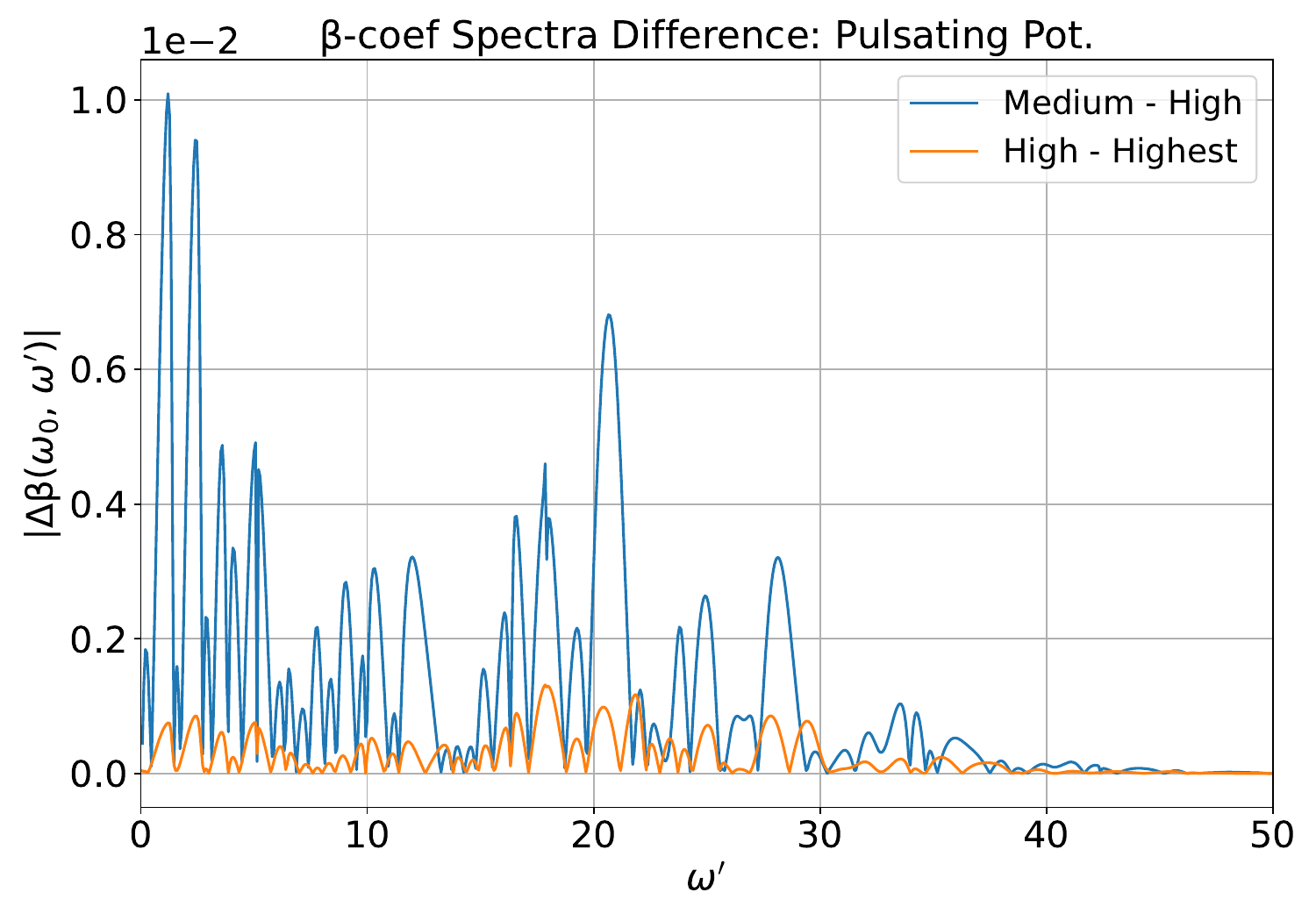}    
    \caption{Absolute value of the differences of the $\alpha_{\omega\omega'}$ and $\beta_{\omega\omega'}$ coefficients with different resolution runs for a pulsating potential scenario. As we increase resolution, the differences in the Bogoliubov coefficients decrease, which indicates the $\alpha_{\omega\omega'}$ and $\beta_{\omega\omega'}$ spectra are converging with increasing resolution and the results are reliable.}
    \label{fig:Bogspectra_ConvTest}
\end{figure}

%Even though convergence is not perfect, these plots show it is overall very reasonable.  
To understand how the signal is converging at $\scri^+$ we also plot the pointwise convergence of Re$(\phi_{\omega_0})$ there for the pulsating potential scenario, as can be seen in Figure \ref{fig:PointwiseConvScri}. Despite the overall good convergence of our evolved signal, it is important to check how the Bogoliubov coefficient analysis posterior to the simulation is affected. Therefore, we plot the differences between resolutions of the $\alpha_{\omega\omega'}$ and $\beta_{\omega\omega'}$ coefficients spectra in Figure \ref{fig:Bogspectra_ConvTest}. For increasing resolution, the differences between the $\alpha_{\omega\omega'}$ and $\beta_{\omega\omega'}$ coefficient spectra become smaller. Low resolution data was excluded, as previous tests demonstrate this regime does not exhibit convergence. These spectra plots shown were obtained exclusively for a pulsating potential. 
This scenario was selected because the shaking potential barrier typically requires longer evolution times for the signal to reach $\scri^+$, making it less suited for detailed convergence studies.
Thus the conclusion is that the signal at $\scri^+$ is indeed converging and so is the Bogoliubov coefficient spectra. We can therefore confidently say that particle creation is taking place.


\subsection{Frequency spectra for different initial frequencies $\omega_0$}

\begin{figure}[h!]
    \centering
    \includegraphics[width=0.9\linewidth]{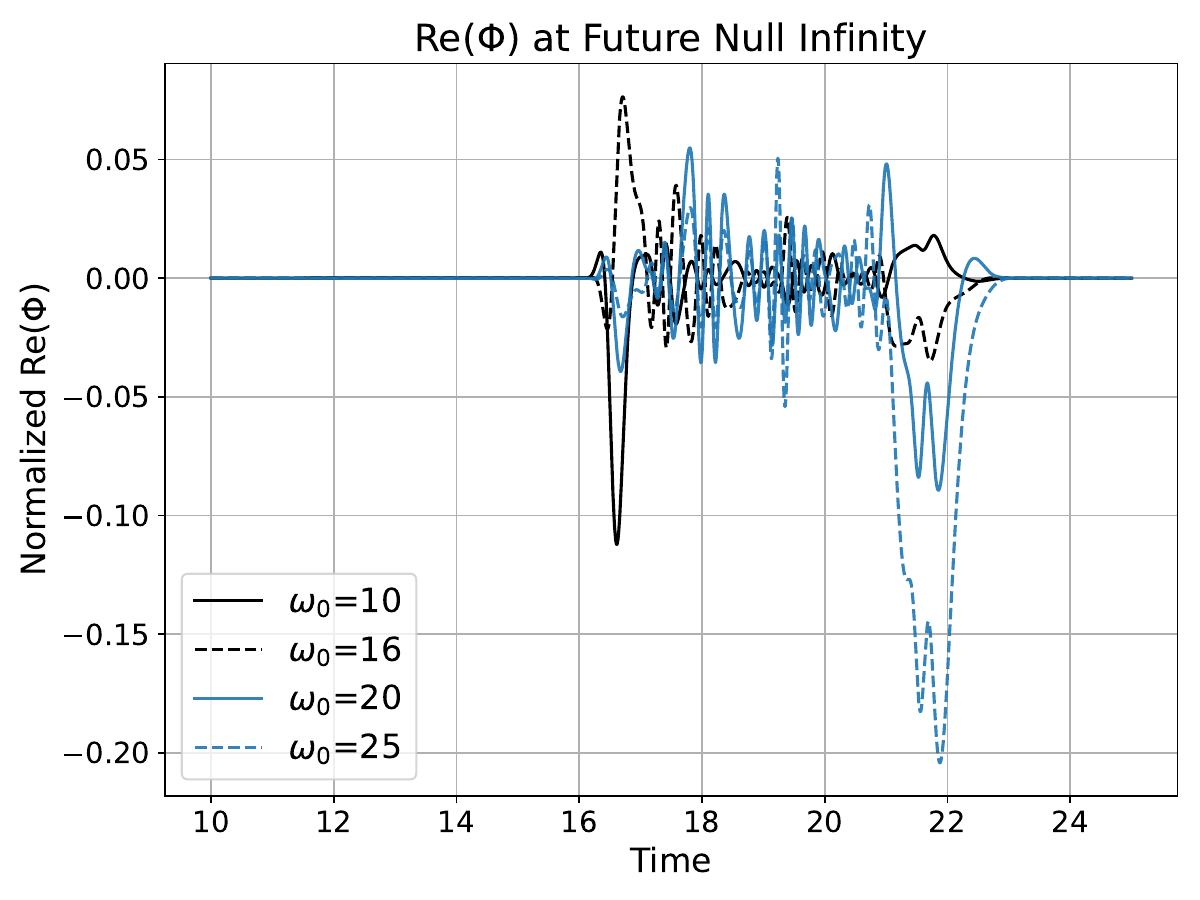}
    \includegraphics[width=0.9\linewidth]{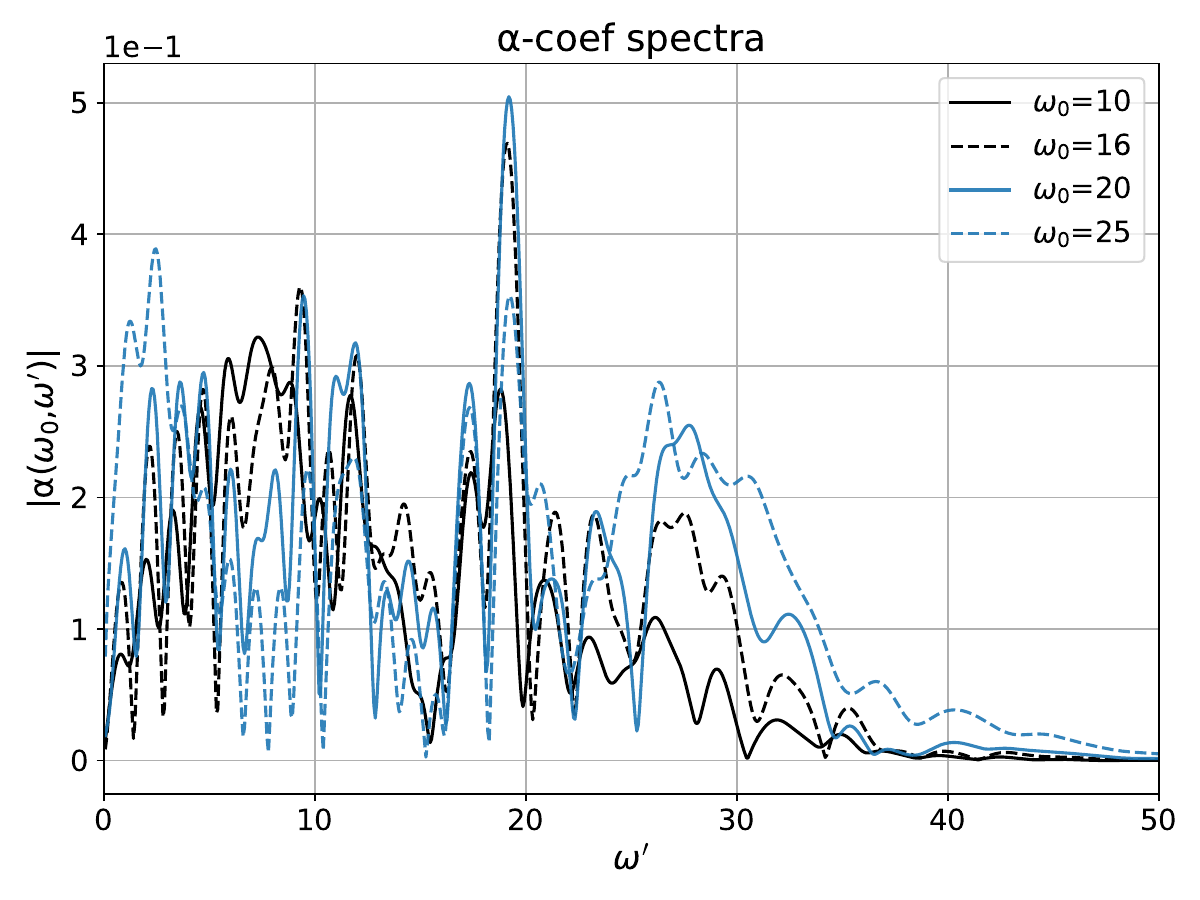}
    \includegraphics[width=0.9\linewidth]{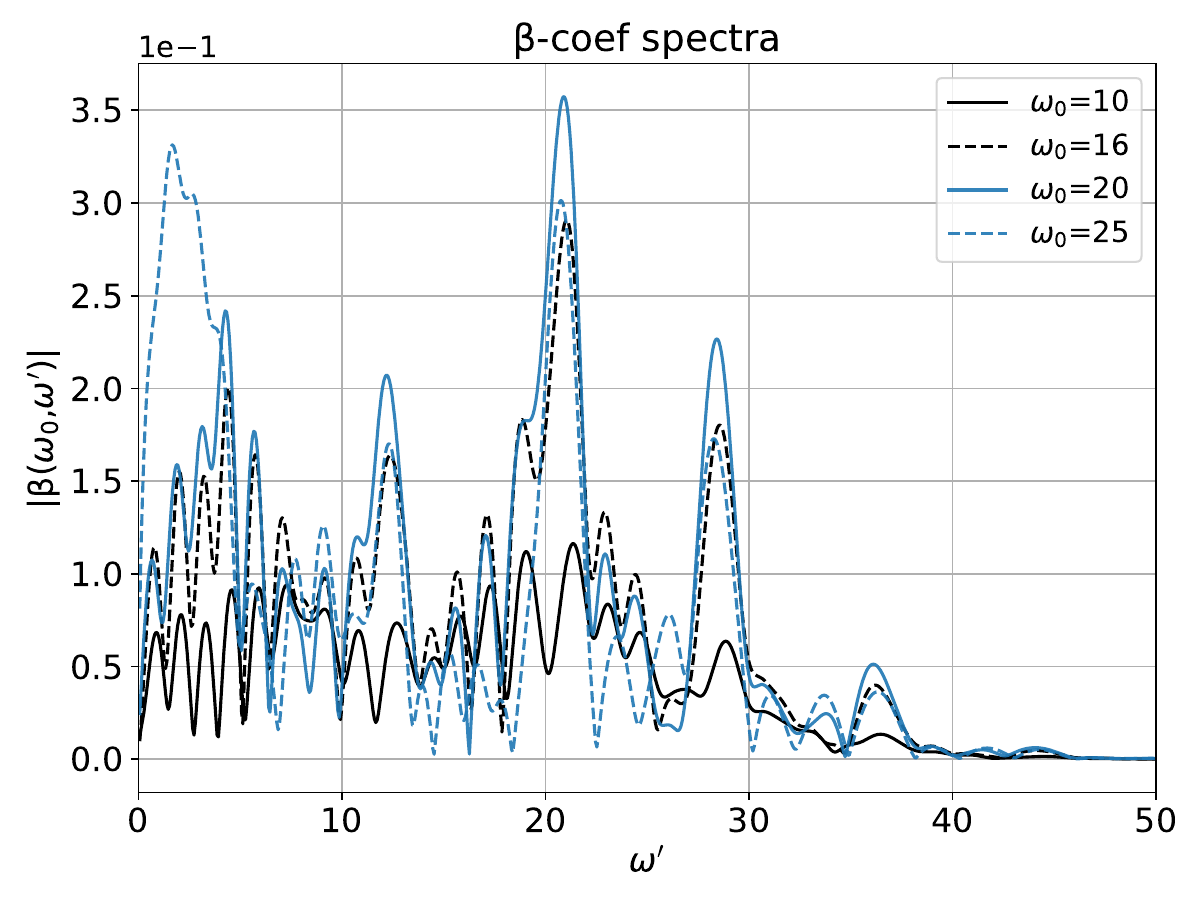}    
    \caption{Real part of the scattered signal $\phi_{\omega_0}$ received at $\scri^+$ for a pulsating potential barrier, obtained with different initial frequencies $\omega_0 \in\{10, 16,20,25\}$ (top panel). The corresponding Bogoliubov coefficients $\alpha_{\omega_0 \omega'}$ (middle panel) and $\beta_{\omega_0 \omega'}$ (bottom panel) are shown as functions of $\omega'$. The $\beta_{\omega_0 \omega'}$ spectrum is non-negligible in all cases, indicating that particle creation occurs. \adr{All the spectra tend to zero quickly as $\omega'\to \infty$, ensuring that the total particle number (\ref{totalN}) will be finite.}}
     %The figure illustrates how particle creation differs for different initial frequencies of the incoming signal.}
    \label{fig:ManyInitialFreqBogSpectra}
\end{figure}

In addition to analyzing the different dynamical scenarios, it is also interesting to investigate how the results depend on the frequency $\omega_0$ of the initial signal~\eqref{Eq:initialPhiPi}. To this end, Fig.~\ref{fig:ManyInitialFreqBogSpectra} shows the Bogoliubov coefficients for several initial frequencies $\omega_0$ in the case of a pulsating potential barrier. While the figure is not straightforward to interpret quantitatively, it illustrates that mode excitation at future null infinity depends sensitively on the frequency of the incoming mode at past null infinity. For example, the line corresponding to $\omega_0=10$ exhibits no particular features, whereas the $\beta_{\omega_0\omega'}$ coefficients for the other choices of $\omega_0$ display clear  resonant peaks at $\omega \approx 12, 21, 28$.

Table~\ref{tab:AlphaBetaConditionFreq} provides a check of the unitarity condition, $\int d\omega' (|\alpha_{\omega_0\omega'}|^2 - |\beta_{\omega_0\omega'}|^2)=1$ for the different initial frequencies considered. This confirms that the total number of particles created, Eq.~\eqref{totalN}, remains finite, and provides a sanity check on the reliability of the numerical computations.

%In addition to studying the different dynamical scenarios, it is of interest to investigate how the results are affected by variations on the frequency of the initial signal~\eqref{Eq:initialPhiPi}. For this particular reason, we show in Figure \ref{fig:ManyInitialFreqBogSpectra} the results for the Bogoliubov coefficients  for different initial frequencies $\omega_0$ for the case of a pulsating potential barrier. Although not easily interpretable, the figure illustrates how the phenomenon of particle creation is indeed different for different frequencies of the incoming signal. While the blue line corresponding to a frequency of the given signal of 10 does not show any specific features, the $\beta_{\omega\omega'}$ coefficients for the other cases clearly show common resonant peaks at $\omega\approx12,21,28$. \adr{Table \ref{tab:AlphaBetaConditionFreq} shows a test on the unitarity condition $\int d\omega' (|\alpha_{\omega_0\omega'}|^2 - |\beta_{\omega_0\omega'}|^2)=1$ for the different initial frequency modes studied. This ensures that the total number of particles created (\ref{totalN}) is finite.} 
\begin{table}[h!]
    \centering
    \begin{tabular}{|c|c|}
        \hline
        $\omega_0$ & $\int d\omega'|\alpha_{\omega_0\omega'}|^2 - |\beta_{\omega_0\omega'}|^2$ \\
        \hline
        10 & 1.0008 \\
        \hline
        16 & 1.0670 \\
        \hline
        20 & 0.9993 \\
        \hline
        25 & 0.9959 \\
        \hline
    \end{tabular}
    \caption{\adr{Test on the unitarity condition $\int d\omega' (|\alpha_{\omega_0\omega'}|^2 - |\beta_{\omega_0\omega'}|^2)=1$  for different initial frequency modes $\omega_0$ for the case of the pulsating potential scenario.}}
    \label{tab:AlphaBetaConditionFreq}
\end{table}

\section{Conclusions and future prospects} 
\label{Section: Conclusion}
In this paper we investigated the phenomenon of particle creation through the numerical evolution of scalar field modes from $\scri^-$ to $\scri^+$, obeying the massless KG equation on a fixed Minkowski background and subject to  effective, time-dependent  potentials. Providing given data through $\scri^-$ and extracting the evolved signal at $\scri^+$ was made possible by the use of hyperbolidal slices in the numerical evolution --- spacelike hypersurfaces that asymptotically reach null infinity. The effect of \adr{a dynamical background on the field modes of the quantum field} is studied through the implementation of a spherically symmetric \adr{effective} potential $V(t,r)$, which is made oscillatory in its amplitude $V_0=V_0(t)$ or location $r_0=r_0(t)$. To {verify the reliability of our numerical} results, we also propagated the scalar field modes with vanishing and static potentials, for which particle creation is not expected. The analysis was performed through the computation of the Bogoliubov coefficients $\alpha_{\omega_0,\omega'}$ and $\beta_{\omega_0,\omega'}$ for several values of the initial frequency $\omega_0$. The shaking potential barrier constitutes the most challenging scenario, as the signal builds up inside the potential and takes a long time to leak out, requiring significantly longer simulations. % with extra storage requirements.\par

\par

The results obtained on particle creation were in line with the {theoretical} expectations. The $\alpha_{\omega\omega'}$  spectra for the stationary scenarios --- vanishing and static potentials --- exhibit only a single peak around the initial frequency of the signal, although slightly deviated. Their $\beta_{\omega\omega'}$ spectra, even though not {identically} zero, are two orders of magnitude below the $\alpha_{\omega\omega'}$ coefficients. The resulting non-zero values are most likely due to numerical and normalization errors and related effects. This became apparent when comparing the spectra for the dynamical scenarios --- pulsating and shaking potentials --- whose normalized $\alpha_{\omega\omega'}$ and $\beta_{\omega\omega'}$ spectra reveal new modes of equivalent order of magnitude. Furthermore, we were able to ensure that the results are reliable by confirming, on each case, that the condition $\int d\omega'(|\alpha_{\omega_0,\omega'}|^2-|\beta_{\omega_0,\omega'}|^2)=1$ holds, with only slight deviations due to numerics.
\par

Pointwise convergence tests indicate reasonably good overall convergence, with noticeable loss being observed mainly in the near potential region ($0<r<2$). The translation process between $\scri^-$ and $\scri^+$ hyperboloidal foliations also induces minor loss of convergence, which is then propagated throughout the rest of the evolution. Because this process involves interpolating and reconstructing the signal, this loss is not unexpected. \avv{So far} only \textit{Neville interpolation} and \textit{B-splines} \avv{have been tested, mainly due to their reliability and ease of use and implementation, so} it remains unclear if any other interpolation algorithm might provide improved convergence behaviour.\par

The toy model entails some challenges that need to be taken in consideration if one wishes to use it on more complex scenarios. First is the obvious translation process {between the ingoing and outgoing hyperboloidal foliations}, as one has to ensure, taken into account the hyperboloidal slices and the width of the initial signal, that the main body of the signal can be recovered with the largest possible recovered grid. This can be challenging when working with spacetimes that scatter the field, as it is uncertain whether the signal can be recovered at a $t_p$ instant before scattering has occurred. Another challenge is the sensitivity of the KG product with the width of the \textit{out} modes' signal, making their normalization procedure more complicated. In this model it is necessary to rescale the product in order to cancel its dependence on the width and, ultimately, ensuring that $\int d\omega'(|\alpha_{\omega_0,\omega'}|^2-|\beta_{\omega_0,\omega'}|^2)=1$ holds. This contrasts sharply with the simplicity of the theoretical calculations where, in principle, ensuring the use of an orthonormal basis for the modes is enough to guarantee reliable results. This disparity is likely due to the impossibility of numerically representing infinite signals and the necessity of working with a compact support scheme.

For further work, one could first investigate ways to make translation between slices smoother and simpler, while also maintaining or improving convergence.  
Very recent work~\cite{penrosetime} proposes an alternative way to translate, which relies on the use of three different hyperboloidal foliations that exactly overlap for specific slices. Including their approach in our setup is a good future plan, as it may provide improved results in terms of accuracy and convergence. Regarding the physical system being simulated, it would be valuable to first reproduce and test this setup on a fixed Schwarzschild background, possibly adding a time-dependent potential to the background to study the dynamical effects on particle creation. Another possibility is to experiment with the mass of the Schwarzschild black hole, making it oscillate harmonically in time for a certain interval. 
%Then subsequently we could extend the setup to a dynamic spacetime corresponding to a gravitational collapse scenario, such as black hole formation. \avv{I think we possibly don't need to go into that much detail, as I recall this has already been mentioned in the intro, right?} In this last case, as mentioned before, evolving the scalar field on a dynamic background would mean solving both the Klein-Gordon equation and the \pbb{Einstein Field Equations} simultaneously. This can prove to be computationally expensive and their coupling might increase the complexity of the setup. However, if the scalar field's influence on the spacetime geometry is negligible, which is a reasonable assumption during gravitational collapse, it might be sufficient to model the dynamic background using extracted data from previous collapse simulations. \avv{I would shorten the bit before here and keep the essence of the following sentence.} 
Then subsequently we could extend the setup by evolving the scalar field on a dynamic spacetime, such as one arising from gravitational collapse. In such cases, if the scalar field’s influence on the geometry is negligible -- a reasonable assumption during collapse -- the background could be modeled using data from existing simulations. It is then essential to ensure that hyperboloidal slices can be constructed to adapt properly to the evolving spacetime geometry.
%In such cases, however, one has to ensure that hyperboloidal slices can be constructed to adapt properly to the evolving spacetime geometry.
\par

%%%%%%%%%%%%%%%%%%%%%%%%%%%%
{\bf \em Acknowledgments.} 
%%%%%%%%%%%%%%%%%%%%%%%%%%%%
ADR acknowledges support through {\it Atraccion de Talento Cesar Nombela} grant No 2023-T1/TEC-29023, funded by Comunidad de Madrid (Spain); as well as   financial support  via the Spanish Grant  PID2023-149560NB-C21, funded by MCIU/AEI/10.13039/501100011033/FEDER, UE.
AVV thanks the Fundac\~ao para a  Ci\^encia e Tecnologia (FCT), Portugal, for the financial support to the Center for Astrophysics and Gravitation (CENTRA/IST/ULisboa) through the Grant Project~No.~UIDB/00099/2020. Funding with DOI 10.54499/DL57/2016/CP1384/CT0090 is also graciously acknowledged. 
This work was supported by the Universitat de les Illes Balears (UIB); the Spanish Agencia Estatal de Investigación grants PID2022-138626NB-I00, RED2022-134204-E, RED2022-134411-T, funded by MICIU/AEI/10.13039/501100011033 and the ERDF/EU; and the Comunitat Autònoma de les Illes Balears through the Conselleria d'Educació i Universitats with funds from the European Union - NextGenerationEU/PRTR-C17.I1 (SINCO2022/6719) and from the European Union - European Regional Development Fund (ERDF) (SINCO2022/18146).

\bibliography{references}

\end{document}